\documentclass[aps,prx,reprint,
superscriptaddress,floatfix,longbibliography,footinbib]{revtex4-1}
\usepackage{mathtools}
\usepackage{amsfonts}
\usepackage{amsmath}
\usepackage{amssymb}
\usepackage{physics}
\usepackage{float}
\usepackage{bbold}
\usepackage[version=4]{mhchem}
\usepackage{hyperref}
\usepackage{xcolor}
\usepackage{lipsum}
\usepackage{xspace} 
\hypersetup{
   colorlinks=true,
   linkcolor=blue,
   filecolor=blue, 
   citecolor=blue,
   urlcolor=blue,
}
\usepackage[doipre={DOI:~}]{uri}

\newcommand{\ex}[1]{{\langle #1 \rangle}}
\newcommand{\pos}{\mathbf{r}}
\newcommand{\xFM}{$x$FM\xspace}
\newcommand{\yFM}{$y$FM\xspace}
\newcommand{\mbf}{\mathbf}
\newcommand{\J}{\mathcal J}
\newcommand{\K}{\mathcal K}
\newcommand{\F}{\mathcal F}
\newcommand{\qx}{q^{xx}}
\newcommand{\qy}{q^{yy}}
\newcommand{\qs}{q^*}
\newcommand{\qb}{q^{i}_{\alpha\beta}}
\newcommand{\Eq}[1]{\langle {#1}\rangle_{q}}
\newcommand{\Eqstar}[1]{\langle {#1}\rangle_{q^*}}
\newcommand{\GM}[1]{{\hat{\Lambda}}^{(#1)}}

\begin{document}

\title{Directly observing replica symmetry breaking in a vector quantum-optical spin glass}

\author{Ronen M.~Kroeze}
\affiliation{Department of Physics, Stanford University, Stanford CA 94305, USA}
\affiliation{E.~L.~Ginzton Laboratory, Stanford University, Stanford, CA 94305, USA}
\author{Brendan P.~Marsh}
\affiliation{E.~L.~Ginzton Laboratory, Stanford University, Stanford, CA 94305, USA}
\affiliation{Department of Applied Physics, Stanford University, Stanford CA 94305, USA}
\author{David Atri Schuller}
\affiliation{E.~L.~Ginzton Laboratory, Stanford University, Stanford, CA 94305, USA}
\affiliation{Department of Applied Physics, Stanford University, Stanford CA 94305, USA}
\author{Henry S.~Hunt}
\affiliation{Department of Physics, Stanford University, Stanford CA 94305, USA}
\affiliation{E.~L.~Ginzton Laboratory, Stanford University, Stanford, CA 94305, USA}
\author{Alexander N.~Bourzutschky}
\affiliation{Department of Physics, Stanford University, Stanford CA 94305, USA}
\affiliation{E.~L.~Ginzton Laboratory, Stanford University, Stanford, CA 94305, USA}
\author{\\Michael Winer}
\affiliation{Joint Quantum Institute, Department of Physics, University of Maryland, College Park, Maryland 20742, USA}
\author{Sarang Gopalakrishnan}
\affiliation{Department of Electrical and Computer Engineering, Princeton University, Princeton NJ 08544, USA}
\author{Jonathan Keeling} 
\affiliation{SUPA, School of Physics and Astronomy, University of St. Andrews, St. Andrews KY16 9SS, United Kingdom}
\author{Benjamin L.~Lev}
\affiliation{Department of Physics, Stanford University, Stanford CA 94305, USA}
\affiliation{E.~L.~Ginzton Laboratory, Stanford University, Stanford, CA 94305, USA}
\affiliation{Department of Applied Physics, Stanford University, Stanford CA 94305, USA}

\date{\today}

\begin{abstract} 

Spin glasses are quintessential examples of complex matter.  Although much about their order remains uncertain, abstract models of them inform, e.g., the classification of combinatorial optimization problems, the magnetic ordering in metals with impurities, and artificial intelligence---where they form a mathematical basis for neural network computing and brain modeling. We demonstrate the ability of an active quantum gas microscope to realize a spin glass of a novel driven-dissipative and vector form.  By microscopically visualizing its glassy spin states, the technique allows us to directly measure replica symmetry breaking and the resulting ultrametric hierarchical structure. Ultrametricity is known to be emergent in models of evolution, protein folding, climate change, and infinite-range equilibrium spin glasses; this work shows it to be directly observable in a physically realized system.  

\end{abstract}

\maketitle

Infinite-range spin glasses are described by the celebrated Parisi solutions to the equilibrium Sherrington--Kirkpatrick (SK) model~\cite{Sherrington1975smo,Stein2013sga,Charbonneau2023sgt}.  This theoretical treatment predicts a rugged free-energy landscape across the high-dimensional spin configuration space. Ergodicity is broken due to quenched disorder and geometric frustration among the spins:  When cooled, exact system copies may relax into distinct regions or states.  These copies, called replicas, can under dynamical evolution result in nonidentical, yet highly correlated, spin configurations (not merely under a global spin flip). Surprisingly, the spin overlaps among replicas exhibit an ultrametric, tree-like structure that emerges below the critical temperature $T_c$ separating the SK spin glass from the paramagnet~\cite{Mezard1987sgt}. This is replica symmetry breaking (RSB) and ultrametricity, the phenomena Giorgio Parisi discovered in the 1980s. Their broad relevance to complex systems earned him the 2021 Nobel Prize~\cite{Parisi2023nlm}. 

We present the results from an experimental system that enables the first visualization of the ultrametric structure of a glassy system in nature, demonstrating that this is indeed a physical property of matter.  We use an \textit{active} quantum gas microscope~\cite{Kroeze2023hcu} to simultaneously create a spin glass and microscopically study its spin structure.  Traditional quantum gas microscopes image atoms directly interacting with each other~\cite{Bakr2009aqg}. Our active version uses a multimode cavity to recycle the imaging light to create photon-mediated, infinite-range interactions among atomic gases trapped inside the cavity.  Multimode cavity emission provides site-resolved imaging of spins, which are motional pseudospin states of the atomic gases. Optical tweezers place the gases at  locations where the cavity field induces glassy states through the interplay of disordered and frustrated spin interactions.  The resulting spin glass is of an unusual vector nature described below.

\begin{figure}[t!]
    \centering\includegraphics[width=\linewidth]{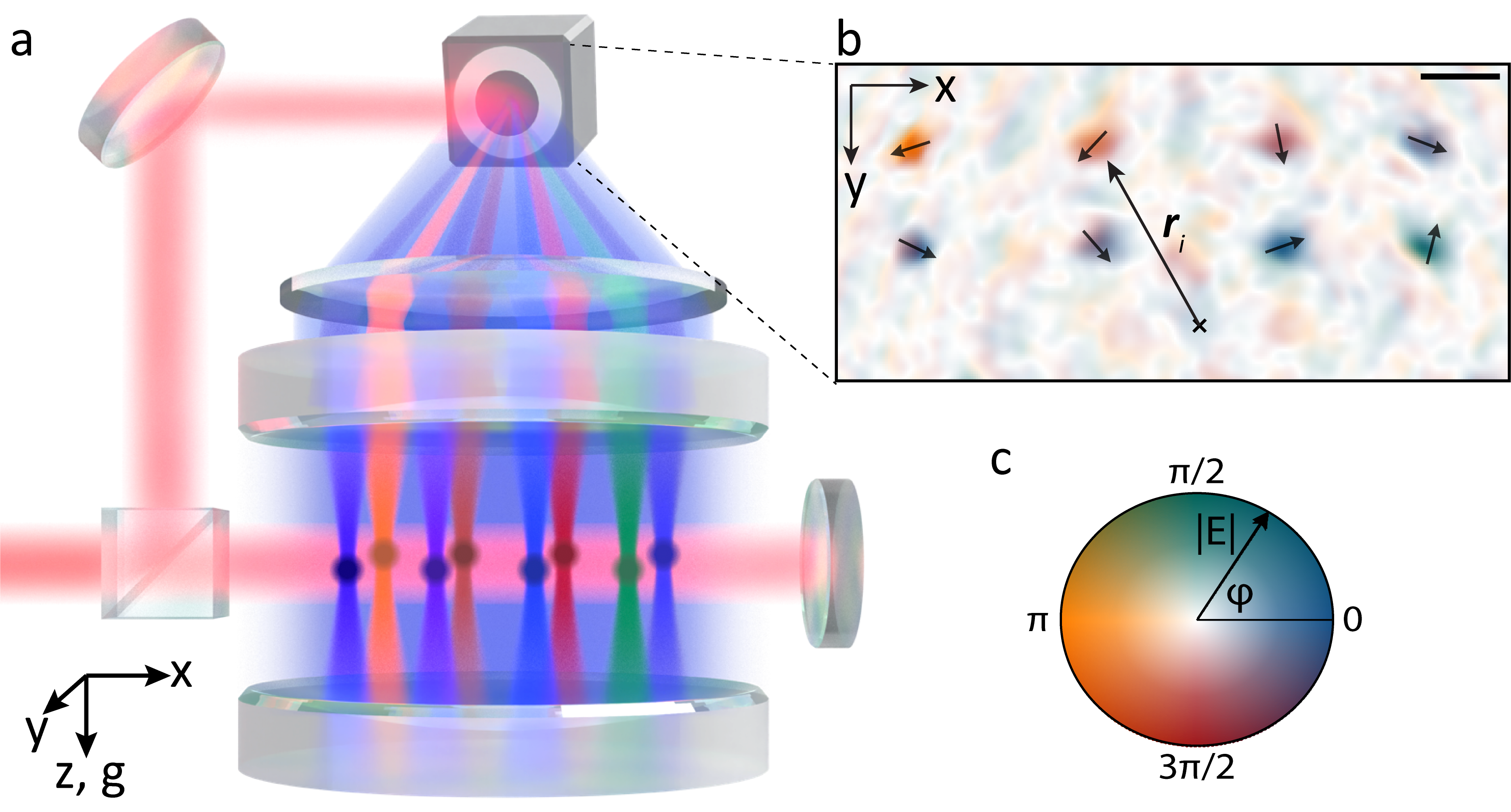}
    \caption{Experimental system and spin detection. (a) Sketch of the confocal multimode cavity QED apparatus. Atomic gases (eight colored balls) at vertex positions $\mathbf{r}_i$ of the network are pumped with a transverse field (red) and scatter light into local (multicolored) and nonlocal (light blue) components of the cavity field.  Neither the mirror-image field components at $-\mathbf{r}_i$ nor the lasers trapping the atomic gases are shown. A portion of the pump mixes with cavity emission and illuminates a camera for holographic imaging; Ref.~\cite{Supp} discusses the analysis required to account for optical aberrations and spin-angle inhomogeneities. (b) Example of a reconstructed hologram of cavity emission for a coupled network of eight atomic gases. The eight bright spots are the local fields coupled to each gas. The arrows here and below indicate the spin vector direction $\theta$ derived from the electric field phase $\varphi$. The scale bar length here and below is $w_0$. (c) Color map for electric fields, where opacity indicates its amplitude $\lvert \text{E} \rvert$ and hue indicates phase $\varphi$. }
    \label{fig1}\vspace{-3mm}
\end{figure}

Continuous photonic driving sustains spin glass states stabilized by dissipation.  This extends spin glass physics beyond the simulation of known equilibrium models by creating an intrinsically driven-dissipative nonequilibrium glass. As discussed in Ref.~\cite{Marsh2024ear}, each real-time spin trajectory obtainable from microscope emission is equivalent to the time evolution of spin states of a system replica. This provides direct access to physical spin correlation distributions among replicas from sets of identically prepared experiments. We measure spin correlations from steady-state images and find that nonthermal replica order parameter distributions arise, adding to the rich nonequilibrium phenomena already found in glasses, such as aging~\cite{Fischer1991sg,Stein2013sga}.

Spin frustration and disorder have been separately realized in controllable systems using trapped ions~\cite{Kim2010qso,Islam2013eaf} and cavity-coupled atoms~\cite{Sauerwein2023ers}. We simultaneously achieve both using multimode cavity QED, as suggested in~\cite{Gopalakrishnan2011fag,Gopalakrishnan2012emo,Vaidya2018tpa,Marsh2021eam,Marsh2024ear} and discussed further in~\cite{Strack2011dqs,Rotondo2015rsb,Erba2021sgp}. RSB has been reported in nonlinear optical devices like random lasers~\cite{ghofraniha2015eeo,gomes2016ool,pierangeli2017oor,Antenucci2015gpd,Niedda2023uco}. However, these devices lack microscopic disorder control and individual spin readout.  In such devices, only indirect measurements of the spin-correlations emerging from RSB are possible, precluding any demonstration of ultrametricity. (Incipient glassiness has also been reported in networks of superconducting circuits~\cite{Harris2018pti,King2023qcd}.)  

Figure~\ref{fig1}a sketches the confocal multimode cavity QED system that forms the microscope; Ref.~\cite{Supp} discusses the apparatus in detail.  Briefly, a pair of mirrors, separated by a distance equal to their radius of curvature, confine many electromagnetic modes at nearly the same frequency. An array of optical dipole traps place $n = 8$ ultracold gases of $^{87}$Rb atoms at different locations $\mathbf{r}_i$ within the cavity midplane $z=0$.  These form the $n$ vertices of the spin network. Each vertex contains $N=2.3(1)\times10^5$ atoms within a radius $\sigma_A\approx4$~$\mu$m centered at $\mathbf{r}_i$.  They are evaporatively cooled to slightly below the Bose-condensation temperature~\cite{Supp}---matter-wave coherence is unimportant to this work. 

A standing-wave transverse pump at $\lambda =780$~nm and oriented along $\hat{x}$ scatters photons off the atoms in each vertex and into a superposition of cavity modes that is peaked at that location~\cite{Vaidya2018tpa,Kroeze2023hcu}. When we linearly increase the pump's power ${\propto}\Omega^2$ through the superradiant threshold of a Hepp-Lieb-Dicke transition~\cite{Mivehvar2021cqw}, the atoms in each vertex $i$ spontaneously break translation symmetry by spatially ordering into a density wave along $\hat{z}$. The atoms choose a spatial phase $\theta_i$ that maximizes the local field amplitude at the vertex's position. However, the field of the confocal cavity is not perfectly localized and has a weaker, long-distance tail that is quasirandom in space~\cite{Vaidya2018tpa,Guo2019spa,Guo2019eab}. When there are two vertices in the cavity, the nonlocal, long-range part of the field due to vertex 1 explicitly breaks the translational symmetry for vertex 2 and vice-versa. This leads to an interaction $J_{12}$ between the symmetry-breaking phases of the two vertices~\cite{Guo2019spa,Guo2019eab}; extending to many vertices directly follows. 

Because the phases $\theta_i$ lie on a circle, we model each as an $XY$-vector spin. The connection to a vector spin model is made by writing the photon-mediated interaction energy in terms of collective spin components $S_i^x=\cos{\theta_i}$ and $S_i^y=\sin{\theta_i}$:
\begin{equation}\label{HamSpin}
E_\text{int}=-\sum\limits_{i,j=1}^n\big[J_{ij}\left(S_i^xS_j^x-S_i^yS_j^y\right) + K_{ij}\left(S_i^xS_j^y+S_i^yS_j^x\right)\big],
\end{equation}
where $J_{ij}\approx J_0\cos{R_{ij}}$, $J_0=N^2g_0^2\Omega^2/(8\Delta_A^2|\Delta_C|)$, and $R_{ij}=2\pos_i\cdot\pos_j/w_0^2$. The $K_{ij}\approx (4\sigma_A^2/w_0^2) J_0 R_{ij}\sin{R_{ij}}$ term arises from the finite extent of each vertex in the imperfect confocal cavity.   The total system energy is that of a transverse-field vector spin model with a transverse field strength proportional to $E_\text{r}$, the atomic recoil energy; see~\cite{Supp} for the full quantum Hamiltonian description.

Equation~\eqref{HamSpin} applies in the far-detuned limit, where the pump is red-detuned by $\Delta_C = -2\pi{\cdot}60$~MHz from the near-degenerate cavity resonances; both cavity and pump are $\Delta_A = -2\pi{\cdot}97.3$~GHz from the atomic resonance. The waist $w_0$ of the cavity's fundamental mode is 35~$\mu$m, and $g_0$ is the single-atom coupling strength to this mode.  The multimode single-atom cooperativity in the dispersive limit is $C_\text{mm} = 110$~\cite{Kroeze2023hcu}. Thus, we can now see how the quantum-optical system acts as an active quantum gas microscope: The intracavity field mediates strong atomic interactions, while the cavity's large effective numerical aperture allows us to holographically image the intracavity field at $z=0$ with a resolution of 1.7~$\mu$m~\cite{Kroeze2023hcu}. This is smaller than the width $\sigma_A$ of each vertex.  Figure~\ref{fig1}b presents an example image taken at a pump power 25\% above threshold, well within the spin-ordered phase.  

We now discuss the nonequilibrium steady-state phase portrait realized by Eq.~\eqref{HamSpin}.  At small $\Omega$, the system is in a normal (i.e., paramagnetic) phase that weakly scatters incoherent light into the cavity.  Superradiant scattering ensues at large pumping strength, and the spins order either as a ferromagnet or spin glass depending on the values of $J_{ij}$.  For $K_{ij}$ sufficiently smaller than $J_{ij}$, as is the case here, Eq.~\eqref{HamSpin} favors an $x$- ($y$-)ferromagnet when all $J_{ij}>0$ (${<}0$) or a spin glass when $J_{ij}$ are disordered. By disorder, we mean $J_{ij}$'s that are both randomly signed and of random magnitude between $\pm J_0/N=\pm2\pi{\cdot}2.0(1)$~kHz.  (This interaction timescale is several-times shorter than the few-hundred-$\mu$s interval between reaching threshold and beginning holographic imaging.)

Ferromagnets differ from paramagnets and spin glasses in their nonzero magnetization order parameter $m^\mu = \frac{1}{n}\sum\limits_{i=1}^n\langle S_i^\mu\rangle$, where $\langle\cdot\rangle$ indicates time average and, for vector spins, we define $\mu \in \{x,y\}$.  Distinguishing between the paramagnet and spin glass requires another order parameter, the Edwards-Anderson spin overlap $q^{\mu\nu}_\text{EA}=\frac{1}{n}\sum\limits_{i=1}^n\langle S_i^{\mu}\rangle\langle S_i^{\nu}\rangle$~\cite{Edwards1975tos}. This overlap is always zero for a paramagnet but is nonzero for the ferromagnet and spin glass due to their frozen spin order. The $q^{\mu\nu}_\text{EA}$ order parameter may be extended to account for overlaps among replicas via $q_{\alpha\beta}^{\mu\nu}=\frac{1}{n}\sum\limits_{i=1}^n\langle S_i^{\alpha,\mu}\rangle\langle S_i^{\beta,\nu}\rangle$, where $\alpha$ and $\beta$ are replica indices~\cite{Charbonneau2023sgt}.  In the following we focus on two linear combinations of this overlap: $Q_{\alpha\beta} \equiv q^{xx}_{\alpha\beta} + q^{yy}_{\alpha\beta}$ and $R_{\alpha\beta} \equiv  q^{yy}_{\alpha\beta}- q^{xx}_{\alpha\beta}$.

Before we present measurements of spin-overlap distributions, we discuss how these overlaps distinguish different phases. For an $XY$-vector ferromagnet, the spin-overlap distribution lies on the boundary of the square spanned by the allowed values of the overlap matrices, $|Q_{\alpha\beta}|\leq Q_{\alpha\alpha}$ and $|R_{\alpha\beta}|\leq |R_{\alpha\alpha}|$.
Both $ Q_{\alpha\alpha}$ and $|R_{\alpha\alpha}| = 1$ near $T=0$ and in the measurements presented below. By contrast, a spin glass that exhibits RSB has nonzero overlaps in the interior of this square. This is the signature manifestation of RSB~\cite{Charbonneau2023sgt}:  Even for identical initial conditions, thermal (or quantum) fluctuations during spin ordering drive replicas into distinct, energetically disconnected regions of the rugged free-energy landscape. This breaks ergodicity, yielding many non-spin-flip-symmetric states that may be accessible to different replicas.  Structure emerges from broken symmetry~\cite{Chaikin1995poc}, and that which arises in spin glasses like that of the SK model is the ultrametric structure of overlap distances $D_{\alpha\beta} = 1-|Q_{\alpha\beta}|$~\cite{Rammal1986ufp}.  In other words, RSB implies that the $D_{\alpha\beta}$ satisfy the strong triangle inequality, $D_{\alpha\gamma} \leq \max\{D_{\alpha\beta},D_{\beta\gamma}\}$, among any triplet of replicas.   

We now discuss the existence of RSB in the aforementioned vector spin model before presenting direct experimental evidence for this. We will then return to discuss differences arising from the driven-dissipative version we experimentally study. First, we perform parallel-tempering Monte Carlo (PTMC) simulations~\cite{Swendsen1986rmc,Hukushima1996emc} based on Eq.~\eqref{HamSpin} that capture the equilibrium properties of the model at varying system sizes. As shown in Ref.~\cite{Supp}, this yields a phase diagram with RSB at low temperature similar to that of the SK spin glass model held in equilibrium. The existence of RSB is further supported by an analytical replica analysis of the equilibrium version of the vector model; see Ref.~\cite{Supp}. We provide the replica symmetric solution and show that it becomes unstable below a critical temperature $T_c$. A recursion relation for $k$-step RSB is derived and numerically solved for $k=1$. This reveals spontaneous RSB at this $T_c$. 

\begin{figure}[t!]
    \centering \includegraphics[width=\linewidth]{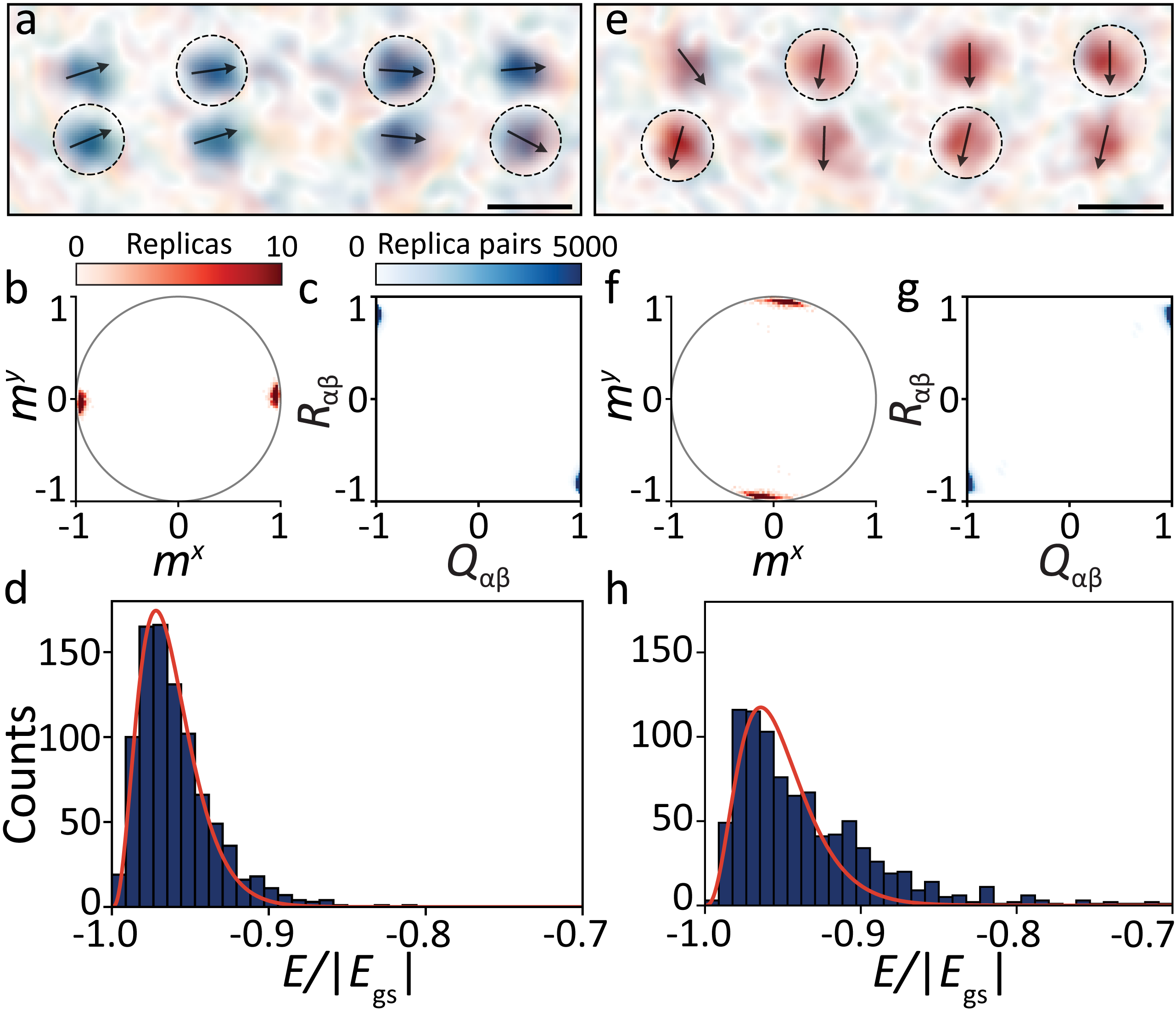}
    \caption{Demonstration of two ferromagnetic states. (a--d) Data in support of an \xFM-state arising from vertices positioned to realize an effective $J_{ij}>0$ network.  (a) Representative reconstructed image showing  a cavity field with vertex phases of a spin configuration ordered along $\hat{x}$.  A local $\pi$ gauge rotation was applied to the vertices indicted by dashed circles~\cite{Supp}.  (b)  Magnetization distribution $m^\mu$.  (c) Distribution of overlap components $Q_{\alpha\beta}$ and $R_{\alpha\beta}$ between pairs of replicas. (d) Histogram of the energy of each \xFM replica, normalized to the magnitude of their ground-state energies. Fitting with a Maxwell-Boltzmann distribution (red line) yields an effective temperature $T/T_c=0.080(8)$.   (e--h) Similar data for an effective $J_{ij}<0$ network realizing a \yFM state. (h) $T/T_c=0.10(1)$.}
    \label{fig2}
\end{figure}

\begin{figure*}[t!]
    \centering
    \includegraphics[width=\textwidth]{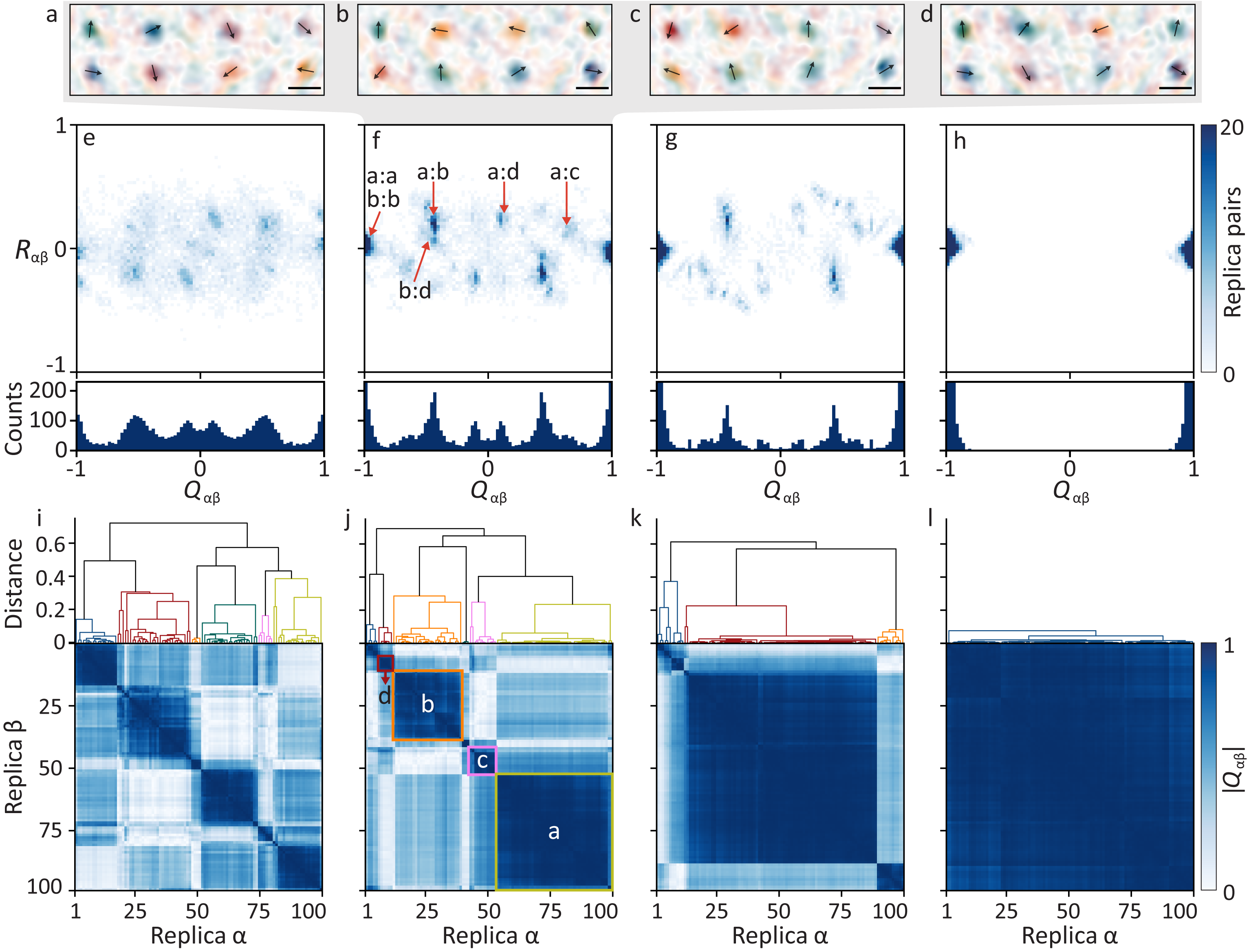}
    \caption{Spin-overlap distributions of the vector spin glass versus ramp rate. (a--d) Reconstructed images of the four most commonly found spin configurations for the same representative disordered $J$. These spin states are found in 45\%, 26\%, 9\%, and 5\% of the 100-total experimental replicas, resp.  (e--h) Overlap distributions versus ramp rate for the same disordered $J$.  The rate of increase of the pump lattice depth is, in $E_\text{r}/\text{ms}$: (e), 1100; (f), 45; (g), 23; and (h), 11.  The mean energies $\langle E_\text{int}\rangle$ of these states, normalized by the ground-state energy $|E_\text{gs}|$, are: (e), -0.75(1); (f), -0.84(1); (g), -0.88(1); and (h), -0.926(4). Panel (f) also indicates the particular replica pairs from panels (a--d) contributing to each major peak. The peaks are denoted $\alpha{:}\beta$ using the panel tags a, b, c, and d as replica labels. Shown below are the 1D marginal distributions; i.e., histograms of the number of replica pairs at each value of $Q_{\alpha\beta}$. (i-l) Hierarchically clustered $|Q_{\alpha\beta}|$ matrices with family-tree-like dendrograms are shown for each ramp rate.  The states in panels (a--d) are contained in the four distinct clusters outlined with squares in panel (j).  The squares are color-coded to match the associated branch of the family-tree-like dendrogram drawn above. The distance of each branch is calculated based on the average of overlaps within the limb below it.  Color-coding is used for families below a distance of 0.4, which corresponds to clusters with spin configurations separated by approximately two spin flips.  }
    \label{fig3}
\end{figure*}

Figures~\ref{fig2}a,e present reconstructed images showing $x$- and $y$-ferromagnets, resp.; we will now denote these as \xFM and \yFM.  Repeating the experiment with identical positions allows us to compile a set of 900 replicas.  We construct the magnetization and spin-overlap distributions shown in Figs.~\ref{fig2}b,c,f,g.  These exhibit the expected near-maximal magnetization and overlap `goalposts' of ferromagnets at $q^{xx}_{\alpha\beta}=\pm1$ ($q^{yy}_{\alpha\beta}=\pm1$) for \xFM (\yFM). The slight tilt of the \yFM magnetization, as well as the non-maximal $q^{yy}_{\alpha\beta}$'s, are caused by the $K$ term~\cite{Supp}.  Figures~\ref{fig2}d,h show energy distributions of the \xFM and \yFM replicas, resp.  While the minimum ferromagnetic energy is rarely found, the effective temperature is well within the ordered phase. (The paramagnet-to-ferromagnet $T_c$ is discussed in Ref.~\cite{Supp}.)  The similarity to an equilibrium distribution is clear despite the likely presence of technical noise and Landau-Zener transitions between nearby states at threshold. Cavity emission above threshold is known to couple the system to an effective thermal bath~\cite{Marsh2021eam,Torre2013kaf}, and the FM's near-thermal behavior might imply that a thermal-like, steady-state equilibrium has been reached. 

\begin{figure*}
    \centering
    \includegraphics[width=\linewidth]{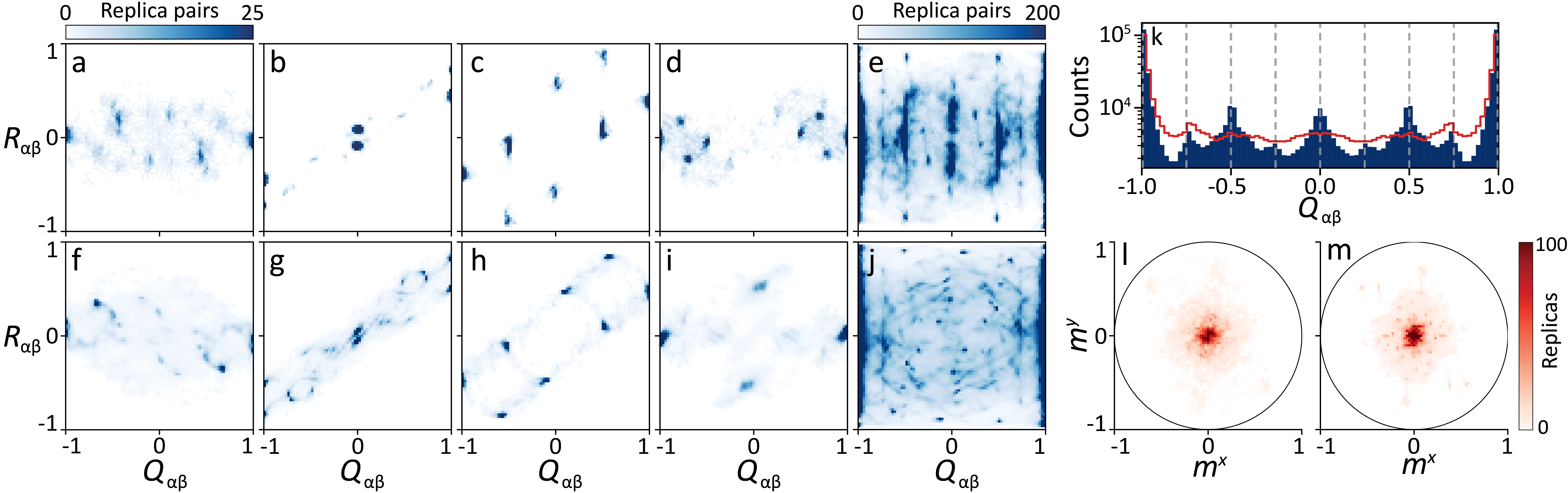}
    \caption{Comparison of experimental and simulated spin overlaps and Parisi order parameters. (a-d) Each of these experimental overlap distributions is realized by moving the vertices to different sets of positions that realize four different disorder realizations $J$. Each of these overlaps are derived from 100 experimental replicas and share the same color scale. (e) The experimental Parisi order parameter distribution formed by averaging 123 overlap distributions, each with a different disordered $J$-matrix. (f--i) Numerical simulations using a mean-field trajectory technique with experimentally measured $J$ and $K$s in panels (a-d). (j) Numerical simulation of the Parisi order parameter distribution using $J$-matrices calculated from observed positions. (k) 1D marginal histogram of the Parisi order parameter distribution in panel (e). Dashed lines indicate allowed overlap values for a binarized (Ising) 8-spin system. The red trace is the 1D marginal histogram from panel (j). (l) Experimental and (m) numerical magnetization distributions averaged over the same experimental ensemble as above.}
    \label{fig4}
\end{figure*}

We now create a spin glass by moving the vertices to realize a disordered $J$.  Figures~\ref{fig3}a--d show four distinct replicas of the same glassy $J$.  These exemplify the multitude of configurations accessible within the glassy rugged landscape.  The spin-overlap distributions in Figs.~\ref{fig3}e--h are compiled from 100  experimental replicas each but are taken with different pump ramp rates; i.e., 
the coefficient relating the pump's power, ${\propto} \Omega^2$, to time. The rate used for panel (f) is the same as that used for the images in panels (a--d), as well as the ferromagnetic data in Fig.~\ref{fig2}.  This ramp rate is sufficiently slow to observe overlap peaks at $Q_{\alpha\beta}= \pm 1$, in addition to multiple $\alpha\neq\beta$ peaks in the distribution's interior. This structure is consistent with the presence of RSB.

A ramp rate increase by more than 24$\times$ blurs these peaks while reducing the overlap goalposts, as seen in Fig.~\ref{fig3}e. Presumably, more Landau-Zener transitions lead to a glass with higher-mean-energy $\langle E_\text{int} \rangle$ that explores more configurations in the rugged landscape. (We have not been able to ramp fast enough to image the paramagnet.) Conversely, slowing the ramp by a factor of two creates sharper, more sparse overlap peaks, as shown in Fig.~\ref{fig3}g. Because we have a finite-sized system, RSB disappears when further slowing the ramp, producing distributions with lower $\langle E_\text{int} \rangle$. As shown in Fig~\ref{fig3}h, the ground-state spin configuration dominates the distribution when $\langle E_\text{int} \rangle$ drops below the first excited state. 

We now provide evidence for nascent ultrametric structure in the $Q_{\alpha\beta}$. Hierarchical clustering is performed to iteratively combine the replicas into groups that minimize the average distance $1 - |Q_{\alpha\beta}|$ within those groups~\cite{Murtagh2011afh}.  This allows the visualization of ultrametric relationships as block-diagonal structures in $|Q_{\alpha\beta}|$~\cite{Rammal1986ufp}. This is shown in Figs.~\ref{fig3}i--l; see Ref.~\cite{Bar-Joseph2001fol} for the leaf sorting algorithm. For example, Fig.~\ref{fig3}i shows at least six clusters. Off-diagonal rectangles of lighter shade provide evidence for multistep RSB structure; the three distinct shades indicate that at least three states contribute to the spin glass order~\cite{Parisi1980aso,Stein2013sga}.  (The clustering hierarchy becomes trivial for the slow-ramp case in panel (l).) 

Ultrametric structure may also be visualized as a family-tree-like dendrogram~\cite{Rammal1986ufp,Stein2013sga}. These are shown above the $|Q_{\alpha\beta}|$ matrices. Replicas are the leaves at the lowest generation level.  The overlap between two replicas is approximately the value of the distance of the topmost branch connecting them. Limbs are colored to denote near-kin families, which are the emergent clusters of $|Q_{\alpha\beta}|$. A quantitative analysis based on the strong triangle inequality~\cite{Katzgraber2009uac} is presented in~\cite{Supp}.

Figures~\ref{fig4}b--d show overlap distributions for three additional realizations of disordered $J$-matrices. They differ from both each other and the overlap in Fig.~\ref{fig3}f (reproduced in panel~\ref{fig4}a), demonstrating the strong dependence of spin overlaps on disorder realization. This is expected of a glass, where the energy gaps and rugged-landscape topography can sensitively depend on disorder details. 

While the purpose of the microscope is not to simulate the equilibrium SK model, it is instructive to compare the experimental overlaps with those of the intrinsically equilibrium-based PTMC simulations of the SK model.  We find that there is no simulated temperature at which all experimentally found features of RSB also appear in the PTMC simulations, presumably because it cannot capture the experiment's driven-dissipative dynamics; see Ref.~\cite{Supp}. While all peaks are reproduced at high temperature, they are blurred due to thermal fluctuations.  Moreover, overlap peaks are sharper at low temperature and some are missing because they arise from higher-energy configurations.  

Seeking closer theory-experiment correspondence, we now compare our experimental overlap distributions in Figs.~\ref{fig4}a--d to a different simulation method, one that accounts for the driven-dissipative effect of the superradiant phase transition.  That is, we numerically calculate the dynamics using a stochastic Lindblad master equation under a mean-field approximation that uses the same pump ramp schedule as the experiment~\cite{Supp}. These trajectory simulations employ the $J$ and $K$-matrices derived from experimental observations for each disorder realization, as explained in~\cite{Supp}. The results appear in panels (f--h); some are in qualitative agreement with the experimental ensemble, but others, such as in panel (i), are far off. These simulations also do not fully capture the experimental spin glass overlaps.  Noise in the coupling matrices does not qualitatively change this assessment~\cite{Supp}. Evidently, future theoretical work is needed to better understand the novel driven-dissipative spin glass realized by the active quantum gas microscope.

Specific disorder details cease to play a role in the large system-size limit of non-glassy systems.  By contrast, the values of $q_{\alpha\beta}^{\mu\nu}$ are different for different disorder realizations of $J$, implying a lack of self-averaging.  A quantity that is independent of microscopic disorder details appears if one takes the disorder average $\langle q^{\alpha\beta}_{\mu\nu}\rangle_J$ over all $J$ realizations. This is the Parisi order parameter~\cite{Elderfield1984tpo,deAlmeida1978tis,Toulouse1981mft,Charbonneau2023sgt}, which exhibits a smooth distribution of $Q_{\alpha\beta}$ between the  $\pm Q_{\alpha\alpha}$ goalposts and whose shape would be determined by temperature via the Gibbs measure for equilibrium spin glasses~\cite{Stein2013sga}.

To measure the Parisi order parameter, we average the experimental overlap distributions in Fig.~\ref{fig4}a--d, plus 119 others (each compiled from 100 experimental replicas).  The result is shown in Fig.~\ref{fig4}e; see~\cite{Supp} for assessment of ensemble randomness and convergence. The 1D marginal distribution in Fig.~\ref{fig4}k exhibits continuous support between the overlap goalposts, similar to that known from solutions to the equilibrium SK Ising spin glass~\cite{Stein2013sga}~\cite{Supp}. However, we observe several peaks between the goalposts rather than a smooth distribution.  Overlaid on the plot are $n+1=9$ equally spaced vertical dashed lines set at the allowed overlap-value locations that would be found in an Ising network of $n=8$ vertices. That the peaks in the data match these positions suggests that the vector spins are weakly binarized along the $Q$ quadrature. 

The Parisi order parameter from the mean-field trajectory simulations is shown in Fig.~\ref{fig4}j. It reproduces some of this striped behavior, as is clear from the marginal shown in red in Fig.~\ref{fig4}k. The PTMC simulation provided in Ref.~\cite{Supp} is even less similar to the experimental data. As with the overlap distributions, it seems that neither an equilibrium (PTMC) nor a nonequilibrium (mean-field trajectory) simulation method faithfully reproduces all details of the experimental Parisi order parameter distribution:  Either the inherently nonequilibrium or the not-quite-mean-field nature of the driven-dissipative experimental spin glass cannot yet be theoretically captured. We note that an unidentified easy-axis anisotropy or nonlinearities arising during the imaging sequence might also play roles~\cite{Supp}. 

A spin glass should also have zero average magnetization.  The experimental disorder-averaged vector magnetization is shown in Fig.~\ref{fig4}l, and it is centered around $\mathbf{m}=0$ as expected.  In contrast to overlaps, simulations similar to those performed above do capture the magnetization distribution width; see Fig.~\ref{fig4}m.

That clear signatures of RSB and ultrametricity can be experimentally observed in a frustrated spin network as small as $n=8$ is surprising.  This fact is supported by finite-size scaling simulations of both equilibrium and nonequilibrium models; see Ref.~\cite{Supp}.  Nevertheless, the active quantum gas microscope will not be limited to this system size:  Optical tweezers can now realize arrays of thousands of atom traps~\cite{Manetsch2024ata} and incorporating this technology should enlarge the spin glass by orders of magnitude. Sending tweezer light directly through the multimode cavity mirrors will allow individual spin sites to be addressed.  This will enable the recall of associative memories stored in the spin glass~\cite{Hopfield1982nna,Hopfield1986cwn,amit1989mbf,Hertz1991itt,Bahri2020smo,Guo2021aol,Marsh2021eam}, or the local addressing of spins for microscopically imaging the spatiotemporal response of spins and their domains, a capability not yet experimentally achieved.

Other types of driven-dissipative spin glasses can be realized using the active quantum gas microscope.  Non-confocal multimode cavities~\cite{Guo2019eab} can realize Ising spin models, and short-range, RKKY-like interactions~\cite{Ruderman1954iec} are possible through a marginally more sophisticated transverse pumping scheme~\cite{KroezeThesis23}.  The latter will enable the experimental exploration of non-mean-field spin glasses, where the theoretical determination of whether some form of RSB describes their order remains an important unresolved problem in statistical physics~\cite{Stein2013sga,Contucci2013pos}. Finally, inducing Rydberg blockade~\cite{Peyronel2012qno,Jia2018asi} in each of the spin network vertices can realize effective spin-1/2 degrees of freedom while preserving high light-matter coupling strength~\cite{Marsh2024ear,Marsh2024rydberg}.  Glassy quantum spin dynamics would then be accessible to microscopic experimental observation
and would provide access to novel quantum states and dynamics~\cite{Gopalakrishnan2011fag,Strack2011dqs,Marsh2024ear,Hosseinabadi2024qci,Tikhanovskaya2024edo,Lang2024rsb} that may find application in quantum neuromorphic computational devices~\cite{Barahona1982otc,Lucas2014ifo}.

We thank Surya Ganguli and Helmut Katzgraber for discussions and Zhendong Zhang for experimental assistance.  We are grateful for funding support from the Army Research Office (Grant \#W911NF2210261), NTT Research, and the Q-NEXT DOE National Quantum Information Science Research Center. J.K. and S.G. were supported in part by the International Centre for Theoretical Sciences (ICTS) through participating in the program, Periodically and quasi-periodically driven complex systems  (code: ICTS/pdcs2023/6). B.M.~acknowledges funding from the Stanford QFARM Initiative and the NSF Graduate Research Fellowship. A.B.~and M.W.~thank the Stanford Graduate Fellowship and the Joint Quantum Institute Fellowship for support, resp. A portion of the computing for this project was performed using the Stanford Sherlock cluster.
Primary data generated in the current study is publicly available in the Harvard Dataverse Repository \doi{10.7910/DVN/JFKKFQ}.

\clearpage
\onecolumngrid
\section*{Supplementary Information}
\tableofcontents

\renewcommand{\theequation}{S\arabic{equation}}
\renewcommand{\thefigure}{S\arabic{figure}}
\setcounter{figure}{0}
\setcounter{equation}{0} 

\section{Experimental methods}\label{sec:methods}

\subsection{BEC preparation}\label{sec:bec}
Atom cooling and trapping follows Refs.~\cite{Kollar2015aac,Guo2019spa} with additional steps taken to produce eight atomic clouds that serve as the spin network vertices. We employ time-multiplexed RF signals produced by voltage-controlled oscillators to drive acousto-optical modulators to dynamically shape the trap. Specifically, we begin by creating a cloud of $2.5(2)\times10^6$ ${}^{87}$Rb atoms at a temperature of $548(8)$~nK held in two dithered, crossed optical dipole traps. Subsequently, the dithered drive signals are adiabatically deformed into a step-ladder waveform with 2 and~4 steps, resp., resulting in the initial atomic cloud being split into a $2{\times}4$ grid of smaller, well-separated clouds. The amplitudes of the steps in the dither drive determine the vertex positions, while the duty cycle for each step controls the relative atomic population per vertex. The population in each resulting gas cloud is balanced to contain ${\sim}2.3(1)\times10^5$ atoms.   Each has a temperature of $440(60)$~nK measured with time-of-flight imaging. This temperature is near the critical temperature for Bose-Einstein condensation (BEC), resulting in a BEC fraction of ${\sim}11\%$. The relatively large thermal fraction causes a shift of the superradiant phase transition point~\cite{Piazza2013bcv}, but does not further affect superradiance. Ensembles of atoms are used to realize a collective coupling large enough to yield spin dynamics faster than heating. Improvements to the cavity mirror coatings will allow us to study vertices with fewer atoms in each, resulting in larger network sizes.  Additionally, Rydberg dressing can be employed to render each vertex a quantum spin-1/2 degree of freedom while maintaining similarly rapid dynamics~\cite{Marsh2024ear,Marsh2024rydberg}.

Some calibration experiments require isolating a single vertex, such as measuring trap frequencies and atomic shape, or for imaging the nonlocal field from a signal vertex. This is accomplished by first preparing the full network of eight vertices, then switching the RF drives to a single tone, instead of the time-multiplexed tones. This allows us to remove the trapping potential at all vertices but the site of interest. We then wait 400~ms to let the atoms in the other vertices fall under the force of gravity. The time-multiplexed RF drive is then reactivated to recover the precise original trap shape, but with only a single vertex populated.

Trap frequencies for each vertex are measured by first isolating the vertex of interest according to the above procedure. The atoms in the trap are then weakly excited to stimulate a sloshing mode. Momentum oscillations from time-of-flight imaging reveal the trap frequency. We observe vertex-to-vertex dependence of the trapping frequencies.  This is caused by the way we create the traps by dithering the trapping beams between spatial locations.  The beam spends more time in some locations than others to balance the atom populations across the vertices. However, this also changes the trap frequencies and depths: Typical frequencies are $[\omega_x,\omega_y,\omega_z]=2\pi\times[296(10), 170(27), 170(8)]$~Hz, where the standard deviation is assessed over all eight sites and two different position configurations. Error below is also given as standard deviation.

The atomic shape at each vertex is described by a bimodal distribution. Thermal atoms contribute a Gaussian density component while the BEC fraction of the gas contributes a Thomas-Fermi component. To simplify calculations that involve the density profile, we approximate the bimodal distribution in the cavity transverse plane by an isotropic 2D Gaussian of 4-$\mu$m standard deviation along the trap directions.  This width is determined by a least-squares fit to the full 2D bimodal distribution on the cavity midplane.

The $1/e$ lifetime of the atoms in an 8-vertex configuration is typically around 2~s, limited by background and three-body collisions. The transverse pump subjects the atoms to additional heating and dephasing stemming from spontaneous emission. This reduces the $1/e$ lifetime to $318(16)$~ms for a pump lattice depth of approximately $3.8E_\text{r}$. Threshold is typically around $45E_\text{r}$, and the superradiant emission decays with a $1/e$ timescale of $3.5(3)$~ms when holding the pump lattice depth at $1.25\times$ threshold.

\subsection{Cavity and pump laser}
We employ a near-confocal cavity~\cite{Kollar2015aac} with a free spectral range (FSR) of $2\pi{\cdot}15.02980(8)$~GHz; its length and mirror radius of curvature are both 1~cm.  The finesse is approximately $5{\times} 10^4$ and the single-mode cooperativity is $C=2g_0^2/\kappa\Gamma\approx 5.2$. The single-atom, multimode light-matter coupling strength is $\sqrt{M}g_0$, where $g_0=2\pi{\cdot}$1.47~MHz and $M\approx21$ is the multimode enhancement factor~\cite{Kroeze2023hcu,Vaidya2018tpa}. The corresponding dispersive multimode single-atom cooperativity is $C_\text{mm} = 2Mg_0^2/\kappa\Gamma =$110~\cite{Kroeze2023hcu}, where the cavity and atomic linewidths are $\kappa =2\pi{\cdot}$137~kHz and $\Gamma =2\pi{\cdot}$6.1~MHz.  

\begin{figure}
    \centering
    \includegraphics[width=0.4\columnwidth]{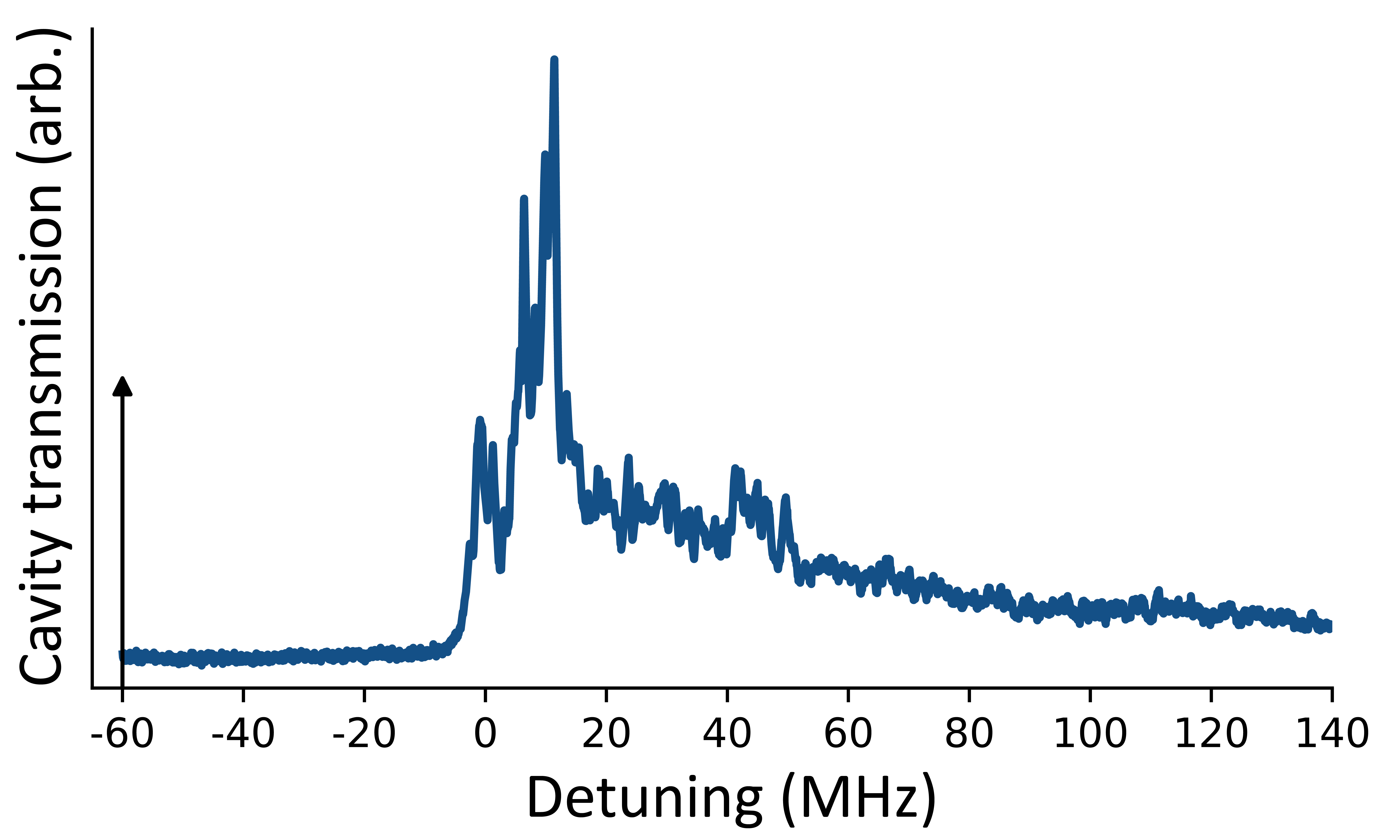}
    \caption{Transmission spectroscopy of the confocal cavity used in this work. A spot of waist 32~$\mu$m aimed 80~$\mu$m radially from the cavity center is used to probe the cavity. The detuning of the transverse pump beam is indicated by the black arrow.}
    \label{fig:spectrum}
\end{figure}

The cavity length is stabilized using the Pound-Drever-Hall technique with a laser coupled to the cavity at 1560~nm~\cite{Kollar2015aac}. The remaining light is amplified and frequency-doubled to provide the 780-nm transverse pump light. It is detuned from the $5{}^{2}S_{1/2}\lvert2,-2\rangle$ to $5{}^2P_{3/2}\lvert3,-3\rangle$ transition by $\Delta_A/2\pi=-97.3$~GHz.  The pump is detuned from the cavity resonance by $\Delta_C/2\pi=-60$~MHz from the lowest frequency peak in the cavity transmission spectrum. An example cavity spectrum is shown in Fig.~\ref{fig:spectrum} for a longitudinal probe beam of waist $0.9w_0$ injected to the side of the cavity center. 

\begin{figure}
    \centering
    \includegraphics[width=0.4\columnwidth]{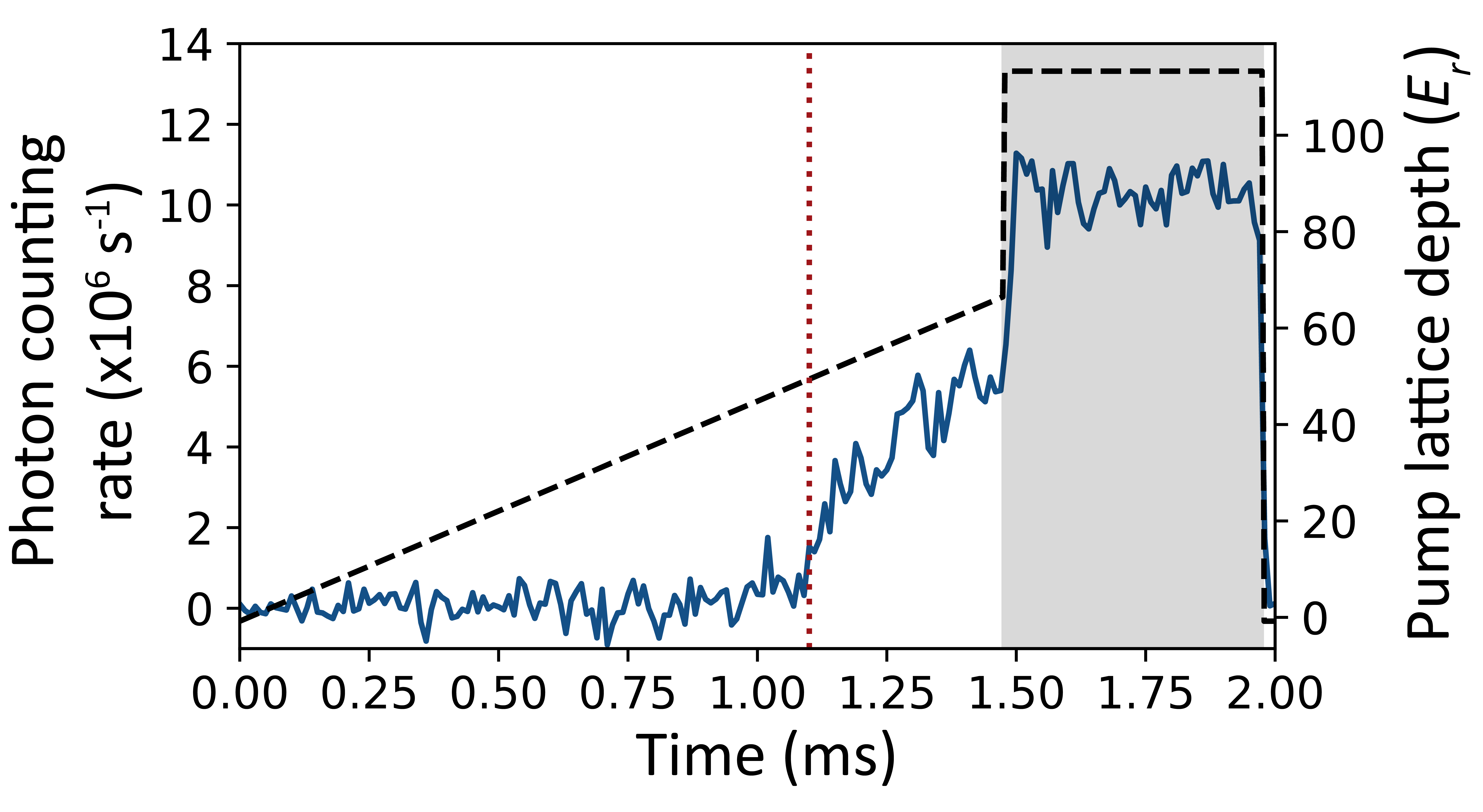}
    \caption{Cavity emission intensity (left axis, blue solid line) detected by a single-photon counter versus time after subtracting background counts. The transverse pump power---in units of lattice depth (right axis, black dashed line)---is ramped linearly through the superradiant threshold. Threshold is demarcated by a sudden increase in cavity emission, indicated by the red dotted line. After ramping to $1.25\times$ the threshold power, the state is read-out for $500$~$\mu$s, as indicated by the gray shaded area. During the read-out process, the pump power is rapidly increased and the local oscillator and cavity emission expose the camera.}
    \label{fig:rampschedule}
\end{figure}

The pump laser is retroreflected to produce a standing-wave optical lattice at the location of the atoms. The lattice depth is calibrated using Kapitza-Dirac diffraction of a BEC. During the experiment, and unless otherwise noted, the depth of the pump lattice linearly increases by $45E_\text{r}/\text{ms}$ for 1.5~ms.  No Mott insulating phase arises at the $\Delta_C$ we employ~\cite{Landig2016qpf}. After reaching $1.25\times$ the critical pump power, the holographic image is taken by rapidly increasing the pump power to $113E_\text{r}$ and holding there for 500~$\mu$s for readout of the spin state. The pump schedule is illustrated in Fig.~\ref{fig:rampschedule} with an example of the measured cavity emission recorded on a single-photon counter. We have independently verified that this imaging process does not alter the spin-organized state: Any dynamics induced by the readout would result in reduction of fringe contrast for the evolving vertices. We do not observe such signal reduction, other than for all vertices simultaneously, which we attribute to reduced atom number in that particular experimental shot. Additionally, we measure the phase of the (spatially integrated) intracavity field using a temporal heterodyne measurement. These measurements show no observable phase difference between the moment just prior to the readout versus during the readout. Finally, we note that fluctuations in overall atom number, measured to be 6\%, result in fluctuations of a few tens of microseconds in the time spent in the superradiant phase before reading-out the spin state. We do not believe these small time fluctuations have a meaningful effect on the observed dynamics or overlap distributions.

\subsection{Holographic imaging}

Detection of the spin state that emerges in the superradiant phase is achieved through holographic imaging of the emitted cavity field, as in our previous work~\cite{Kroeze2018sso,Guo2019spa,Guo2019eab,Guo2021aol}. The emitted field is mixed with a local oscillator (LO) derived from the transverse pump and focused onto an electron-multiplied charge-coupled-device camera. The cavity field and LO strike the camera at slightly different angles to produce a phase-sensitive interference pattern, realizing a spatial heterodyne measurement. Sections~\ref{sec:readout} and~\ref{processing} provide details about how the detected images are processed to extract the density wave (DW) phase for each vertex. 

Optical components between the cavity and camera can cause phase aberrations in the detected holographic images. This aberration is characterized by positioning a single gas cloud of atoms at the cavity center and pumping the system above the superradiant threshold. The emitted cavity field for this particular configuration of atoms should have a completely flat phase front. Any additional phase must be due to aberrations. This becomes a phase mask that is subtracted from all holographic images.

\subsection{Processing of holographic images and spin-state extraction}\label{sec:readout}

\begin{figure*}[t!]
    \centering
    \includegraphics[width=\linewidth]{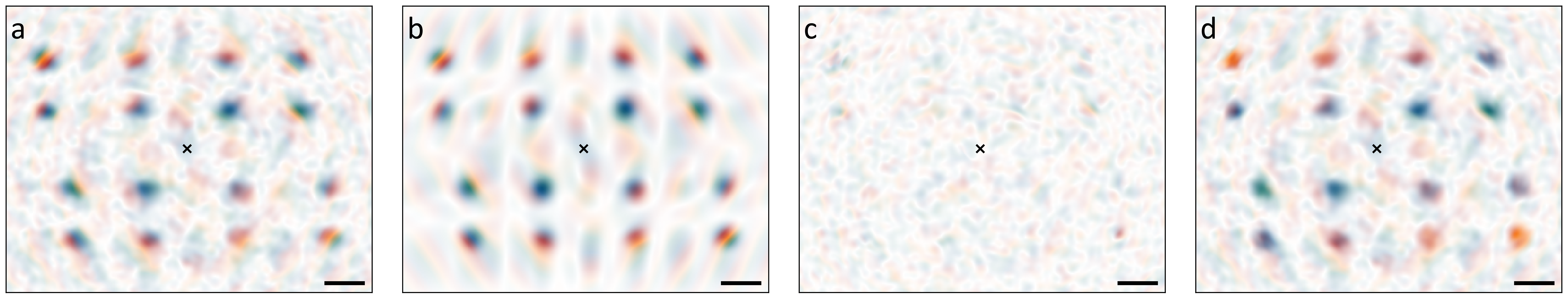}
    \caption{Processing of holographic images. In all panels, the scale bar indicates $w_0=35$~$\mu$m and cavity center is marked by a cross. (a) Raw holographic image of an emitted cavity field for the processed image presented in Fig.~1b of the main text. Each vertex scatters a localized spot, with a mirror-image spot on the opposite side of the cavity. The nonlocal component of the field produces a weaker background field present throughout the image. (b) Optimal fit obtained from the model given by Eq.~\eqref{eq:fullField}. (c) The residual between the measured and fitted images in panels (a) and (b), resp. The color scale is the same as in panel (a). (d) The resulting gradient-corrected holographic image. The upper half of this image is identical to that presented in Fig.~1b of the main text.}
    \label{fig:imageprocessing}
\end{figure*}

We now describe how we extract the spin configuration from a recorded holographic image. As we show in Sec.~\ref{sec:Efield}, the phase of the emitted cavity field is directly related to the spin configuration. Phase extraction is accomplished through a fit by least-squares regression to the full field-of-view of the emitted cavity field. We account for finite-size effects by allowing for DW phase gradients. A discussion of the origin of these gradients is presented in Sec.~\ref{theory:finite-size} and Sec.~\ref{sec:gradients}. 

As we derive in Sec.~\ref{sec:Efield}, the total field is a sum over the fields generated by each vertex,
\begin{equation}\label{eq:fullField}
    \Phi^F(\mathbf{r}) = e^{i\phi}\sum\limits_{i=1}^n A_i\left[e^{-i\theta_i}\Phi_\text{local}(\pos;\pos_i) + e^{i\theta_i}\Phi_\text{non}(\pos;\pos_i)\right],
\end{equation}
where the superscript $F$ denotes that this is the forward-propagating field emitted from the cavity, $\phi$ is a global phase, $A_i$ is the amplitude of the field and $\theta_i$ is the DW phase of vertex $i$, which is at position $\pos_i$. While the emitted cavity field and LO have a stable phase within the time span of a single experimental shot, there is a slow drift of the LO phase from shot-to-shot that results in a different global phase $\phi$ that must be fit for each image. The local and nonlocal field generated by vertex $i$ are given by Eqs.~\eqref{eq:Phi_local} and~\eqref{eq:Phi_non}. These further contain the Gaussian width of the vertex $\sigma_A$ and the DW phase gradient $\mbf{g}_i$. Equation~\eqref{eq:fullField} is fit to the electric field obtained from holographic imaging to determine the global parameters $\phi$ and $\sigma_A$ as well as the parameters for describing each vertex, $\{A_i,\theta_i,\pos_i,\mbf{g}_i\}$. The spin state corresponds to the DW phases $\theta_i$. The extracted spin states are not random due to noise, as can be seen by the specific overlap structure in Figs.~3 and 4 of the main text. Nonetheless, technical noise in the spin readout results in an effective noise temperature. From the phase fluctuations within each local spot, we estimate the readout noise to have $0.09$~radians standard deviation. This corresponds to a noise temperature of $0.018T_c$ (cf.~the fitted temperatures in Fig.~2 of $0.08(8)T_c$ and $0.10(1)T_c$).

An example hologram is shown in Fig.~\ref{fig:imageprocessing}a with the corresponding best fit result in panel (b) and the residual in panel (c). To speed-up the fitting procedure, we downsample the observed electric field by averaging blocks of $3\times3$ pixels into single superpixels. Comparing the results from downsampled fits to full fits reveals no systematic bias, with a root-mean-square difference between these spin configurations of less than 0.002~radians per spin across a set of 30 images. 

The images in Figs.~1--3 of the main text are corrected to remove the global phase $\phi$ as well as the gradient in the phase across each vertex.  The goal is to leave only the average phase. The gradient-correction term is computed as the difference of two images; the fitted field, such as in Fig.~\ref{fig:imageprocessing}b, and the fitted field with the gradient parameters $\mathbf{g}_i$ set to zero for the local field terms. This correction is applied to the observed image in Fig.~\ref{fig:imageprocessing}a to produce the gradient-corrected image in Fig.~\ref{fig:imageprocessing}d. The images of ferromagnetic spin configurations in Fig.~2 of the main text have also been subjected to a gauge transformation:  A subset of the vertices receive a local $\pi$ phase shift; see Sec.~\ref{sec:gauge} for an explanation.  

\subsection{Atomic position calibration}

The positions of the atoms are confined to the $z=0$ midplane of the confocal cavity. In order to locate this position, we do an interferometric measurement of an intracavity lattice injected longitudinally. Specifically, we address the cavity with two different wavelengths separated by an integer multiple $\Delta q$ of the FSR. The difference in spatial phase between the lattices created with these two wavelengths changes as a function of $z$ as $\Delta q \frac{\pi}{2L}(2z + L)$ and can be measured by a multipulse Kapitza-Dirac measurement.

This kind of measurement relies on accurate knowledge of the longitudinal spatial phase of the intracavity lattice. This is challenging in a confocal cavity, since the longitudinal character depends strongly on the transverse mode content of the light field being injected~\cite{Guo2019spa,Guo2019eab}. Instead, we perform this measurement with the length of the cavity set to a single-mode configuration, where the TEM$_{00}$ mode, which has a known longitudinal spatial phase dependence, can be isolated and addressed. By accounting for the known change in cavity length between this single-mode configuration and the confocal configuration employed in this work, we are then able to position the atoms to within 10~$\mu$m of the midplane.

To transversely position the atoms in the $z=0$ midplane, we rely on the local-field emission of the confocal cavity, which can be imaged with high resolution~\cite{Kroeze2023hcu}. We extract the position(s) by fitting the superradiant cavity emission as described in Sec.~\ref{sec:readout}. 

\begin{figure}[t!]
    \centering
    \includegraphics[width=\linewidth]{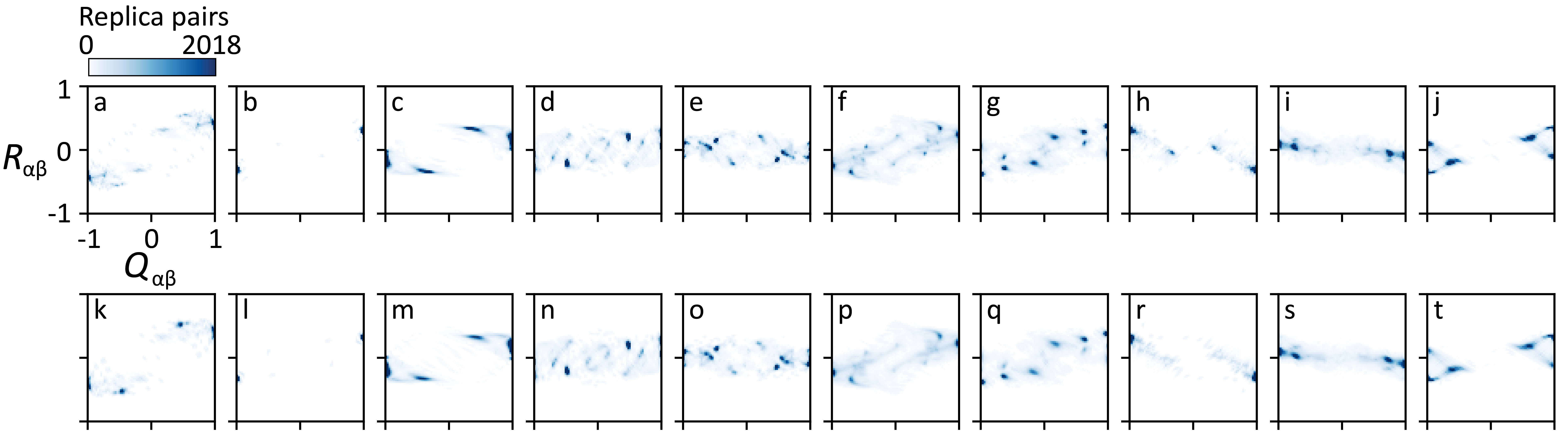}
    \caption{Sensitivity of overlap distributions to atom number and position fluctuations. (a-j) Noise-free overlap distributions simulated using the mean-field trajectory framework of Sec.~\ref{sec:mf}. (k-t) Corresponding simulated overlap distributions with realistic atom number and position noise added. This does not cause any qualitative differences.}
    \label{fig:sensitivity}
\end{figure}

\subsection{Data processing}\label{processing}

We perform the following data processing procedure after fitting the cavity fields to reconstruct the spin state; the reconstruction is described above in Sec.~\ref{sec:readout}. We exclude experimental shots where the fitter does not provide a reasonable set of atomic vertex positions based on approximately known trap positions. This occurs in $\sim$8\% of the experimental shots. These shots are characterized by below-average atom number.

For a given position configuration, we require 100 experimental replicas (after filtering) to produce a sufficiently converged overlap distribution. See Sec.~\ref{sec:boot} for a discussion of convergence of the overlap in terms of the number of replicas. Any additional replicas are discarded to keep constant the number per disorder-instance at exactly 100.

Comparing all observed positions to the target positions, we find that the fluctuations in vertex position are distributed as an isotropic Gaussian. The shot-to-shot position fluctuations have a standard deviation of 0.73~$\mu$m. This position noise induces a small amount of variance in the quenched disorder in the coupling matrix from shot-to-shot. Likewise, fluctuations in atom number between vertices have the same effect. These relative atom number fluctuations are estimated to be less than 7\% from analyzing the relative brightness in the superradiant images. To assess the effect of these fluctuations on the overlap distribution, we use the MF trajectory technique described in Sec.~\ref{sec:mf}. Each replica (trajectory) has a slightly different $J$ and $K$ matrix due to the position and atom number noise we add. The resulting overlap distribution is compared to the distribution with no noise. Results are presented in Fig.~\ref{fig:sensitivity}, where the top row present the noise-free overlap distributions, the bottom row shows the noisy distributions, and each column is a different disorder instance. There is no qualitative difference among them, showing that this noise does not qualitatively affect the visibility of RSB in the system.

\section{Confocal cavity QED spin model}\label{theory}

For simplicity, the atoms in each of the $n=8$ atomic gas clouds are approximated as being in a BEC, allowing each ensemble to be described by an atomic field operator $\hat{\psi}_i(\mbf{x})$. After elimination of the atomic excited state, the Hamiltonian is:  
\begin{align}
    \hat{H}&=-\sum\limits_\mu \Delta_\mu \hat{a}_\mu^\dagger \hat{a}_\mu + \sum\limits_{i=1}^n \hat{H}_{A,i} + \sum\limits_{i=1}^n \hat{H}_{LM,i},\label{eq:Htotal}\\
    \hat{H}_{A,i}&=\int d^3\mbf{x} \hat{\Psi}_i^\dag(\mbf{x})\left(-\frac{\nabla^2}{2m}+V(\mbf{x}) 
    \right)\hat{\Psi}_i(\mbf{x}),\label{eq:Hatomic}\\
    \hat{H}_{LM,i}&=\frac{1}{\Delta_A}\int d^3\mbf{x} \hat{\Psi}_i^\dag(\mbf{x})\hat{\Phi}_T^\dag(\mbf{x})\hat{\Phi}_T^{}(\mbf{x})\hat{\Psi}_i^{}(\mbf{x}).
    \label{eq:Hlightmatter}
\end{align}
The individual cavity modes are indexed by $\mu$.  The pump detuning is $\Delta_\mu$ and photon annihilation operators are $\hat{a}_\mu$. The externally applied trapping potential is given by $V(\mbf{x})$. The trap potential contains 8 minima that define the locations of the vertices of the network. The minima occur at locations $\mbf{x}_i$ and are anisotropic with trap frequencies $\omega_{x,y,z}$. We neglect $s$-wave scattering between atoms because it results in only a small shift of the superradiance threshold~\cite{Baumann2010dqp}.

The light-matter Hamiltonian $\hat{H}_{LM,i}$ contains the total light field $\hat{\Phi}_T(x)=\Omega\cos(k_rx)+g_0\hat{\Phi}(x)$. The first term is the standing-wave transverse pump, where the Rabi strength is $\Omega$, $k_r=2\pi/\lambda$ is the recoil momentum, and $\lambda$ is the wavelength of the light. The second term contains the total cavity field operator $\hat{\Phi}(\mbf{x})$; the  single-mode, single-photon interaction strength is $g_0$. The cavity field is expressed in terms of individual modes as 
\begin{equation}\label{eq:cavfield_modes}
    \hat{\Phi}(\mbf{x}) = \sum\limits_\mu \hat{a}_\mu \Xi_\mu(\mbf{r})\cos\left(k_rz-\theta_\mu\right).
\end{equation}
This expression assumes that $\abs{z}$ is much smaller than the mode's Rayleigh range (5~mm in this confocal cavity), which is satisfied since typical $z$ are less than 10~$\mu$m. A mode-dependent phase shift $\theta_\mu=n_\mu\pi/4$ serves to satisfy boundary conditions in the cavity, where $n_\mu=l+m$. The transverse-mode function near the cavity midplane is
\begin{equation}
\Xi_\mu(\mbf{r})=H_l\left(\sqrt{2}x/w_0\right)H_m\left(\sqrt{2}y/w_0\right)\exp(-\pos^2/w_0^2),
\end{equation}
where $H_n(x)$ are the Hermite polynomials.
We use $\mbf{r}$ to denote the in-plane part of the general coordinate $\mbf{x}$. Only families of modes with all even or all odd $n_\mu$ are simultaneously resonant; we focus on an even resonance, and so all sums over $\mu$ are implicitly restricted to $n_\mu=0\mod2$. This results in cavity modes that alternate in longitudinal shape between $\cos(k_rz)$ and $\sin(k_rz)$~\cite{Guo2019eab,Guo2019spa}.

The atomic degrees of freedom develop a DW of wavevector $k_r$ at the superradiant transition. This can be described by writing the atomic field operators in a basis containing the DW excitations, 
\begin{equation}\label{eq:ansatz}
    \hat{\Psi}_i(\mbf{x}) = \sqrt{E(\mbf{x}-\mbf{x}_i)}\left[\hat{\psi}_{0,i} + 2\hat{\psi}_{c,i}\cos(k_rz)\cos(k_rx)+ 2\hat{\psi}_{s,i}\sin(k_rz)\cos(k_rx)\right].
\end{equation}
The operators $\hat{\psi}_{\sigma,i}$, where $\sigma\in\{0,c,s\}$, are independent bosonic modes with canonical commutation relations $[\hat{\psi}_{\sigma,i}^{},\hat{\psi}_{\tau,j}^\dag]=\delta_{\sigma\tau}\delta_{ij}$ and $[\hat{\psi}_{\sigma,i}^{},\hat{\psi}_{\tau,j}^{}]=[\hat{\psi}_{\sigma,i}^\dag,\hat{\psi}^\dag_{\tau,j}]=0$. These operators describe excitations of the BEC ground state, cosine DW, or sine DW, respectively. The envelope function $E(\mbf{x})$ is assumed to be identical for each vertex. The position of vertex $i$ is $\mbf{x}_i = (\mbf{r}_i,z_i)$, where we take all vertices to be in the midplane $z_i=0$ and $\mbf{r}_i$ denotes the 2D position in the midplane. Normalization of the field operators is achieved by choosing an envelope function for which $\int d^3\mbf{x}E(\mbf{x})=1$. We note that this atomic ansatz ignores higher-order corrections involving Mathieu functions that describe organization in a deep lattice~\cite{Guo2021aol}. This results in a renormalization of only the superradiant threshold, which is unimportant to this work.

To proceed, we insert the expansion of the atomic fields into Eqs.~\eqref{eq:Hatomic} and~\eqref{eq:Hlightmatter} and evaluate the integrals. We assume that the extent of $E(\mbf{x})$ is large compared to $\lambda$, allowing us to drop fast-oscillating terms. For $\hat{H}_{A,i}$, this results in 
\begin{equation}
    \hat{H}_{A,i} = 2E_r\left(\hat{\psi}_{c,i}^\dag \hat{\psi}_{c,i}^{} +\hat{\psi}_{s,i}^\dag\hat{\psi}_{s,i}^{} \right) + E_{\text{trap},i}\left(\hat{\psi}_{0,i}^\dag \hat{\psi}_{0,i}^{}+\hat{\psi}_{c,i}^\dag \hat{\psi}_{c,i}^{} +\hat{\psi}_{s,i}^\dag\hat{\psi}_{s,i}^{}\right) ,\label{eq:Ha_simplified}
\end{equation}
where $E_r=\hbar^2k_r^2/(2m)$ is the recoil energy and $E_{\text{trap},i}=\int d^3\mbf{x}V(\mbf{x})E(\mbf{x}-\mbf{x}_i)$ is the trap energy.
Evaluation of the light-matter coupling produces three terms,
\begin{equation}\begin{split}\label{eq:Hlm_simplified}
    \hat{H}_{LM,i} = & \frac{\Omega^2}{2\Delta_A}\left(\hat{\psi}_{0,i}^\dag \hat{\psi}_{0,i}^{} +\frac{3}{2}\hat{\psi}_{c,i}^\dag \hat{\psi}_{c,i}^{}+\frac{3}{2}\hat{\psi}_{s,i}^\dag \hat{\psi}_{s,i}^{}\right) \\
    &+\frac{g_0\Omega}{2\Delta_A}\sum\limits_\mu (\hat{a}_\mu+\hat{a}_\mu^\dag) \left[\cos(\theta_\mu)\hat{\chi}_{c,i} + \sin(\theta_\mu)\hat{\chi}_{s,i}\right]\int d^2\mbf{r}\rho(\mbf{r}-\mbf{r}_i)\Xi_\mu(\mbf{r})\\
    &+\frac{g_0^2}{2\Delta_A}\hat{\psi}_{0,i}^\dag \hat{\psi}_{0,i}^{}\sum\limits_{\mu,\nu}\hat{a}_\mu^\dagger\hat{a}_\nu\cos(\theta_\mu-\theta_\nu)\int d^2\mbf{r}\rho(\mbf{r}-\mbf{r}_i)\Xi_\mu(\mbf{r})\Xi_\nu(\mbf{r})+O(\hat{a}^\dagger\hat{a}\hat{\psi}_{c/s,i}^\dag \hat{\psi}_{c/s,i}),
\end{split}\end{equation}
where $\hat{\chi}_{c/s,i} = \hat{\psi}_{0,i}^{}\hat{\psi}_{c/s,i}^\dag + \hat{\psi}_{0,i}^\dag\hat{\psi}_{c/s,i}$ are Hermitian operators and $\rho(\mbf{r})=\int dz E(\mbf{x})$ is the transverse density profile of the atomic ensemble. The first line describes a pump-induced light shift while the second line describes coupling between the atomic DW and cavity light. The last line describes a multimode dispersive shift due to both the ground-state atomic gas and the DW components. The latter terms are not written out because we ignore the dispersive shift for simplicity in what follows. 

The total Hamiltonian has a set of $n$ conserved quantities corresponding to the total number of atoms in each ensemble. The observables ${N}_i = \hat{\psi}_{0,i}^\dag \hat{\psi}_{0,i}^{} + \hat{\psi}_{c,i}^\dag \hat{\psi}_{c,i}^{} + \hat{\psi}_{s,i}^\dag \hat{\psi}_{s,i}^{}$ correspond to the number of atoms in the $i$'th vertex and are conserved by $\hat{H}$, leading to the U$(1)$ symmetry of the overall phase of the atomic field $\hat{\Psi}_i(\mbf{x})$. The system also possesses two independent $\mathbb{Z}_2$ symmetries. The first symmetry corresponds to the transformation $\hat{\psi}_{c,i}\to -\hat{\psi}_{c,i}$ for all $i$, in addition to $\hat{a}_\mu\rightarrow-\hat{a}_\mu$ for all modes $\mu$ coupling to the cosine DW ($n_\mu = 0\mod4$). The other symmetry is identical but for the $\hat{\psi}_{s,i}$ operators and the corresponding cavity modes ($n_\mu = 2\mod4$). We further discuss the role of these symmetries in Sec.~\ref{sec:props}.

The photonic degrees of freedom can now be adiabatically eliminated. This approximation is valid in the limit $\abs{\Delta_\mu} \gg \frac{g_0\Omega}{\abs{\Delta_A}},\frac{Ng_0^2}{\abs{\Delta_A}},E_r$, where the cavity field dynamics are much faster than the atomic motion. Elimination proceeds by writing the equation of motion for $\hat{a}_\mu$ based on Eqs.~\eqref{eq:Htotal}, \eqref{eq:Ha_simplified}, and \eqref{eq:Hlm_simplified} and including cavity dissipation with photon loss rate $2\kappa$. Setting the time derivative to zero yields
\begin{equation}\label{eq:adiabatic_mode}
    \hat{a}_\mu = \frac{g_0\Omega}{2\Delta_A(\Delta_\mu+i\kappa)}\sum_{i=1}^n \hat{R}_{\mu,i}\int d^2\mbf{r}\rho(\mbf{r}-\mbf{r}_i)\Xi_\mu(\mbf{r}),
\end{equation}
where we define $\hat{R}_{\mu,i} = \cos(\theta_\mu)\hat{\chi}_{c,i} + \sin(\theta_\mu)\hat{\chi}_{s,i}$. This expression for $\hat{a}_\mu$ is substituted back into the equations of motion for the atomic fields to find an atom-only Hamiltonian with a light-mediated interaction. Up to an overall energy shift, the atom-only Hamiltonian is given by
\begin{equation}\begin{split}
    \hat{H}_\text{atom-only} = &\left(2E_r+\frac{\Omega^2}{4\Delta_A}\right)\sum_{i=1}^n \left(\hat{\psi}_{c,i}^\dag \hat{\psi}_{c,i}^{} +\hat{\psi}_{s,i}^\dag\hat{\psi}_{s,i}^{} \right) \\&+ \frac{g_0^2\Omega^2}{4\Delta_A^2}\sum\limits_{i,j=1}^n\sum\limits_\mu \frac{\hat{R}_{\mu,i}\hat{R}_{\mu,j}\Delta_\mu}{\Delta^2_\mu+\kappa^2}\int d^2\mbf{r}d^2\mbf{r'}\rho(\mbf{r}-\mbf{r}_i)\rho(\mbf{r'}-\mbf{r}_j)\Xi_\mu(\mbf{r})\Xi_\mu(\mbf{r'}).
\end{split}\end{equation}
The second line is the light-mediated interaction $\hat{H}_\text{int}$, and  can be simplified by defining a position-dependent interaction function in $2{\times}2$ matrix form: 
\begin{equation}
    \mathcal{D}(\mbf{r},\mbf{r'}) = \Delta_C\sum\limits_\mu\frac{\Xi_\mu(\mbf{r})\Xi_\mu(\mbf{r'})}{\Delta_\mu+i\kappa}\begin{bmatrix}\cos^2(\theta_\mu) & \cos(\theta_\mu)\sin(\theta_\mu) \\ \cos(\theta_\mu)\sin(\theta_\mu) & \sin^2(\theta_\mu) \end{bmatrix}.
\end{equation}
The interaction term can then be written as
\begin{equation}\label{eq:Hint}
    \hat{H}_\text{int} = \frac{g_0^2\Omega^2}{4\Delta_A^2\Delta_C}\sum\limits_{i,j=1}^n \begin{bmatrix} \hat{\chi}_{c,i} \\ \hat{\chi}_{s,i} \end{bmatrix}^\intercal \int d^2\mbf{r}d^2\mbf{r'}\rho(\mbf{r}-\mbf{r}_i)\rho(\mbf{r'}-\mbf{r}_j) \Re\big[\mathcal{D}(\mbf{r},\mbf{r'})\big]\begin{bmatrix} \hat{\chi}_{c,j} \\ \hat{\chi}_{s,j} \end{bmatrix}.
\end{equation}
We now evaluate the interaction function $\mathcal{D}$ explicitly using the known Green's function of the harmonic oscillator,
\begin{equation}\label{eq:Greens}
    G(\mbf{r},\mbf{r'},\varphi)\equiv\sum\limits_\mu\Xi_\mu(\mbf{r})\Xi_\mu(\mbf{r'})e^{-n_\mu\varphi} = \frac{e^\varphi}{2\sinh(\varphi)}\exp\left[-\frac{(\mbf{r}-\mbf{r'})^2}{2w_0^2\tanh(\varphi/2)}-\frac{(\mbf{r}+\mbf{r'})^2}{2w_0^2\coth(\varphi/2)}\right].
\end{equation}
We consider the limit of a perfectly degenerate confocal cavity, for which $\Delta_\mu=\Delta_C$ for all $\mu$. Recalling that we consider an even-parity confocal cavity, for which summations over $\mu$ include only modes with $n_\mu= 0\mod 2$, we find
\begin{equation}
    \mathcal{D}(\mbf{r},\mbf{r'}) = \frac{1}{2(1+i\kappa/\Delta_C)} \begin{bmatrix}
        G^+(\mbf{r},\mbf{r'},0) + G^+(\mbf{r},\mbf{r'},i\pi/2) & 0 \\
        0 & G^+(\mbf{r},\mbf{r'},0) - G^+(\mbf{r},\mbf{r'},i\pi/2) \end{bmatrix},
\end{equation}
where $G^+(\mbf{r},\mbf{r'},\varphi) = \frac{1}{2}\left[G(\mbf{r},\mbf{r'},\varphi) + G(\mbf{r},-\mbf{r'},\varphi)\right] = \frac{1}{2}\left[G(\mbf{r},\mbf{r'},\varphi) + G(\mbf{r},\mbf{r'},\varphi + i\pi)\right]$. This then results in
\begin{equation}
    \hat{H}_\text{int} = -\sum\limits_{i,j=1}^n\begin{bmatrix} \hat{\chi}_{c,i} \\ \hat{\chi}_{s,i} \end{bmatrix}^\intercal \begin{bmatrix}
        J_{ij}^\text{local} + J_{ij}^\text{non} & 0 \\
        0 & J_{ij}^\text{local} - J_{ij}^\text{non} \end{bmatrix}\begin{bmatrix} \hat{\chi}_{c,j} \\ \hat{\chi}_{s,j} \end{bmatrix},
\end{equation}
where we have defined the elements of the connectivity matrix as
\begin{align}
    J_{ij}^\text{local} &= \mathcal{J}\int d^2\mbf{r}d^2\mbf{r'}\rho(\mbf{r}-\mbf{r}_i)\rho(\mbf{r'}-\mbf{r}_j)G^+(\mbf{r},\mbf{r'},0), \label{eq:Jloc}\\
    J_{ij}^\text{non} &= \mathcal{J}\int d^2\mbf{r}d^2\mbf{r'}\rho(\mbf{r}-\mbf{r}_i)\rho(\mbf{r'}-\mbf{r}_j)G^+(\mbf{r},\mbf{r'},i\pi/2), \label{eq:Jnon}
\end{align}
and $\mathcal{J} = g_0^2\Omega^2|\Delta_C|/[8\Delta_A^2(\Delta_C^2+\kappa^2)]$.

We first consider the simplest case in which the atomic distribution is point-like before considering finite-size effects in subsequent sections. The atomic density profiles are then given by $\rho(\mbf{r}-\mbf{r}_i)=\delta(\mbf{r}-\mbf{r}_i)$. In this case, evaluation of $G^+(\mbf{r},\mbf{r'},\varphi)$ in the limit $\varphi\to0$ results in $\pi\delta\big[(\mbf{r}-\mbf{r'})/w_0\big] + \pi\delta\big[(\mbf{r}+\mbf{r'})/w_0\big]$. We consider distinct vertex locations $\mbf{r}_i$ for which the mirror locations also have no overlap, such that $\mbf{r}_i\neq\pm\mbf{r}_j$ for $i\neq j$. This yields the simplification $J_{ij}^\text{local} = J^\text{local}\delta_{ij}$. The nonlocal term takes the form $J_{ij}^{\text{non}}\propto\cos(2\mbf{r}_i\cdot\mbf{r}_j/w_0^2)$~\cite{Guo2019eab}. Thus, the atom-only Hamiltonian takes the form
\begin{equation}
    \hat{H}_\text{atom-only} = \left(2E_r+\frac{\Omega^2}{4\Delta_A}\right)\sum_{i=1}^n \left(\hat{\psi}_{c,i}^\dag \hat{\psi}_{c,i}^{} +\hat{\psi}_{s,i}^\dag\hat{\psi}_{s,i}^{} \right) - J^{\text{local}}\sum_{i=1}^n \big(\hat{\chi}_{c,i}^2 + \hat{\chi}_{s,i}^2\big) - \sum_{i,j=1}^n J^{\text{non}}_{ij}\big(\hat{\chi}_{c,i}\hat{\chi}_{c,j}-\hat{\chi}_{s,i}\hat{\chi}_{s,j}\big).
\end{equation}
The first term imposes an energy cost on the formation of an atomic DW. In the absence of other terms, or at zero pump power, the ground state of the system is thus a BEC with no DW. The second and third terms, which scale with the pump power, represent the light-mediated interaction. Competition between these and the first term drive a transition into the superradiant phase, in which the atoms can organize into a complex pattern of DWs to minimize the energy.

The atom-only Hamiltonian can be recast using Gell-Mann operators to describe the quantum limit of the system. The atomic ansatz in Eq.~\eqref{eq:ansatz} approximates each atom as a three-level system, with each atom in either the BEC state or the state with cosine or sine DW modulation. Each atom can thus be represented by the SU$(3)$ Gell-Mann matrices $\{\lambda^k\}$, where $k\in 1,\ldots,8$. Because the atoms within a vertex all couple symmetrically to the rest of the system, the only operators entering into the Hamiltonian are collective sums of Gell-Mann operators. We denote these operators as $\hat{\Lambda}^{(k)}_i\equiv \sum_{j=1}^{N_{i}}\hat{\lambda}_{ij}^k$, where $\hat{\lambda}_{ij}^k$ is the Gell-Mann operator for the $j$'th atom in vertex $i$. The collective Gell-Mann operators $\hat{\Lambda}_i^{(k)}$ satisfy the same commutation relations as the $\lambda^k$ matrices. The collective Gell-Mann operators are related to the atomic field operators $\bm{\hat{\psi}_i}=(\hat{\psi}_{0,i},\hat{\psi}_{c,i},\hat{\psi}_{s,i})^\intercal$ through the Jordan-Schwinger map $\hat{\Lambda}^{(k)}_i=\bm{\hat{\psi}_i}^\intercal \lambda^k \bm{\hat{\psi}_i}$. Performing the map on the atom-only Hamiltonian  yields, up to an overall energy shift:
\begin{equation}
    \label{final_hamiltonian}
    \hat{H}_\text{atom-only} = -\left(E_r+\frac{\Omega^2}{8\Delta_A}\right)\sum_{i=1}^n\left(\hat{\Lambda}_i^{(3)}+\frac{1}{\sqrt{3}}\hat{\Lambda}_i^{(8)}\right)-\sum_{i=1}^n J_{ii}^\text{local}\left[\big(\hat{\Lambda}_i^{(1)}\big)^2+\big(\hat{\Lambda}_i^{(4)}\big)^2\right]-\sum_{i,j=1}^n J_{ij}^{\text{non}}\left(\hat{\Lambda}_i^{(1)} \hat{\Lambda}_j^{(1)} - \hat{\Lambda}_i^{(4)} \hat{\Lambda}_j^{(4)} \right).
\end{equation}
Note that the squared operators $(\hat{\Lambda}_i^{(k)})^2$ do not simplify for any $N_i>1$. The reason is the same as for the case of products of SU$(2)$ collective spin operators. These do not simplify except in the spin-$1/2$ limit where the spin operators, corresponding to Pauli matrices, form a complete basis.

\subsection{Finite-sized atomic distributions}\label{theory:finite-size}

Two important effects follow from considering the finite size of the atomic distributions. The first is that the interaction matrix changes into a form that normalizes the divergent $J^{\text{local}}$ term. Second is that one must account for the possibility that atoms within a vertex adopt a DW whose phase is nonuniform across the vertex. This leads to a new effective coupling term between $\hat{\psi}_{c,i}$ and $\hat{\psi}_{s,j}$ operators.

To account for a spatially dependent DW phase within a vertex, the atomic ansatz is generalized to 
\begin{equation}\label{eq:ansatz2}
    \hat{\Psi}_i(\mbf{x}) = \sqrt{E(\mbf{x}-\mbf{x}_i)}\left[\hat{\psi}_{0,i} + 2\hat{\psi}_{c,i}\cos[k_rz+\vartheta_i(\mbf{r})]\cos(k_rx)+ 2\hat{\psi}_{s,i}\sin[k_rz+\vartheta_i(\mbf{r})]\cos(k_rx)\right],
\end{equation}
where $\vartheta_i(\mbf{r})$ describes the spatial dependence of the DW phase. Derivation of the atom-only Hamiltonian with this modified atomic ansatz proceeds in an identical fashion. An explicit expression for $\hat{H}_{\text{int}}$ can be obtained under the condition that the phase deviation functions $\vartheta_i(\mbf{r})$ are sufficiently small across the atomic vertices. Specifically, we consider a functional form that contains only the first-order correction, a linear gradient of the phase, using $\vartheta_i(\mbf{r})=\mbf{g}_i\cdot(\mbf{r}-\mbf{r}_i)$. In this case, we simplify trigonometric expressions involving $\vartheta_{i}(\mbf{r})$ using the small-angle approximations $\cos[\vartheta_i(\mbf{r})]\approx 1$ and $\sin[\vartheta_i(\mbf{r})]\approx \mbf{g}_i\cdot(\mbf{r}-\mbf{r}_i)$. We also ignore the contribution of the DW phase gradient to the kinetic energy as it is much smaller than $E_r$. The interaction Hamiltonian is then
\begin{equation}
    \hat{H}_\text{int} = -\sum\limits_{i,j=1}^n\begin{bmatrix} \hat{\chi}_{c,i} \\ \hat{\chi}_{s,i} \end{bmatrix}^\intercal \begin{bmatrix}
        J_{ij}^\text{local} + J_{ij}^\text{non} & K_{ij} \\
        K_{ij} & J_{ij}^\text{local} - J_{ij}^\text{non} \end{bmatrix}\begin{bmatrix} \hat{\chi}_{c,j} \\ \hat{\chi}_{s,j} \end{bmatrix},
\end{equation}
where the new symmetric matrix $K_{ij}$ describes a cross-coupling between $\hat{\chi}_c$ and $\hat{\chi}_s$ operators. Within this small-angle approximation, the integral expressions for $J_{ij}^\text{local}$ and $J_{ij}^\text{non}$ are unchanged and thus are still given by Eq.~\eqref{eq:Jloc} and Eq.~\eqref{eq:Jnon}, respectively. Elements of the $K$ matrix are given by
\begin{equation}\label{eq:K}
    K_{ij} = \mathcal{J}\int d^2\mbf{r}d^2\mbf{r'}\rho(\mbf{r}-\mbf{r}_i)\rho(\mbf{r'}-\mbf{r}_j)G^+(\mbf{r},\mbf{r'},i\pi/2)\left[\mbf{g}_i\cdot(\mbf{r}-\mbf{r}_i) + \mbf{g}_j\cdot(\mbf{r'}-\mbf{r}_j)\right].
\end{equation}
This results in an atom-only Hamiltonian given by
\begin{align}\label{eq:HquantumK}
    \hat{H}_\text{atom-only} &= \left(2E_r+\frac{\Omega^2}{4\Delta_A}\right)\sum_{i=1}^n \left(\hat{\psi}_{c,i}^\dag \hat{\psi}_{c,i}^{} +\hat{\psi}_{s,i}^\dag\hat{\psi}_{s,i}^{} \right) - \sum_{i=1}^n J_{ii}^\text{local} \big(\hat{\chi}_{c,i}^2 + \hat{\chi}_{s,i}^2\big) \\ &\quad - \sum_{i,j=1}^n J^{\text{non}}_{ij}\big(\hat{\chi}_{c,i}\hat{\chi}_{c,j}-\hat{\chi}_{s,i}\hat{\chi}_{s,j}\big)-\sum_{i,j=1}^n K_{ij}\left(\hat{\chi}_{c,i}\hat{\chi}_{s,j}+\hat{\chi}_{s,i}\hat{\chi}_{c,j}\right)\nonumber.
\end{align}
The quantum model, including the new $K$ matrix coupling term, is then derived through the same Jordan-Schwinger mapping. The associated atom-only Hamiltonian is  
\begin{align}\label{eq:HquantumKgellmann}
    \hat{H}_\text{atom-only} &= -\left(E_r+\frac{\Omega^2}{8\Delta_A}\right)\sum_{i=1}^n\left(\hat{\Lambda}_i^{(3)}+\frac{1}{\sqrt{3}}\hat{\Lambda}_i^{(8)}\right)-\sum_{i=1}^n J^{\text{local}}_{ii}\left[\big(\hat{\Lambda}_i^{(1)}\big)^2+\big(\hat{\Lambda}_i^{(4)}\big)^2\right]\\&\quad-\sum_{i,j=1}^n J_{ij}^{\text{non}}\left(\hat{\Lambda}_i^{(1)} \hat{\Lambda}_j^{(1)} - \hat{\Lambda}_i^{(4)} \hat{\Lambda}_j^{(4)} \right)-\sum_{i,j=1}^n K_{ij}\left(\hat{\Lambda}_i^{(1)} \hat{\Lambda}_j^{(4)} + \hat{\Lambda}_i^{(4)} \hat{\Lambda}_j^{(1)} \right)\nonumber.
\end{align}

The integral expressions for the coupling matrices can be exactly evaluated for Gaussian atomic densities $\rho(\mbf{r}) = \exp(-\mbf{r}^2/2\sigma_A^2)/(2\pi\sigma_A^2)$. Integration results in the expressions
\begin{align}
    \label{Jlocalformula}
    J^\text{local}_{ij} &= \frac{\mathcal{J}w_0^2}{8\sigma_A^2}\left[\exp(-\frac{(\pos_i-\pos_j)^2}{4\sigma_A^2}) + \exp(-\frac{(\pos_i+\pos_j)^2}{4\sigma_A^2})\right],\\
    \label{eq:Jij_non}
    J_{ij}^\text{non} &= \mathcal{J}\frac{w_0^2}{w_\text{eff}^2} \exp(-\frac{2\sigma_A^2}{w_\text{eff}^2}\frac{\pos_i^2+\pos_j^2}{w_0^2})\cos(\frac{2\pos_i\cdot\pos_j}{w_\text{eff}^2}),\\
    K_{ij} &=  -\frac{2\mathcal{J}\sigma_A^2}{w_\text{eff}^2}\exp(-\frac{2\sigma_A^2}{w_\text{eff}^2}\frac{\pos_i^2+\pos_j^2}{w_0^2})\label{eq:Khalfway}\\ &\quad\times \left[\frac{w_0^2}{w_\text{eff}^2}\left(\pos_j\cdot\mbf{g}_i+\pos_i\cdot\mbf{g}_j\right)\sin(\frac{2\pos_i\cdot\pos_j}{w_\text{eff}^2})+\frac{2\sigma_A^2}{w_\text{eff}^2}\left(\pos_i\cdot\mbf{g}_i + \pos_j\cdot\mbf{g}_j\right)\cos(\frac{2\pos_i\cdot\pos_j}{w_\text{eff}^2})\right]\nonumber,
\end{align}
where $w_\text{eff} = w_0\sqrt{1+4\sigma_A^4/w_0^4}$. Thus, due to the finite size of the atomic distribution, the $J^{\text{local}}$ term is broadened from a delta function for point-like particles, to a distribution with finite width and amplitude. The $J^{\text{non}}$ term is less significantly impacted:  The finite size of the atomic distribution primarily sets a long-wavelength Gaussian envelope over the nonlocal interaction.

In summary, the finite size of the atomic distribution allows gradients to appear in the DW phase of each vertex. A coarse-grained model is obtained by integrating over the area of the atomic distribution to recover a point-like vertex model at the cost of introducing a new term to the effective energy. This new term, described by the $K$-matrix, is proportional to the area of the atomic density.

\subsection{Intracavity field}\label{sec:Efield}
To find the intracavity field, we use the expression for $\hat{a}_\mu$ in the adiabatic approximation. Accounting for the DW phase gradient, this is
\begin{equation}
    \hat{a}_\mu = \frac{g_0\Omega}{2\Delta_A(\Delta_\mu+i\kappa)}\sum_{i=1}^n \int d^2\mbf{r}\left(\cos[\theta_\mu+\vartheta_i(\mbf{r})]\hat{\chi}_{c,i}+\sin[\theta_\mu+\vartheta_i(\mbf{r})]\hat{\chi}_{s,i}\right)\rho(\mbf{r}-\mbf{r}_i)\Xi_\mu(\mbf{r}).
\end{equation}
Note the appearance of the phase deviation functions $\vartheta_i(\mbf{r})$ inside the integral, compared to Eq.~\eqref{eq:adiabatic_mode}. We can then insert this expression into Eq.~\eqref{eq:cavfield_modes}. Relying on the Green's functions, we have
\begin{equation}\begin{split}
    \hat{\Phi}(\mbf{x}) = \frac{g_0\Omega}{4\Delta_A(\Delta_C+i\kappa)}\sum\limits_i\int d^2\mbf{r'} \rho(\mbf{r'}-\mbf{r}_i)\bigg[&G^+(\mbf{r},\mbf{r'},0)\Big(\cos[k_rz+\vartheta_i(\mbf{r'})]\hat{\chi}_{c,i} + \sin[k_rz+\vartheta_i(\mbf{r'})]\hat{\chi}_{s,i}\Big) \\
    &+ G^+(\mbf{r},\mbf{r'},i\pi/2)\Big(\cos[k_rz-\vartheta_i(\mbf{r'})]\hat{\chi}_{c,i} - \sin[k_rz-\vartheta_i(\mbf{r'})]\hat{\chi}_{s,i}\Big) \bigg].
\end{split}\end{equation}
The light emitted from the cavity on one side, say propagating along $+z$, is the component of the field with spatial dependence $e^{ik_rz}$. This forward-propagating part is
\begin{equation}\label{eq:quantumfield}
    \hat{\Phi}^F(\mbf{r}) = \frac{g_0\Omega}{8\Delta_A(\Delta_C+i\kappa)}\sum\limits_i \big[\Phi_\text{loc}(\mbf{r};\mbf{r}_i) (\hat{\chi}_{c,i}-i\hat{\chi}_{s,i}) + \Phi_\text{non}(\mbf{r};\mbf{r}_i) (\hat{\chi}_{c,i} + i\hat{\chi}_{s,i})\big],
\end{equation}
where we defined
\begin{align}
    \Phi_\text{local}(\mbf{r};\mbf{r}_i)&=\int d^2\mbf{r'} \rho(\mbf{r'}-\mbf{r}_i)G^+(\mbf{r},\mbf{r'},0)e^{i\vartheta_i(\mbf{r'})},\\
    \Phi_\text{non}(\mbf{r};\mbf{r}_i)&=\int d^2\mbf{r'} \rho(\mbf{r'}-\mbf{r}_i)G^+(\mbf{r},\mbf{r'},i\pi/2)e^{-i\vartheta_i(\mbf{r'})},
\end{align}
as the local and nonlocal field arising from a source at position $\mbf{r}_i$ with density distribution $\rho$. We note that the nonlocal field generated by vertex $i$, i.e., $\Phi_\text{non}(\pos;\pos_i)$, allows one to recover elements of the $J$ and $K$ matrices. To average over the vicinity of $\pos_j$, we calculate the integral
\begin{equation}
    I_{ij} = \int d^2\pos \rho(\pos-\pos_j)\Phi_\text{non}(\pos;\pos_i) = \int d^2\pos d^2\mbf{r'}\rho(\pos-\pos_i)\rho(\mbf{r'}-\pos_j)G^+(\pos,\mbf{r'},i\pi/2)\big[1-i\mbf{g}_i\cdot(\pos-\pos_i)\big],
\end{equation}
where we used the linear approximation of $\vartheta_i(\mbf{r'})$ as well as a small-angle expansion, swapped integration variables, and used $G(\pos,\mbf{r'},\varphi)=G(\mbf{r'},\pos,\varphi)$. By comparing to Eqs.~\eqref{eq:Jnon} and~\eqref{eq:K}, it can thus be seen that $J_{ij}^\text{non}/\mathcal{J}=\Re\{I_{ij}\}$ and $K_{ij}/\mathcal{J}=-\Im\{I_{ij} + I_{ji}\}$.

Assuming a Gaussian density and again using the linear approximation for $\vartheta_i(\mbf{r'})=\mbf{g}_i\cdot(\mbf{r'}-\pos_i)$, we can explicitly evaluate the local and nonlocal fields as
\begin{align}
    \Phi_\text{local}(\mbf{r};\mbf{r}_i)&=\frac{ w_0^2}{4\sigma_A^2}\left(\exp\left[-\frac{(\mathbf{r}-\mathbf{r}_i)^2}{2\sigma_A^2}+i\mathbf{g}_i\cdot(\mathbf{r}-\mathbf{r}_i)\right]+\exp\left[-\frac{(\mathbf{r}+\mathbf{r}_i)^2}{2\sigma_A^2}-i\mathbf{g}_i\cdot(\mathbf{r}+\mathbf{r}_i)\right]\right),\label{eq:Phi_local}\\
    \Phi_\text{non}(\mbf{r};\mbf{r}_i)&=\exp\left[-\frac{2\mathbf{r}^2\sigma_A^2}{w_0^4}-\frac{1}{2}\mathbf{g}_i^2\sigma_A^2\right]\cos\left[\frac{2}{w_0^2}\mathbf{r}\cdot(\mathbf{r}_i-i\mathbf{g}_i\sigma_A^2)\right].\label{eq:Phi_non}
\end{align}

\subsection{Discussion of density-wave gradients}
\label{sec:gradients}
Experimentally, we find that the gradients are related to vertex positions via $\mbf{g}_i= -2\pos_i/w_\text{eff}^2$. We now discuss this behavior in more detail. The relation between gradient $\mbf{g}_i$ and vertex position $\pos_i$ can be understood for individual vertices as a consequence of the nonlocal superradiant emission pattern that it generates: It is energetically favorable for the DW to adopt a gradient that matches the wavevector of the nonlocal field. A vertex at position $\pos_i$, generates a nonlocal field of the form $e^{i\mbf{k}\cdot\pos}+e^{-i\mbf{k}\cdot\pos}$, where the wavevector is $\mbf{k}=\pm2\pos_i/w_\text{eff}^2$. The expression for $\mbf{g}_i$ matches one of these wavevectors.

\begin{figure}[t!]
    \centering
    \includegraphics[width=0.7\textwidth]{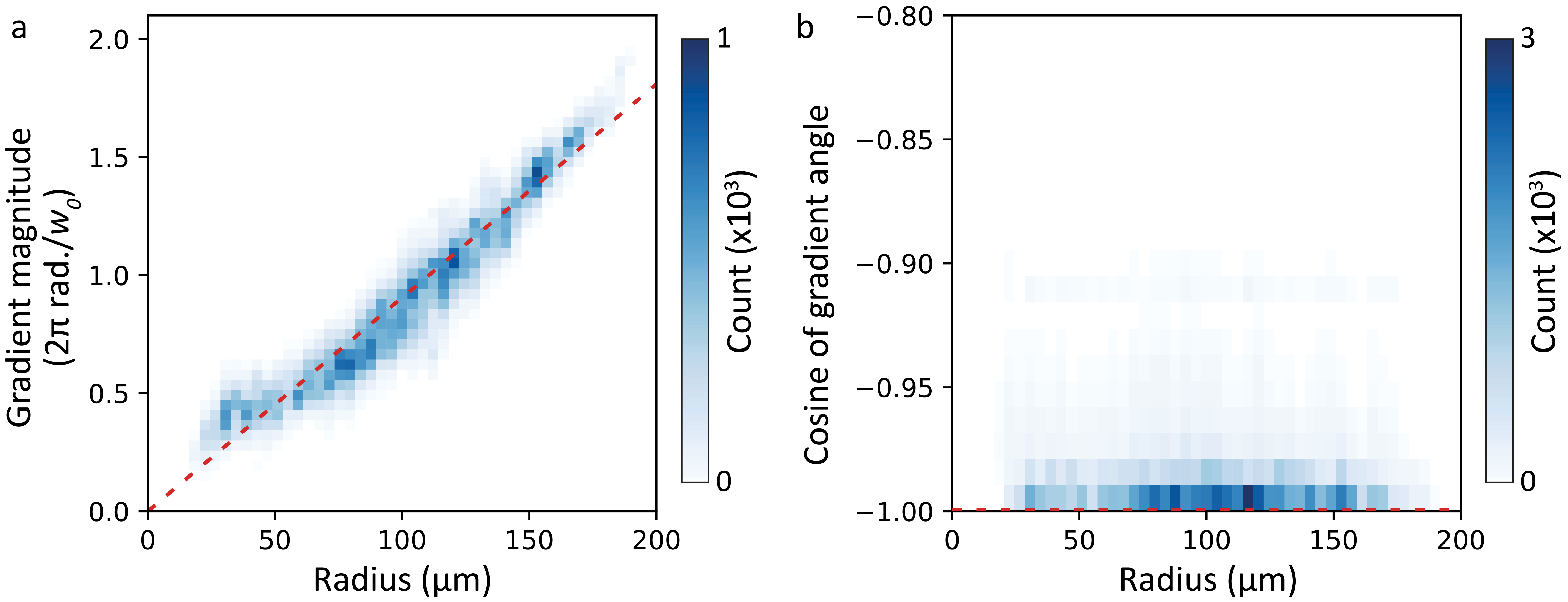}
    \caption{Observed density-wave gradients. (a) Two-dimensional histogram of observed gradient magnitudes and vertex positions. A strong correlation is observed, closely matching the relation $\mbf{g}=-2\mbf{r}/w_0^2$ indicated by the red dashed line. (b) 2D histogram of the gradient directions plotted by taking the cosine of their angle with respect to the position vector. A value of $-1$ ($+1$) indicates radially inward (outward) pointing gradients.}
    \label{fig:gradients}
\end{figure}

For superradiance experiments with multiple vertices, nonlocal fields interfere and the optimal phase evolution across a vertex may differ from this single-vertex optimal gradient. Nonetheless, we experimentally find that the gradients still closely follow this same prediction. We extract the gradients from the full fit as described in Sec.~\ref{sec:readout}. Their magnitude and direction are illustrated in Fig.~\ref{fig:gradients}, where we plot the statistics of the entire dataset, i.e., 12,300 experimental shots with 8 atomic vertices each. Figure~\ref{fig:gradients}a shows the strong correlation between gradient magnitude and distance from cavity center; the red dashed line is the relation $\abs{\mbf{g}}=2\abs{\pos}/w_0^2$. To assess the direction of the gradients, we show in Fig.~\ref{fig:gradients}b the inner product between gradient and position unit vectors, i.e., the cosine of the angle between the gradient direction and vertex position vector. These are all clustered at $-1$, indicating that the gradient points toward the cavity center.

Using the above relation between gradient and position, the $K$-matrix from Eq.~\eqref{eq:Khalfway} can now be simplified to
\begin{equation}
\label{eq:Kij}
    K_{ij} = \frac{2\mathcal{J}\sigma_A^2}{w_\text{eff}^2}\exp(-\frac{2\sigma_A^2}{w_\text{eff}^2}\frac{\pos_i^2+\pos_j^2}{w_0^2})\left[\frac{4w_0^2\,\pos_i\cdot\pos_j}{w_\text{eff}^4}\sin(\frac{2\pos_i\cdot\pos_j}{w_\text{eff}^2})+\frac{4\sigma_A^2(\pos_i^2+\pos_j^2)}{w_\text{eff}^4}\cos(\frac{2\pos_i\cdot\pos_j}{w_\text{eff}^2})\right].
\end{equation}
The form of this is no longer gradient-dependent.  This position-dependent form is used in this work and presented in an approximate form in the main text.

We also performed numerical simulations to verify the formation of phase gradients. Simulated annealing for the semiclassical energy Eq.~\eqref{eq:angleOnly} is used to probe the ground-state spin configuration. We represent each vertex by an ensemble of 100 individual spins that are spatially distributed within the vertex with standard deviation $\sigma_A=4$~$\mu$m. This method directly probes finite-size effects within a vertex, and so the approximate $K$-matrix interaction is not used. We verify that gradients develop within the vertices for systems up to at least 8 vertices. In particular, when there is only one vertex, the gradient closely matches the wavevector of the nonlocal field as predicted above. However, for system sizes $n>1$ we find that the gradient no longer follows this simple prediction, instead forming much less predictable gradients. This can be understood by the fact that the phase gradient now depends on the sum of nonlocal fields from all vertices. Further study is needed to understand the difference between these simulations and experimental observations. Nevertheless, the heuristic model presented above conforms well to our experimental observations.

\subsection{Semiclassical limit}

We use a semiclassical description to describe the experimental situation of many atoms per vertex. Each vertex can be described by a 3-component vector $(S_i^x, S_i^y, S_i^z)$. The spin components are defined in terms of expectation values of the atomic field operators, $S_i^x = \langle \hat{\chi}_{c,i} \rangle/N$,  $S_i^y = \langle \hat{\chi}_{s,i} \rangle/N$, and $S_i^z = \langle \hat{\psi}^\dag_{c,i}\hat{\psi}_{c,i}^{} + \hat{\psi}^\dag_{s,i}\hat{\psi}_{s,i}^{} - \hat{\psi}^\dag_{0,i}\hat{\psi}_{0,i}^{} \rangle/N$. Normalization of the atomic wavefunction constrains the semiclassical spin vector to lie within the unit sphere. Taking the expectation value of Eq.~\eqref{eq:HquantumK} and performing a mean-field decoupling of $\chi$ operators yields the semiclassical energy
\begin{align}\label{eq:Esemiclassical}
     E &= N\left[2E_r+\frac{\Omega^2}{4\Delta_A}\right]\sum_{i=1}^n S_i^z -N^2\sum_{i=1}^n J^{\text{local}}_{ii}\left[\big(S_i^x\big)^2+\big(S_i^y\big)^2\right]-N^2\sum_{i,j=1}^n \left[ J_{ij}^{\text{non}}\left(S_i^x S_j^x - S_i^y S_j^y \right)+ K_{ij}\left(S_i^x S_j^y + S_i^y S_j^x\right) \right].
\end{align}

Far above the superradiant threshold, the transverse field term ${\propto} S_i^z$ plays little role and the spin vectors lie near the $xy$-plane. Approximating the spins as laying fully within the $xy$-plane allows for a polar coordinate representation $S_i^x=s_i\cos\theta_i$ and $S_i^y=s_i\sin\theta_i$, resulting in a simplified energy
\begin{equation}\label{eq:Eint_semiclassical}
    E = -N^2 \sum\limits_{i=1}^n J_{ii}^\text{local}s_i^2 - N^2\sum\limits_{i,j=1}^ns_is_j\left[J_{ij}^\text{non}\cos(\theta_i+\theta_j) + K_{ij}\sin(\theta_i+\theta_j)\right].
\end{equation}
Furthermore, the ratio of local-to-nonlocal interaction prefactors are $J^\text{local}/J^\text{non}\approx10$, using the observed $\sigma_A\approx4$~$\mu$m. The interaction energy is thus dominated by the local energy, which ensures that the length of each spin vector is maximized, $s_i=1$. We can then use an angle-only model
\begin{equation}\label{eq:angleOnly}
    E = -N^2\sum_{i,j=1}^n \left[J_{ij}^\text{non}\cos(\theta_i+\theta_j) + K_{ij}\sin(\theta_i+\theta_j)\right].
\end{equation}
This is equivalent to Eq.~(1) of the main text after a rewriting using spin components $S^x$ and $S^y$ and the dropping of the superscript on $J^\text{non}_{ij}$ for simplicity. In the main text we define $J_0\equiv N^2\mathcal{J}$ and omit the negligible contribution from $\kappa$. 

The expression for the semiclassical intracavity field follows from Eq.~\eqref{eq:quantumfield}, with the above substitutions, and results in 
\begin{equation}\label{eq:forward_field}
    \Phi^F(\mbf{r}) = \frac{Ng_0\Omega}{8\Delta_A(\Delta_C+i\kappa)}\sum\limits_i s_i \big[\Phi_\text{local}(\mbf{r};\mbf{r}_i) e^{-i\theta_i} + \Phi_\text{non}(\mbf{r};\mbf{r}_i) e^{i\theta_i}\big].
\end{equation}
Hence it is clear that the information about the DW phase $\theta_i$ for each atomic vertex is encoded in the light-field we detect using holographic imaging.

We verify that the spin vectors indeed lie close to the $xy$-plane at the equator through numerical simulations. Equilibrium distributions of the semiclassical energy in Eq.~\eqref{eq:Esemiclassical} are simulated via parallel-tempering Monte Carlo.  The simulation uses vertex positions and widths identical to those measured in the experiment to generate $J_{ii}^\text{local}$, $J_{ij}^\text{non}$, and $K_{ij}$. Within the semiclassical approximation, we find the superradiant threshold by performing a linear stability analysis of the normal phase. This becomes classically unstable when $\frac{1}{N}\left[2E_r+\frac{\Omega^2}{4\Delta_A}\right] < \max_i \lambda_i$, where $\lambda_i$ are the eigenvalues of the block matrix:
\begin{equation}\label{eq:thresholdmatrix}
    M = \begin{bmatrix}
    J^\text{local} + J^\text{non} & K \\
    K & J^\text{local} -J^\text{non}
    \end{bmatrix}.
\end{equation}
We assume a pump power (${\propto}\Omega^2$) that is a factor 1.25 above the critical coupling strength for the superradiant transition.  The radius of spins in their equilibrium states is recorded at a temperature of $0.03 \times \max_i \lambda_i$. We find that the average total spin radius is ${\geq}0.97$ with a standard deviation ${\leq}3.7\%$ across an ensemble of disorder realizations. This implies that the spins are close to the boundary of the unit sphere. Additionally, the average radius in the $xy$-plane ranges between 0.87 and 0.94 across disorder realizations, with a standard deviation ${\leq}6.7\%$. Other than a global rescaling, we can therefore approximate $s_i \approx 1$ for all $i$, which implies that approximating the semiclassical energy expression with the angle-only model is valid in this regime. 

Finally, we discuss two additional terms that could be added to this energy. One could introduce a 780-nm intracavity field via longitudinal pumping of the cavity. Use of a digital micromirror device or spatial light modulator enables phase-sensitive local addressing of vertices; the former was demonstrated in~\cite{Guo2021aol}. This would result in an energy term corresponding to that of a local longitudinal field,
\begin{equation}
    E_\text{longitudinal} = \sum\limits_{i=1}^n h_i \cos(\theta_i - \phi_i),
\end{equation}
where $h_i$ is the effective field strength and $\phi_i$ is the spatial phase of the driven intracavity light. This would enable the measurement of magnetic susceptibilities. Additionally, a 1560-nm longitudinal probe would result in an optical potential for the DWs that corresponds to an easy-axis term
\begin{equation}
    E_\text{easy-axis}=\sum\limits_{i=1}^n f_i\cos^2(\theta_i - \varphi_i),
\end{equation}
where $f_i$ is the easy-axis strength and $\varphi_i$ its direction.

\subsection{$\mathbb{Z}_2$ symmetries}\label{sec:props}

The effective energy in Eq.~\eqref{eq:Eint_semiclassical} possesses a global $\mathbb{Z}_2$ symmetry; we now describe its effect on the overlaps. For brevity, we denote the configuration of all spins together as $\bm{\theta} = \{\theta_i\}$ . The transformation $\bm{\theta}\rightarrow\bm{\theta}+\pi$ leaves the energy invariant. As in other superradiance experiments~\cite{Mivehvar2021cqw}, this symmetry is spontaneously broken at the superradiant phase transition. Experimental evidence for this in the spin network is presented in the next section. To understand how this transforms the overlap parameter, we can apply this global Ising spin-flip on replica $\alpha$ but not on replica $\beta$. Then, in terms of $Q,R$, the linear combinations of overlaps matrices defined in the main text, the Ising symmetry results in $(Q,R)\rightarrow(-Q,-R)$. This implies that the overlap distributions are symmetric under a 180 degree rotation.

Before considering finite-size effects in Sec.~\ref{theory:finite-size}, we noted two independent $\mathbb{Z}_2$ symmetries, $\hat{\psi}_{c,i}\rightarrow -\hat{\psi}_{c,i}$ and $\hat{\psi}_{s,i}\rightarrow -\hat{\psi}_{s,i}$ (for all $i$). The joint transformation corresponds to the Ising symmetry discussed above. Applying the transformation of only the sine component results in $\bm\theta \rightarrow -\bm\theta$ for the semiclassical spin configuration. Physically, this corresponds to the reflection of the DWs in the cavity midplane $z=0$. The resulting transformation on the overlap parameters can again be understood by applying this operation on only one replica. This results in $(Q,R)\rightarrow(R,Q)$, i.e., a reflection of the overlap distribution along the diagonal. This symmetry is missing in the observed spin overlap distributions, except in rare cases such as Fig.~4c of the main text (see also Sec.~\ref{sec:alloverlaps}). Indeed, the finite-size Hamiltonian Eq.~\eqref{eq:HquantumK} [or its semiclassical equivalent Eq.~\eqref{eq:Esemiclassical}] is not invariant under this transformation. The symmetry is explicitly broken by the choice of DW gradients $\mbf{g}_i$. In principal, both $\pm\mbf{g}_i$ configurations should lead to low-energy states. In practice we find only one of these sets, as discussed in Sec.~\ref{sec:gradients}. Potential reasons for this symmetry-breaking are the residual $\sim10$-$\mu$m displacement of the atoms from the cavity midplane, remnant cavity mode dispersion, cavity mirror aberrations, or nonlinearities arising from the imaging readout~\cite{Guo2019eab}. The 1560-nm light used to stabilize the cavity length could also introduce a bias via the easy-axis term it induces.

\subsection{Local gauge rotations}
\label{sec:gauge}
We can also consider what happens if we apply the Ising transformation discussed above to only a single spin: $\bm{\theta}\rightarrow\bm{\theta}+\pi \bm{e}_i$, where $\bm{e}_i$ is the $i$'th unit vector. This is a local transformation of spin $i$ that, by itself, does not leave the energy invariant. This can become a local gauge transformation that leaves the energy invariant by making a corresponding change on the $J$- and $K$-matrices, specifically by flipping the sign of each element in the $i$'th row as well as in the $i$'th column. The signs of the diagonal elements $J_{ii}$ and $K_{ii}$ are thus flipped twice and remain the same. These local gauge transformations can be applied to transform a staggered ferromagnet into a conventional one and vice versa, leaving the energy landscape invariant.

\begin{figure*}[t!]
    \centering
    \includegraphics[width=\linewidth]{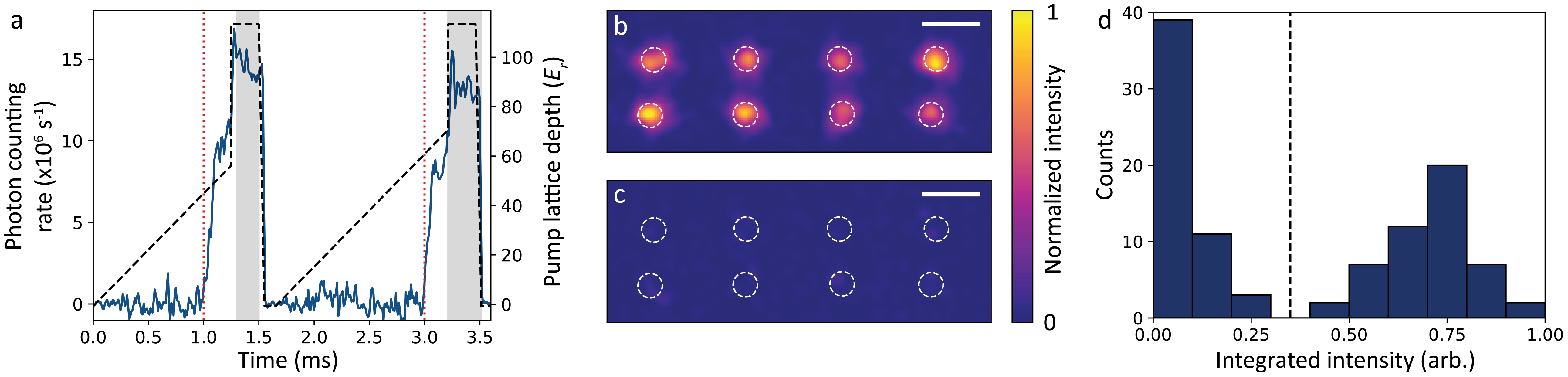}
    \caption{Experimental evidence for the presence of a $\mathbb{Z}_2$ Ising symmetry. (a) Plot of the double ramp sequence accompanied by a typical cavity emission record. The pump power (dashed black line, right axis) is ramped through the superradiant transition twice, with two readout periods shown by gray regions where the pump is rapidly increased to high power. A typical cavity emission pattern (blue, left axis) shows two superradiant emissions with the superradiant thresholds demarcated by red dashed lines. (b) An example intensity profile of a demodulated image in which individual holograms from the two superradiant emissions constructively add to enhance the fringes. (c) An example intensity profile in which the two holograms cancel each other to diminish the fringe contrast. Panels (b,c) are shown on the same color scale. (d) Histogram of the integrated intensities of 103 experimental cycles. The signal is derived by integrating the field intensity within the white dashed circular regions of panels (b,c).  A dashed line at 0.33 separates populations that correspond to holograms wherein the images constructively versus destructively add to each other}.
    \label{fig:Z2}
\end{figure*}

\section{Experimental verification of $\mathbb{Z}_2$ symmetry}\label{sec:Z2}

As discussed in Sec.~\ref{sec:props}, the system contains a $\mathbb{Z}_2$ Ising symmetry in which the spin state $\bm{\theta}$ may be transformed into $\bm{\theta}+\pi$. Unfortunately, it is not possible to distinguish between these two states from a single holographic image, due to the following reasons. The confocal field in Eq.~\eqref{eq:fullField} is invariant under the joint transformation $\bm{\theta}\to\bm{\theta}+\pi$ and $\phi\to\phi+\pi$, where $\phi$ is phase of the LO beam with respect to the cavity emission. While $\phi$ is stable over the course of a single experiment, it is not stable from shot to shot, and thus the two $\mathbb{Z}_2$ related states cannot be distinguished using the readout method described in Sec.~\ref{sec:readout}.

Nevertheless, we are able to perform an independent set of experiments that demonstrate the existence of this Ising symmetry. As in earlier work~\cite{Baumann2011esb,Kollar2017scw,Kroeze2019dsc}, the long-term phase fluctuations of the LO can be circumvented by repeatedly ramping in and out of the superradiant phase within a single experimental shot. We accomplish this using the ramp schedule shown in Fig.~\ref{fig:Z2}a.  This is shown together with a typical cavity emission for a configuration of vertices resulting in a \yFM. We change the cavity detuning to $\Delta_C/2\pi = -40$~MHz to enhance the signal strength for these types of experiments. This change is sufficiently small as to not affect the ferromagnetic behavior of this configuration:  Indeed, only two global minima oriented along $\theta_i=\pm \pi/2$ are commonly found. To demonstrate this discrete symmetry, two ramps and readout periods follow in close succession, with two superradiant emissions separated by approximately 2~ms. The ramp rate is kept the same as described in the main text, while the ramp durations are optimized to balance the integrated intensity of the two superradiant emissions. 

We highlight the use of holographic imaging, rather than temporal heterodyne, to observing spontaneous $\mathbb{Z}_2$-symmetry breaking.  This differs from earlier experiments~\cite{Baumann2011esb,Kollar2017scw,Kroeze2019dsc} and works in the following manner. Holographic fringes from interference between the cavity emission and the LO beam accumulate only during the readout periods, shown as the gray regions in Fig.~\ref{fig:Z2}a. The camera is exposed during the full ramp sequence, which allows these two holograms to add to each other on the camera. The LO phase is sufficiently stable over this short period to allow a signal to develop. The \yFM states of opposite $\mathbb{Z}_2$ symmetry would produce cavity emission patterns $\pi$ out of phase with each other. If the Ising symmetry is broken spontaneously at the superradiant phase transition, then the phases of the two pulses would be completely independent.  Thus, the two sequential holograms sum in a way that is equally likely to enhance or cancel the holographic fringes. Indeed, we observe both enhancement and cancellation as expected from the Ising symmetry. Figures~\ref{fig:Z2}b,c show normalized intensity profiles of the cavity emission for two independent repetitions of the experiment, where Fig.~\ref{fig:Z2}b (Fig.~\ref{fig:Z2}c) demonstrates enhancement (cancellation) of the individual holograms. 
This approach can also be thought of as recording a two-time correlation measurement of the state between the two different pulse sequences; such a method may potentially be further extended as a route to multitime correlations.

The probability to realize identical or opposite Ising states is measured over 103 independent repetitions of the experiment. To quantify the level of fringe enhancement in an image, the intensity of the total demodulated field at the location of the vertices is integrated. The regions of integration are demarcated by dashed circles in Figs.~\ref{fig:Z2}b,c. The histogram of integrated intensity over all experiments is shown in Fig.~\ref{fig:Z2}d. A value of zero corresponds to full cancellation of the holographic fringes; the integrated intensity is normalized by the largest value measured. Two distinct populations are found, corresponding to enhanced versus canceled holographic fringes. Separating the two populations at a value of 0.33 results in 48.5\% (51.5\%) of experiments realizing enhancement (cancellation) of the holograms. This is in close agreement with the 50\% probability expected from spontaneously breaking the Ising symmetry. Because our measurements that involve an individual hologram are not sensitive to the way this symmetry is broken in the experiment, all overlap and magnetization distributions shown in the main text are explicitly symmetrized for clarity.

\begin{figure}[t!]
    \centering
    \includegraphics[width=0.6\textwidth]{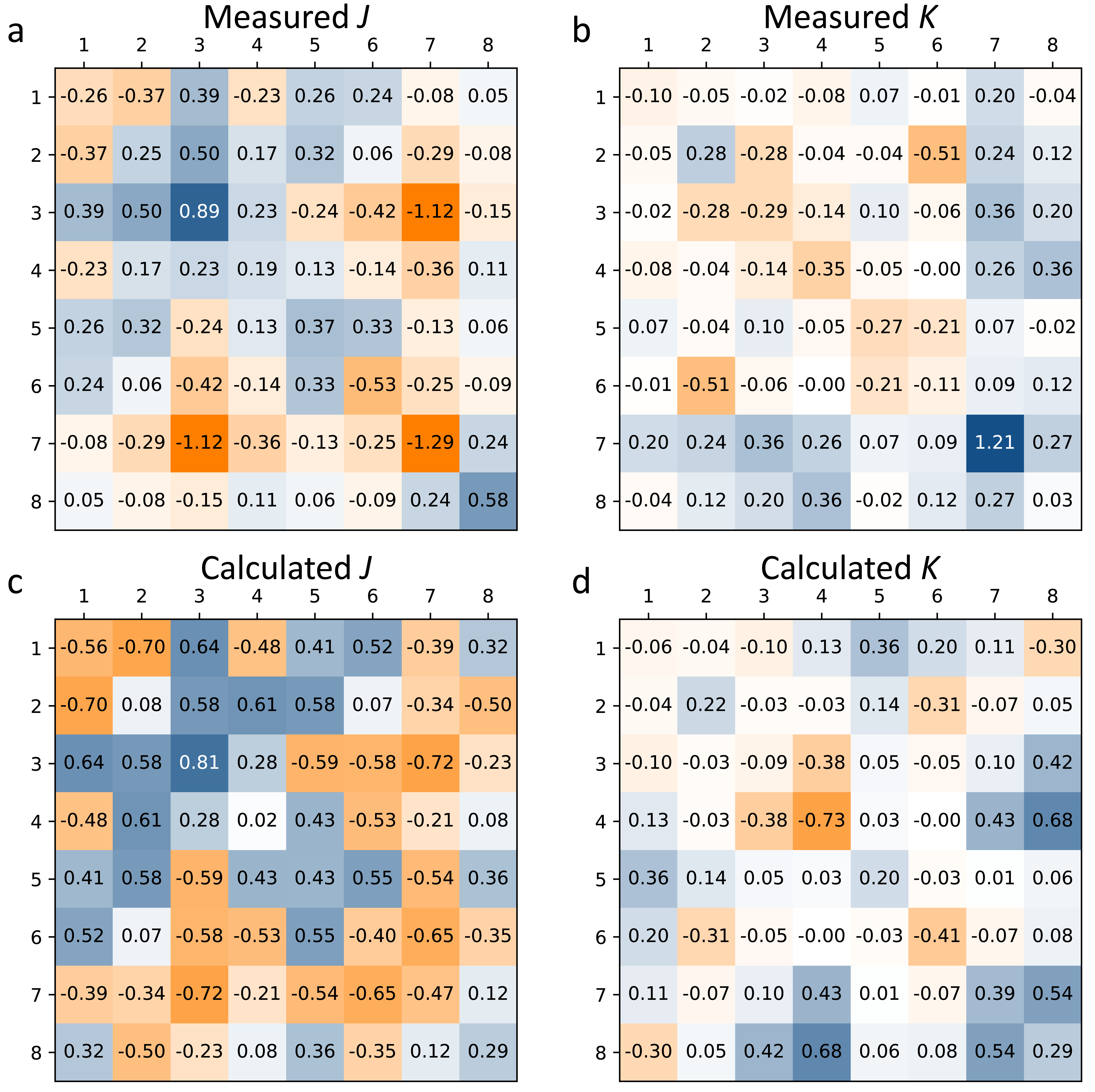}
    \caption{Comparison between measured and calculated interaction matrices for the disorder instance of Fig.~\ref{fig3}. (a,b) Measured $J$ and $K$ matrices, respectively. (c,d) $J$ and $K$ matrices, respectively, calculated using Eqs.~\eqref{eq:Jij_non} and \eqref{eq:Kij} based on vertex position data and measured width of the vertices. Color scale is the real part of the color scheme in Fig.~1c of the main text.}
    \label{fig:JKcomparison}
\end{figure}

\begin{figure}
    \centering
    \includegraphics[width=0.6\textwidth]{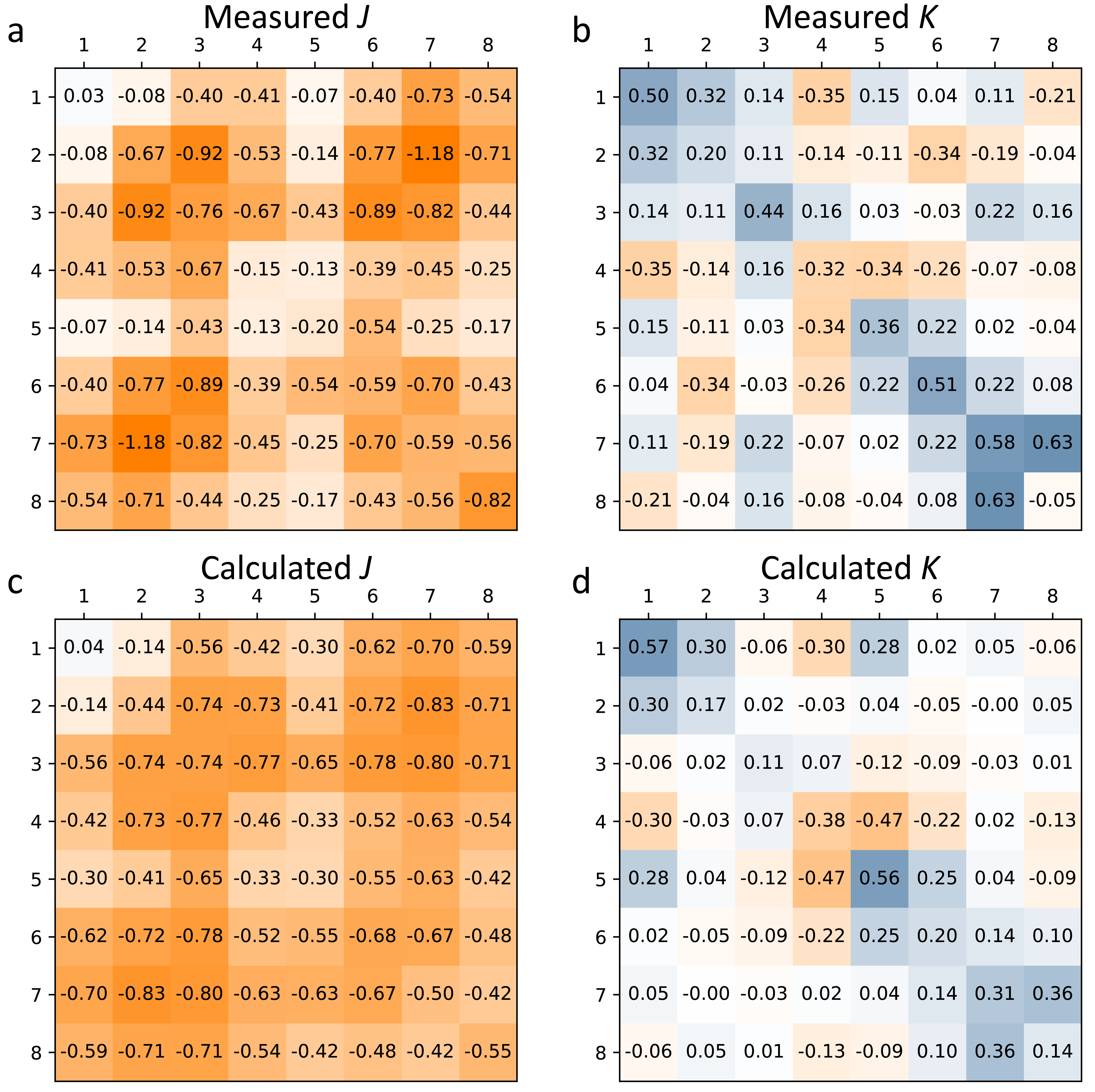}
    \caption{Comparison between measured and calculated interaction matrices for the $y$-ferromagnetic instance of Fig.~\ref{fig2}. (a,b) Measured $J$ and $K$ matrices, respectively. (c,d) $J$ and $K$ matrices, respectively, calculated using Eqs.~\eqref{eq:Jij_non} and \eqref{eq:Kij} based on vertex position data and the measured width of the vertices. Color scale is the real part of the color scheme in Fig.~1c of the main text.}
    \label{fig:JKcomparison_FM}
\end{figure}

\section{Measurement of the $J$ and $K$-matrices}

The $J$ and $K$ matrices can be calculated based on the positions $\pos_i$ of the atomic vertices in a perfect confocal cavity; see Eqs.~\eqref{eq:Jij_non} and~\eqref{eq:Kij}. We also independently measure these matrices, as discussed in this section, to assess deviations from the calculated matrices due to cavity imperfections. The measurement procedure is as follows. For a given position configuration, we isolate a single vertex $i$, as described in Sec.~\ref{sec:bec}. Then we perform a superradiance experiment using only this single vertex and record the intracavity field using holographic imaging. This reveals the intracavity field generated by vertex $i$, which we denote $E_i(\pos)$. Since the interactions are photon-mediated, this directly provides the interaction coupling strength via the theory presented in Sec.~\ref{sec:Efield}. A vertex at position $\pos_j$ experiences total interaction $E_i(\pos_j)$; note that we actually take a Gaussian average around $\pos_j$ to account for the finite vertex width $\sigma_A$. The elements $J_{ij}$ and $K_{ij}$ correspond to the real and imaginary part of this quantity, respectively. The recorded cavity field thus provides the entire $i$'th row of both $J$ and $K$ matrices. 

To reduce imaging noise when performing these interaction matrix calibrations, we average together the fields from approximately 30 repetitions of the same superradiance experiment. Repeating this procedure for each vertex provides the values of independent rows of the $J$ and $K$ matrices. We rescale these rows because the brightness of the hologram is not consistent across the vertex positions: Using the fact that the $J$ and $K$ matrices are symmetric, this rescaling is performed by algorithmically minimizing $\norm{J - J^\intercal}/\norm{J}$. The same rescaling is used for both $J$ and $K$ matrices, and the antisymmetric part of the resulting matrices is discarded.

This measurement process is very time consuming, so we perform this measurement for only a few selected position configurations.  Namely, the two ferromagnetic configurations presented in Fig.~2, as well as the disorder instances in Fig.~3 and Figs.~4a--d. We find good overall  agreement between the measured $J$ and $K$ matrices and those calculated from Eqs.~\eqref{eq:Jij_non} and~\eqref{eq:Kij}. Figures~\ref{fig:JKcomparison} and \ref{fig:JKcomparison_FM} present a direct comparison for the position configuration of Fig.~3 and Fig.~2 of the main text, resp. The latter has been gauge-rotated as discussed in Sec.~\ref{sec:gauge}. The measured matrices can be determined up to only a global rescaling. In Figs.~\ref{fig:JKcomparison} and \ref{fig:JKcomparison_FM}, this scale is chosen to maximize the similarity between the experimental and calculated $J$ matrix.  We reiterate that the calculated matrices assume perfect confocality, which in practice is hampered by, e.g., astigmatism and mirror surface defects~\cite{Kollar2015aac}. The calculated $K$-matrix includes further approximations, hence the difference between measured and calculated $K$-matrices tends to be more significant.

\begin{figure}
    \centering
    \includegraphics{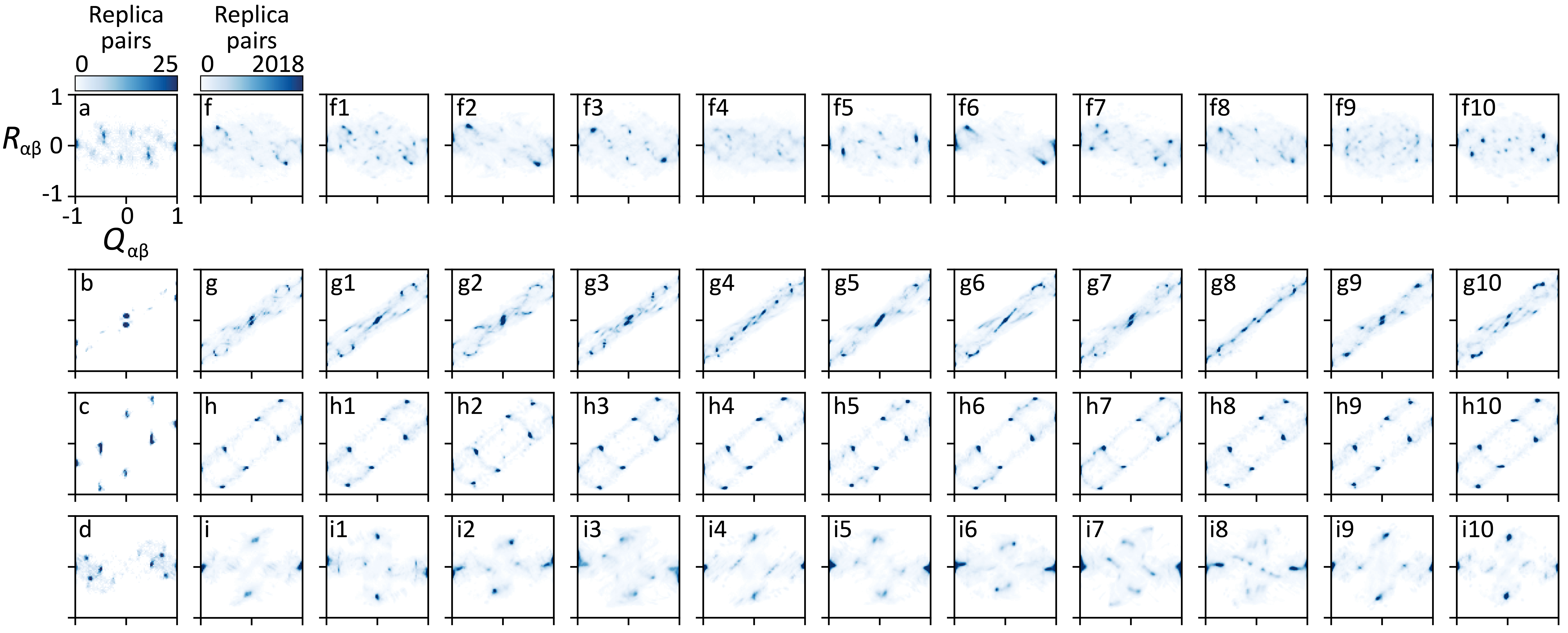}
    \caption{Sensitivity of overlap distributions to measurement errors in $J$ and $K$. (a-d) Experimentally observed overlap distributions, from Fig.~4 of the main text. (f-i) Original mean-field simulated overlap distributions, from Fig.~4 of the main text. (f1-i10) Simulated overlap distributions using bootstrapped interaction matrices. All look qualitatively similar to the original simulations in panels (f-i).}
    \label{fig:JKsensitivity}
\end{figure}

The error for the measured interaction matrices can be estimated by performing bootstrap resampling of the set of superradiance experiments used in each measurement. To assess the sensitivity of the overlap distribution to these measurement errors, we perform mean-field trajectory simulations as described in Sec.~\ref{sec:mf}. We focus on the 4 disorder instances presented in Fig.~4 of the main text and simulate 10 bootstrap disorder instances each. The resulting overlap distributions are shown in Fig.~\ref{fig:JKsensitivity}. The first two columns represent the experimentally observed and originally simulated overlap distributions respectively, copied from the respective panels in Fig.~4 of the main text for convenience. The simulations in the remaining columns, indexed 1 through 10, are all qualitatively similar to the original simulations in the second column, showing that errors in the measurement of $J$ and $K$ are not significant. 

\section{Equilibrium properties of the model}\label{sec:pt}
In this section we describe the properties of a statistical mechanical model with energy given by Eq.~\eqref{HamSpin} in thermal equilibrium. Here, the interaction matrices will be generated in the same way as the SK model, that is all-to-all, symmetric interactions, identically and independently drawn (i.i.d.) from a Gaussian distribution. Specifically, $J_{ij}\sim \mathcal{N}(\bar J/n,\J^2/n)$ and $K_{ij}\sim \mathcal{N}(\bar K/n,\K^2/n)$, where $\mathcal{N}(\mu,\sigma^2)$ is the normal distribution with mean $\mu$ and variance $\sigma^2$. We note that this approach ignores some of the correlations (between elements as well as between $J$ and $K$ matrices) generated by the confocal interactions Eqs.~\eqref{eq:Jij_non} and \eqref{eq:Kij}. The equilibrium properties for our vector spin model using realistic interactions will be subject of future work; see also \cite{Marsh2021eam,Marsh2024ear,Erba2021sgp} for results with the Ising model. We set $\K/\J = 0.5$, which is slightly larger than the ratio found in the ensemble of 123 disorder instances used in this work. Following the same program, we set $\bar K = 0$ and use $\bar J$ to select between ferromagnetic ($\abs{\bar J}\gg 1$) and glassy ($\abs{\bar J}\ll1$) ensembles.

We use parallel-tempering Monte Carlo simulations~\cite{Swendsen1986rmc} to study equilibrium properties of our system. The simulations follow the even/odd deterministic swapping algorithm~\cite{Rozada2019eos,Hukushima1999dfe}, typically with 20 (roughly) geometrically sampled temperatures ranging from $0.1\J$ to $2\J$. States are sampled from the angle-only energy model Eq.~\eqref{eq:angleOnly} to yield equilibrium overlap and magnetization distributions. We typically use $10^6 n$ update steps to evolve a replica at each temperature, each with a local Gaussian proposal function of standard deviation $\pi/8$. Swaps are performed every $n$ steps. We find that this is sufficient to achieve convergence of the overlap distribution for system sizes up to $n = 100$. We use approximately 1000 disorder instances for each simulation.  As in the main text, we primarily focus on $q^{xx}_{\alpha\beta}$ and $q^{yy}_{\alpha\beta}$ overlaps. The simulations yield an overlap distribution as  a two-dimensional histogram for each disorder instance and are aggregated into a Parisi order parameter.

\subsection{Finite-size scaling}

We are interested in the behavior of the order parameter in the thermodynamic limit, where the system size goes to infinity. Simulating large system sizes is difficult, especially for continuous models~\cite{Yavorskii2012ogs}. Instead, we perform simulations for small and intermediate system sizes, up to $n = 100$, and attempt to extrapolate trends to the thermodynamic limit.

\begin{figure}[htbp]
    \centering
    \includegraphics[width=0.85\textwidth]{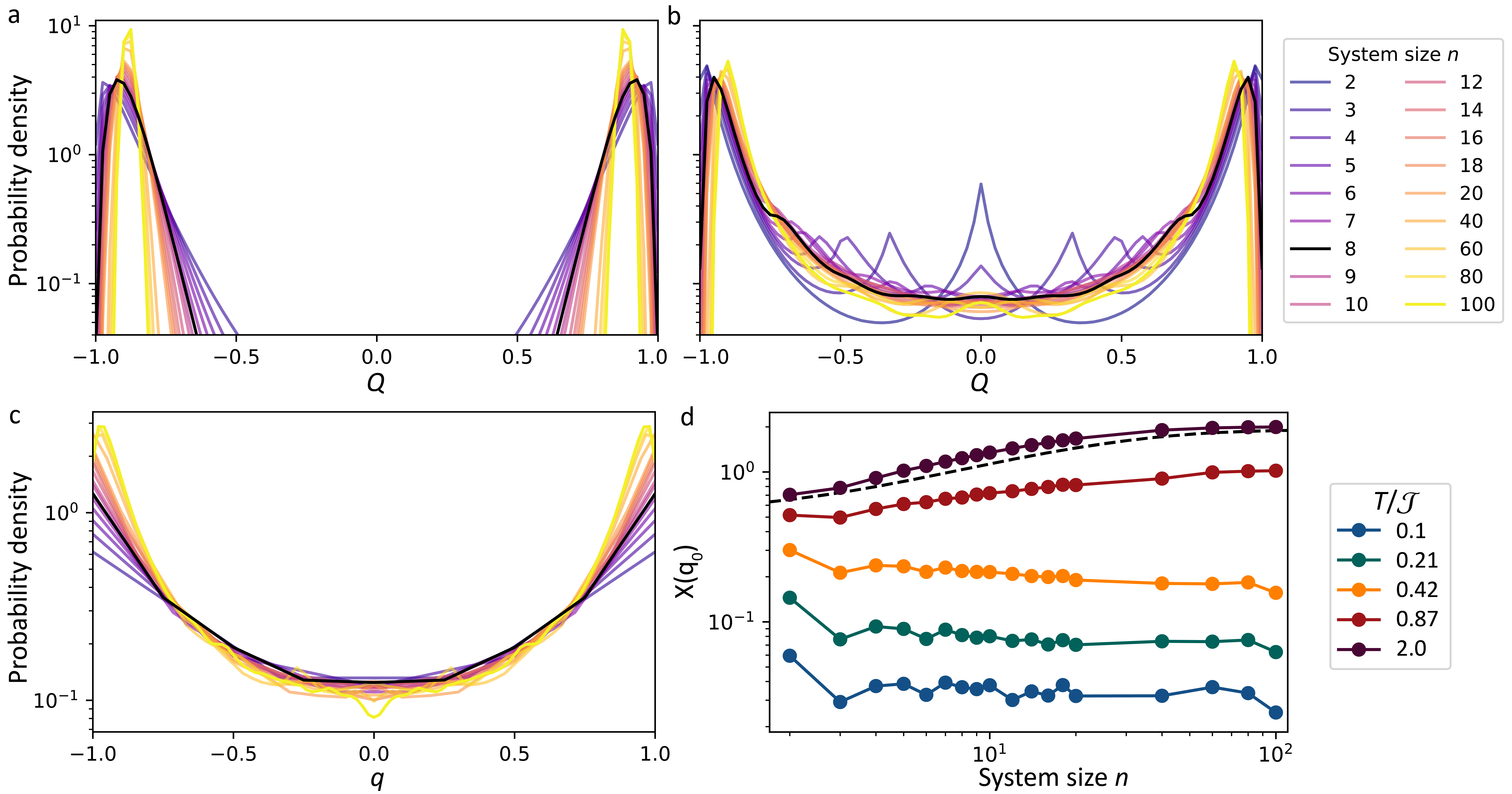}
    \caption{Finite-size scaling analysis of the equilibrium overlap distribution. (a,b) Marginal for $Q$ of the Parisi order parameter distribution versus system size, for (a) a ferromagnetic ensemble and (b) a spin glass at $T/\J=0.21$. $n=8$ is highlighted in black. (c) Parisi order parameter distribution for the Ising SK model at the same temperature. (d) Scaling of $P(Q=0)$ with system size for various temperatures for our vector model. Lack of a significant downward trend demonstrates that the plateau in the Parisi order parameter distribution is robust.}
    \label{fig:PTscaling}
\end{figure}

First we verify ferromagnetic behavior. We do this by setting $\bar J=5$, which is comparable to the ferromagnetic instances used in Fig.~\ref{fig2}. We extract the marginal for $Q$ at a temperature well into the ordered regime, $T/\J = 0.21$. The results for various system sizes are shown in Fig.~\ref{fig:PTscaling}a. We observe the expected Edwards-Anderson goalposts, which narrow into delta functions at $\pm Q_{\alpha\alpha}$ as the system size is increased.

Next we switch to a spin glass ensemble by setting $\bar J=0$. The resulting $Q$-marginals are shown in Fig.~\ref{fig:PTscaling}b and display nonzero overlap between the goalposts for all system sizes studied. We observe sharp peaks in the distribution for small system sizes. These are similar to those found in the experiment, though less pronounced at $n = 8$ than those experimentally observed (cf.~Fig.~\ref{fig4}k). We further note that if we set $\K = 0$ (i.e., remove the $K$-matrix term altogether), then these peaks disappear and the distribution is smooth throughout the interior.

The distributions for $n = 8$ and $n = 100$ appear close, which is qualitatively indicative of a rapid convergence to the large system-size limit. For comparison, Fig.~\ref{fig:PTscaling}c shows the numerically equilibrated Parisi order parameter distribution for the Ising SK model at the same temperature ($T/\J= 0.21$). There, $n=8$ and $n=100$ appear more distinct, particularly in terms of formation of the goalposts.

To quantitatively assess the presence of RSB, we focus on the central ``plateau'' region of the overlap distribution, where, for RSB, one expects a nonzero value.  To avoid issues with numerical sensitivity, the plateau is traditionally assessed by averaging the overlap distribution in the vicinity of $q=0$; see, e.g., Refs.~\cite{Hartmann2000hte,Katzgraber2001mcs}. We generalize this approach for our vector spin model with 2D overlap distributions: Because we are interested in the plateau of the $Q$-marginal, we define
\begin{equation}
    X(q_0) = \frac{1}{2q_0}\int\limits_{-q_0}^{q_0}dQ\int\limits_{-1}^{1}dR\,P(Q,R).
\end{equation}
The results are shown in Fig.~\ref{fig:PTscaling}d for 4 different temperatures, all with $q_0=0.26$. For high temperatures, the paramagnetic overlap distribution can be well captured with an isotropic 2D Gaussian distribution with variance equal to $n$. Then, 
\begin{equation}
    P(Q,R) = \exp(-\frac{Q^2+R^2}{2n}) \Big/ \int\limits_{-1}^1dQ\int\limits_{-1}^1dR \exp(-\frac{Q^2+R^2}{2n}),
\end{equation}
and $X(q_0)$ can be calculated analytically. This is indicated by the dashed black line, and the highest simulated temperature is in good agreement. For temperatures well into the ordered regime, we observe no significant dependence on system size. We estimate error bars by performing a bootstrap analysis and find that these are smaller than the data points shown. We conclude that the vector spin model hosts RSB for these types of $J$ and $K$ couplings. 

\begin{figure}[htbp]
    \centering
    \includegraphics[width=0.85\textwidth]{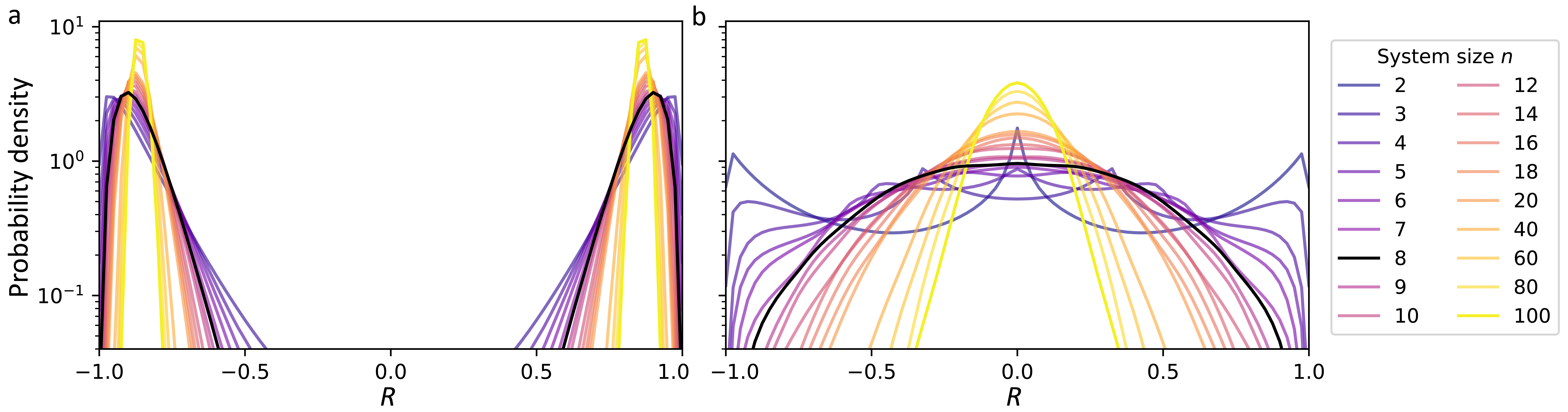}
    \caption{Finite-size scaling analysis of the equilibrium overlap distribution. (a,b) Marginal for $R$ of the Parisi order parameter distribution versus system size, for (a) a ferromagnetic ensemble and (b) a spin glass at $T/\J=0.21$. $n=8$ is highlighted in black.}
    \label{fig:PTscalingR}
\end{figure}

We perform a similar analysis for the $R$-marginal. These are presented in Fig.~\ref{fig:PTscalingR} for the same simulations as in Fig.~\ref{fig:PTscaling}a,b. For the ferromagnet, we again observe goalposts, consistent with the distribution shown in Fig.~\ref{fig2}c. Since $\bar J>0$, we expect $q^{xx}\neq0$ and $q^{yy} = 0$, resulting in $Q = R$. The simulations indeed demonstrate this correlation (not shown). For the spin glass ensemble, the $R$-marginal is shown in Fig.~\ref{fig:PTscalingR}b. Larger system sizes yield an increasingly narrow distribution, suggesting that the $R$-marginal becomes a $\delta$-peak at $R=0$ at large $n$.  This indicates that $q^{xx}_{\alpha\beta} = q^{yy}_{\alpha\beta}$ and can be understood as an absence of any overlap anisotropy. We revisit overlap anisotropy and the absence thereof in Subsec.~\ref{subsec:nematicity}.

We now add a discussion regarding the ability to observe finite-size scaling in the presence of finite-size limitation. ``True" spontaneous symmetry breaking is a thermodynamic limit effect, since it is about the non-commuting limits of long times and large system sizes.  That is, it is about the fact that at infinite system size, there is never any switching between symmetry broken states.   For finite systems, switching can occur, but the timescales may be very long.  Practically speaking, since experiments are finite in time as well as system size, that means that the switching that may occur on very long times is irrelevant. The practical question that matters is whether a system is large enough that one sees symmetry breaking on sufficiently long timescales (i.e., those of the experiment).  The data we show below in Sec.~\ref{MFfinitesize} from finite-size MF simulations gives clear evidence that it is large enough in that sense.

\subsection{Phase diagram}
In the previous subsection we showed the finite-size scaling of a ferromagnet and a spin glass. The transition between these phases is controlled by $\bar J$. We will now attempt to locate the boundaries between these phases, as well as a high-temperature paramagnetic phase. As before, $\K/\J$ is fixed and the phase diagram is mapped by $\bar J$ and $T$. For each $(\bar J,T)$ value, we analyze the marginal distributions for the overlaps $Q$ and $R$, as well as the magnetizations $m^x$ and $m^y$. Each distribution is assessed using the (normalized) Binder ratio~\cite{Binder1981cpf} of the respective quantity; i.e., $g_{X}=\frac{1}{2}\left(3-\frac{\langle X^4\rangle}{\langle X^2\rangle^2}\right)$ for $X=Q,R,m^x$ or $m^y$.  The Binder ratio is commonly used to assess phase boundaries~\cite{Binder2010mcs}. (Note that there are some nuances to keep in mind for vector spin glasses~\cite{Franz2022dti}.) The normalized Binder ratio is zero if the underlying distribution is Gaussian, unity for a coin-flip distribution (i.e., goalposts), and negative for distributions with tails heavier than a Gaussian's. 

\begin{figure}[htbp]
    \centering
    \includegraphics[width=\textwidth]{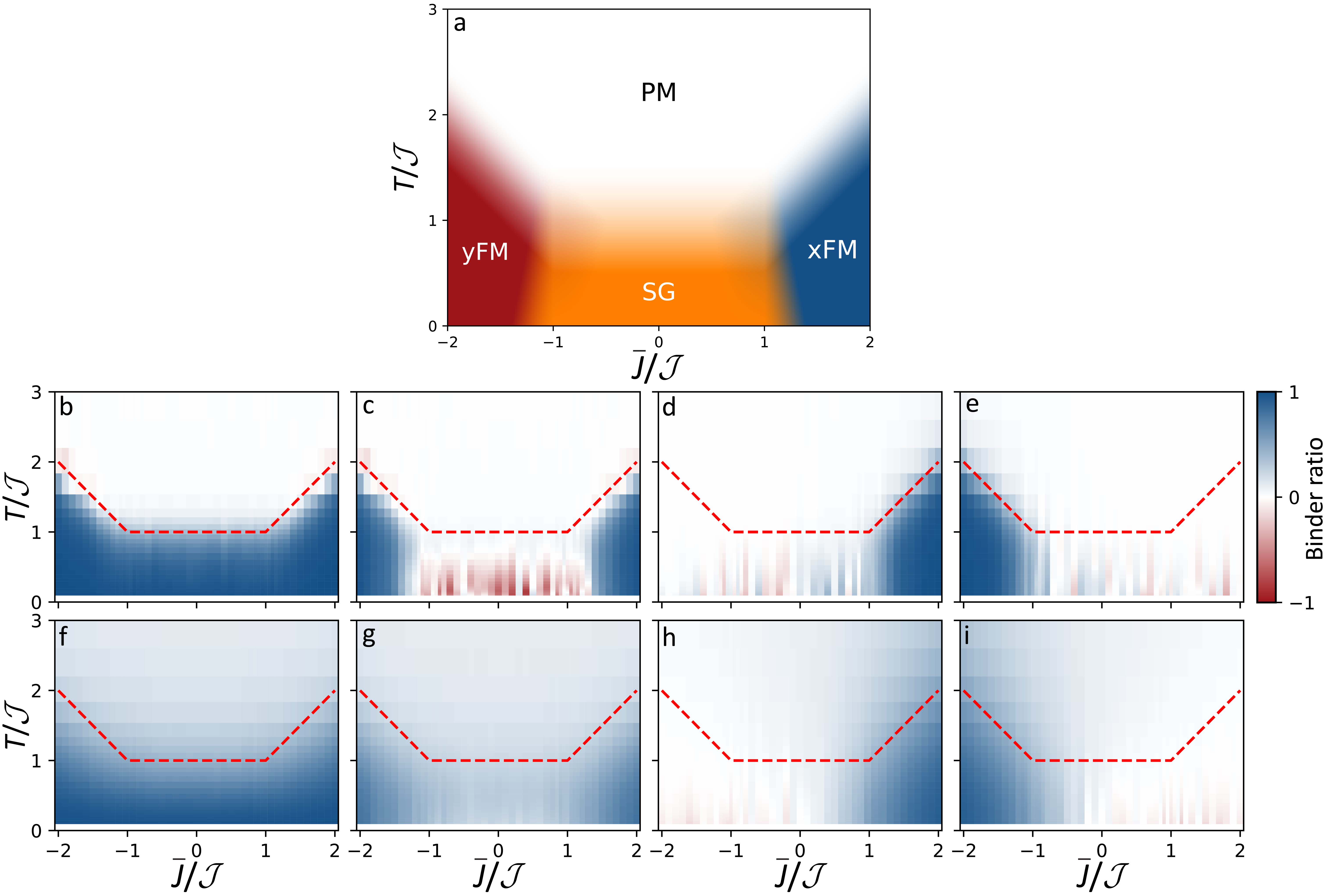}
    \caption{Equilibrium phase diagram of the vector spin glass model obtained using parallel-tempering Monte Carlo simulations. (a) Estimated phase diagram based on panels (b-e), parameterized by the temperature $T$, the mean interaction strength $\bar J$ and the (normalized) standard deviation of interaction strengths $\J$. Four phases are indicated: paramagnet (PM), $x$-ferromagnet (\xFM), $y$-ferromagnet (\yFM), and spin glass (SG). (b-e) Normalized Binder ratios $g$ of marginal distributions for (b) $Q$; (c) $R$; (d) $m^x$; (e) $m^y$, for a system of size $n=128$. The red dashed line shows the boundary of the paramagnetic phase in the Ising SK model in the thermodynamic limit~\cite{Sherrington1975smo}. (f-i) Identical to panels (b-e) but for a smaller system of $n=8$, rendering the crossover more blurred.}
    \label{fig:PTphasediagram}
\end{figure}

We sketch the phase diagram in Fig.~\ref{fig:PTphasediagram}a. This is based on the numerical results for $n=128$ (using 100 disorder instances), for which the Binder ratios are shown in Fig.~\ref{fig:PTphasediagram}b-e. We can clearly identify a paramagnetic phase (PM), characterized by a Binder ratio of zero in all quantities. At sufficiently low temperature but non-zero $\bar J$, we identify the two ferromagnetic phases (\xFM and \yFM) characterized by the Binder ratio of the $x$-and $y$-magnetization, respectively. The central low-temperature region indicates the spin glass phase (SG), where the Binder ratio of $Q$ is positive and those of the magnetizations are zero (or negative). The phase diagram is reminiscent of that of the Ising-SK model for positive mean interactions ($\bar J>0$)~\cite{Sherrington1975smo}, while the negative half of the diagram can be understood from the joint transformation $J\rightarrow-J$, $S^x\rightarrow S^y$, $S^y\rightarrow-S^x$ on the energy given by Eq.~\eqref{HamSpin}. Indeed, the boundary of the paramagnetic phase is similar to the Ising-SK model; this boundary is indicated by the red dashed line in panels (b-i). For the simulated system size, the phase boundaries are not sharp, and a careful finite-size scaling analysis of the Binder ratios would be required to pinpoint the phase transitions. Likewise, we do not indicate the exact location of the boundary between $x$- and $y$-ferromagnets and spin glass. Future work will investigate this boundary in more detail. Such a study can also determine whether mixed phases exist in that region.

For reference, in Fig.~\ref{fig:PTphasediagram}f-i we also show the map of Binder ratios for a system with $n=8$ as in the experiment. As expected, there are no more sharp transitions between the paramagnetic and ordered phases, but their signatures remain clear.

\subsection{Comparison to experimentally realized disorder instances}

\begin{figure}[htbp]
    \centering
    \includegraphics[width=\textwidth]{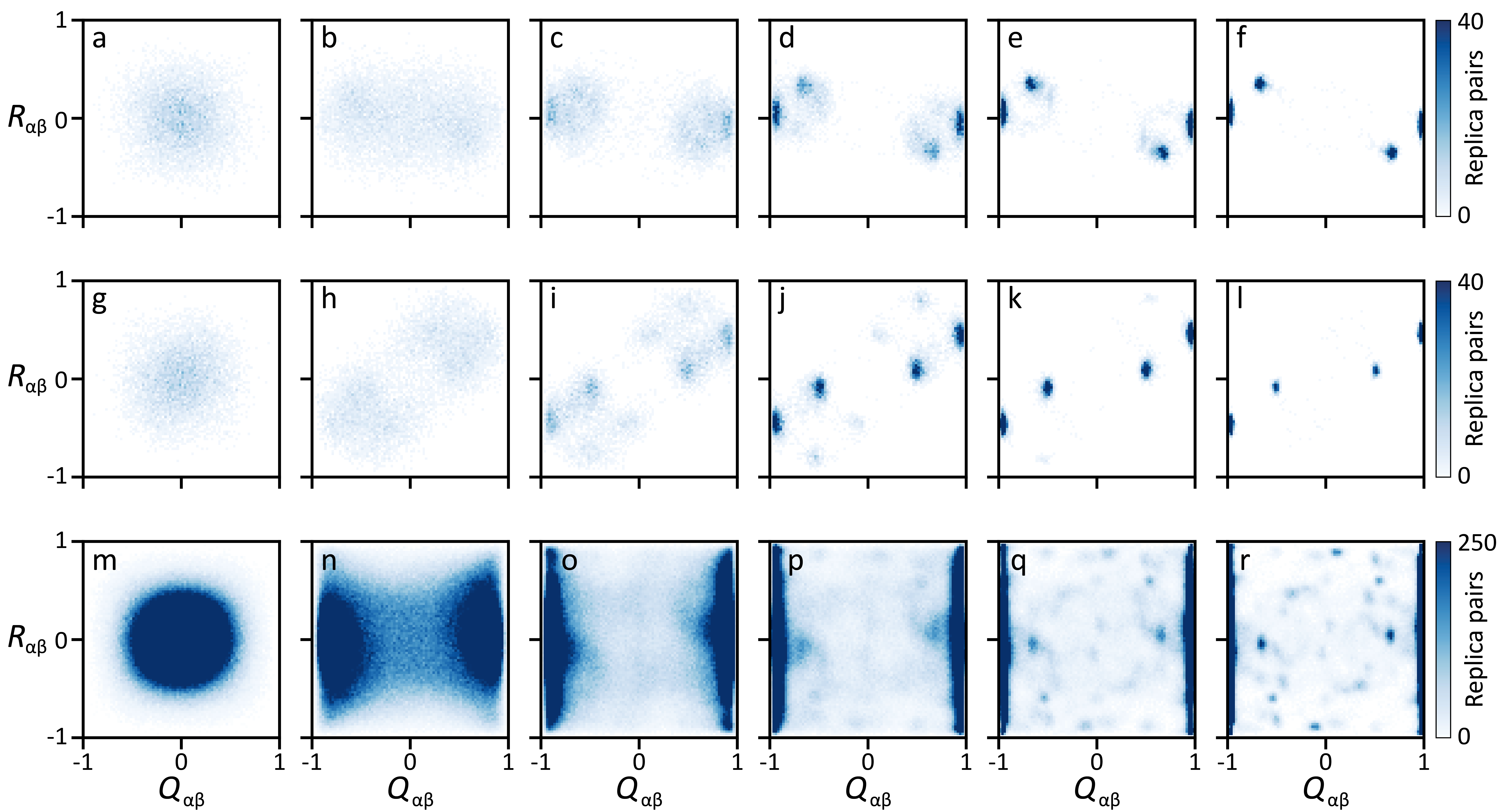}
    \caption{Equilibrium overlap distributions from parallel tempering. (a-f) Numerical spin overlaps for the disorder instance of Fig.~3 (and Fig.~4a) of the main text. (g-l) Numerical spin overlaps for the disorder instance of Fig.~4c of the main text. (m-r) Numerical Parisi order parameter, disorder averaged over all 123 disorder instances of Fig.~4e of the main text. The temperatures, normalized to the maximum magnitude eigenvalue of the (measured) $J$ matrix $\lambda_\text{max}$, are: (a,g,m), 1.11; (b,h,n), 0.34; (c,i,o), 0.19; (d,j,p), 0.11; (e,k,q), 0.06; and (f,l,r), 0.03.}
    \label{fig:PTresults}
\end{figure}

As discussed in the main text, the parallel-tempering simulations fail to accurately match experimentally measured spin-overlap distributions of specific disorder instances. This is illustrated in Fig.~\ref{fig:PTresults}(a-l), where we simulate two of the experimentally realized disorder instances using the measured $J$ and $K$ matrix. We show a temperature progression from the paramagnetic phase into an ordered phase. At no temperature does the overlap look comparable to that experimentally observed in Figs.~3 and 4c of the main text. Likewise, the parallel-tempering simulation shown in Fig.~\ref{fig:PTresults}m--r for the Parisi order parameter distribution lacks key features exhibited by the experimentally found distribution in Fig.~4e. 

\section{Analytical replica calculations for the vector model}
In this section, we will follow standard references~\cite{Castellani2005stf, Panchenko2012tsm,Mezard1987sgt,Charbonneau2023sgt} for performing analytical replica calculations and apply these to the calculation and  analysis of the mean-field free energy for the confocal spin glass model described by Eq.~\eqref{HamSpin}. To make our vector spin model amenable to an analytical approach, we ignore finite-size effects resulting in the $K$-matrix. Furthermore, as in Sec.~\ref{sec:pt}, we will assume that the pairwise interaction strengths $J_{ij}$ are i.i.d.~Gaussian variables with variance $\J^2/n$ and mean zero. Note that the position-dependent interactions generated by the confocal cavity produce a variety of subtle correlations; see also Ref.~\cite{Marsh2021eam}. For the purposes of a replica analysis, we will simply assume that these vanishing correlations have no impact on the thermodynamics.

This section is structured as follows. First we will derive general expressions for the replicated free energy. Then, in subsection~\ref{subsec:repAction}, we will find the replica symmetric (RS) free energy and show that replica order appears at $\beta \J=1$, where $\beta = (k_BT)^{-1}$. Next, in subsection~\ref{subsec:Replicon}, we will analyze the stability of the replica symmetric solution, and find that it becomes unstable at precisely the same temperature $\beta \J=1$. In other words, as soon as replicas develop a nonzero overlap, replica symmetry breaks. In subsection~\ref{subsec:RSBAction}, we derive an expression for the free energy as a function of the overlap matrix $q^{\mu\nu}_{\alpha\beta}$, for $q^{\mu\nu}_{\alpha\beta}$ that obey Parisi's $k$-step replica symmetry breaking ansatz. Next, in subsection~\ref{subsec:1RSB} we find the one-step RSB (1RSB) solution, providing a good approximation of the free energy and overlap in the thermodynamic limit. Finally, in subsection \ref{subsec:nematicity} we show that the system is statistically isotropic, justifying one of our assumptions in the present analysis.

We note that we will use $s$ to denote the number of replicas, since $n$ is already used for denoting the number of atomic vertices. We will be especially interested in the $s\to 0^+$ limit used to calculate the quenched free energy and quenched statistics. We use the ``replica trick''~\cite{Charbonneau2023sgt} to compute the disorder-averaged free energy
\begin{equation}
    -\beta \overline \F=\overline{\log Z}=\lim_{s\to 0^+} \frac{\overline{Z^s}-1}{s},
    \label{eq:sto0}
\end{equation}
where the overline denotes an average over disorder realizations. The fundamental idea behind the replica trick is that $\overline{Z^s}$ is not just the $s$th power of the partition function of one system, it can also be interpreted as the partition function of $s$ replicas of the original system with the same disorder realization. As we will see, the average does not commute with the logarithm, which is a consequence of replica order. At low temperatures $\overline{\log Z}$ and $\log \overline Z$ can then differ by extensive amounts in system size.

We begin by deriving an expression for the replicated free energy of the system with no assumption on the form of the overlap matrices. The partition function for a system of $s$ replicas is
\begin{equation}
    Z^s=\int \frac{d^{sn}\theta}{(2\pi)^{sn}} \exp \left(\beta\sum_{\alpha=1}^s\sum_{i,j=1}^nJ_{ij}(S^{x,\alpha}_iS^{x,\alpha}_j-S^{y,\alpha}_iS^{y,\alpha}_j)\right),
\end{equation}
where $S_i^{x,\alpha}=\cos(\theta_i^\alpha)$ is the $x$-component of the $i$th vector spin in the replica labeled by $\alpha$, and similar for $S_i^{y,\alpha}$.
Using their Gaussian distribution we can integrate out the $J_{ij}$'s to get the disorder average of $Z^s$.
\begin{equation}
\begin{split}
    \overline{Z^s}&=\int \frac{d^{sn}\theta}{(2\pi)^{sn}}\exp \left(\frac{\beta^2\J^2}{n}\sum_{\alpha,\beta=1}^s\sum_{i,j=1}^n(S^{x,\alpha}_iS^{x,\alpha}_j-S^{y,\alpha}_iS^{y,\alpha}_j)(S^{x,\beta}_iS^{x,\beta}_j-S^{y,\beta}_iS^{y,\beta}_j)\right)\\
    &=\int \frac{d^{sn}\theta}{(2\pi)^{sn}} dq^{xx}dq^{xy}dq^{yx}dq^{yy}\exp \left(-\beta\F_\text{full}[q^{\mu\nu},\theta]\right),
    \label{eq:Zs}
\end{split}
\end{equation}
where $q^{\mu\nu}$ are the spin overlaps as defined in the main text, and $\F_\text{full}$ plays the role of a free energy given by
\begin{equation}\begin{split}
    \beta \F_\text{full}=&{n(\beta \J)^2}\sum_{\alpha,\beta=1}^s\left((q^{xx}_{\alpha\beta})^2 - (q^{xy}_{\alpha\beta})^2 -(q^{yx}_{\alpha\beta})^2 + (q^{yy}_{\alpha\beta})^2\right)\\
    &-2(\beta\J)^2\sum_{\alpha,\beta=1}^s\sum_{i=1}^n\left(q^{xx}_{\alpha\beta}S^{x,\alpha}_iS^{x,\beta}_i +q^{xy}_{\alpha\beta}S^{x,\alpha}_iS^{y,\beta}_i+q^{yx}_{\alpha\beta}S^{y,\alpha}_iS^{x,\beta}_i+q^{yy}_{\alpha\beta}S^{y,\alpha}_iS^{y,\beta}_i\right).
\end{split}\end{equation}
The correct saddle point of $\F_\text{full}$ will give the free energy of the replicated system.

Inspired by the experimental data and additional evidence from numerical simulations described in Sec.~\ref{sec:pt}, we will assume that $q^{xy}=q^{yx}=0$ at the saddle point. This reduces $\F_\text{full}$ to two terms, $\F_\text{full} = \F_\text{quad} + \F_\theta$. The first term, quadratic in the overlaps, is given by
\begin{equation}
    -\beta \F_\text{quad}/n=-{(\beta \J)^2}\sum_{\alpha,\beta=1}^s \left[(q^{xx}_{\alpha\beta})^2+(q^{yy}_{\alpha\beta})^2\right],
\end{equation}
while the second term depends on the spin variables $S_i$ (and therefore the angles $\theta$) via
\begin{equation}
    -\beta\F_{\theta}/n=\frac {2(\beta \J)^2}{n} \sum_{\alpha,\beta=1}^s\sum_{i=1}^n\left(q^{xx}_{\alpha\beta}S^{x,\alpha}_iS^{x,\beta}_i+q^{yy}_{\alpha\beta}S^{y,\alpha}_iS^{y,\beta}_i\right).
\end{equation}
As a side note, this is the exact same free energy one would get if the coefficients for $xx$ and $yy$ coupling were picked independently with variance $\J^2/n$.

Since only $\F_\theta$ is angle-dependent, we can rewrite Eq.~\eqref{eq:Zs} as
\begin{equation}\begin{split}
    \overline{Z^s}&=\int dq^{xx}dq^{yy}\exp \left(-\beta\F_\text{quad}\right)\cdot\left[\int \frac{d^s\theta}{(2\pi)^s} \exp\left(2\beta^2\J^2 \sum_{\alpha,\beta=1}^s\left(q^{xx}_{\alpha\beta}S^{x,\alpha}S^{x,\beta}+q^{yy}_{\alpha\beta}S^{y,\alpha}S^{y,\beta}\right)\right)\right]^n\\
    &\equiv \int dq^{xx}dq^{yy}\exp \left(-\beta\F_\text{quad}\right) \left(Z_\theta[q^{xx}_{\alpha\beta},q^{yy}_{\alpha\beta}]\right)^n,
\end{split}\end{equation}
where we defined
\begin{equation}
    Z_\theta[q^{xx}_{\alpha\beta},q^{yy}_{\alpha\beta}] = \int \frac{d^s\theta}{(2\pi)^s} \exp\left(2\beta^2\J^2 \sum_{\alpha,\beta=1}^s\left(q^{xx}_{\alpha\beta}\cos(\theta_\alpha)\cos(\theta_\beta)+q^{yy}_{\alpha\beta}\sin(\theta_\alpha)\sin(\theta_\beta)\right)\right) .
\end{equation}
The total free energy is thus
\begin{equation}
    -\beta \F_{\textrm{full}}\equiv -\beta \F_{\textrm{quad}}+n\log Z_\theta.
\end{equation}


\subsection{The replica-symmetric free energy}
\label{subsec:repAction}
We now consider the case of replica symmetry and evaluate the free energy, expressing it as an integral involving the modified Bessel function $I_0$. The free energy is then found as a function of $q$ by taking the limit $s\to 0^+$. Normally, the free energy of a system is found through minimization of the action. However, within the context of the replica trick, one must always maximize the action to find the free energy~\cite{Baldwin2023rtr,Aizenman2003evp}. This arises from treating the number of replicas $s$ as a continuous variable when taking the limit $s\to0$. While the number of off-diagonal elements of the overlap matrix, $s(s-1)/2$, is positive for any integer $s>1$, it becomes negative as $s$ continuously approaches zero from above. The change in sign thus requires the action be maximized rather than minimized. Upon doing so, we will find that replica order appears at the temperature $\beta \J=1$.

We now perform the $\theta$ integral exactly for replica-symmetric $q^{\mu\nu}$'s. Motivated again by the numerical evidence from Sec.~\ref{sec:pt}, we will first make the assumption $q^{xx}_{\alpha\beta}=q^{yy}_{\alpha\beta}$. This assumption is further justified in Subsec.~\ref{subsec:nematicity}. The replica symmetric ansatz can then be summarized as 
\begin{equation}
    q^{xx}_{\alpha\beta}=q^{yy}_{\alpha\beta}=\frac {1}{2}q+\frac{1}{2}(1-q)\delta_{\alpha\beta}=\begin{cases}
  \frac{1}{2}  & \alpha=\beta, \\
  \frac{1}{2}q   & \alpha\neq \beta,
\end{cases}
\end{equation}
where $q$ is \textit{the} overlap for the replica symmetric case, describing replica order ($q>0$) or absence thereof ($q=0$).  Our goal will be to find the value of $q$ that results in a saddle point. Inserting the replica symmetric ansatz into $\F$, we evaluate the $\theta$ term as
\begin{equation}\begin{split}
    -\beta\bar\F_\theta/n &\equiv \log(Z_\theta) =
    s{\beta^2\J^2} (1-q)+\log \int \frac{d^s \theta}{(2\pi)^s} \exp \left(\beta^2\J^2q\left|\sum_{\alpha=1}^s\cos \theta_\alpha \mathbf{i}+\sin \theta_\alpha \mathbf{j}\right|^2\right)\\
    &=s{\beta^2\J^2}(1-q)+\log \int \frac{dXdY}{4\pi q\beta^2\J^2}\frac{d^s \theta}{(2\pi)^s} \exp \left(-\frac 1{4q\beta^2\J^2}(X^2+Y^2)+\sum_{\alpha=1}^s\left(X\cos \theta_\alpha+Y\sin \theta_\alpha\right) \right)\\
    &=s{\beta^2\J^2}(1-q)+\log \int \frac{dXdY}{4\pi q\beta^2\J^2}\exp \left(-\frac 1{4q\beta^2\J^2}(X^2+Y^2)+s\log I_0(\sqrt{X^2+Y^2}) \right)\\
    &=s{\beta^2\J^2}(1-q)+\log \int_0^\infty \frac{rdr}{ 2q\beta^2\J^2}\exp \left(-\frac 1{4q\beta^2\J^2}r^2+s\log I_0(r) \right).
\end{split}
\label{eq:thetaTerm}
\end{equation}
To go from the first line to the second line, we introduced Hubbard-Stratonovich fields $X$ and $Y$, which act as fictitious magnetic fields on the spins. To go from the second to the third line, we used the definition of the modified Bessel functions of the first kind: $I_n(x)=\frac 1{2\pi}\int_0^{2\pi} \cos (n\theta) e^{x\cos\theta}d\theta$. Going from the third to the fourth line is a transformation from Cartesian to cylindrical coordinates in terms of $X$ and $Y$, and noting that the integrand depends only on the radius $r$.

\begin{figure}
    \centering
    \includegraphics[width=0.6\textwidth]{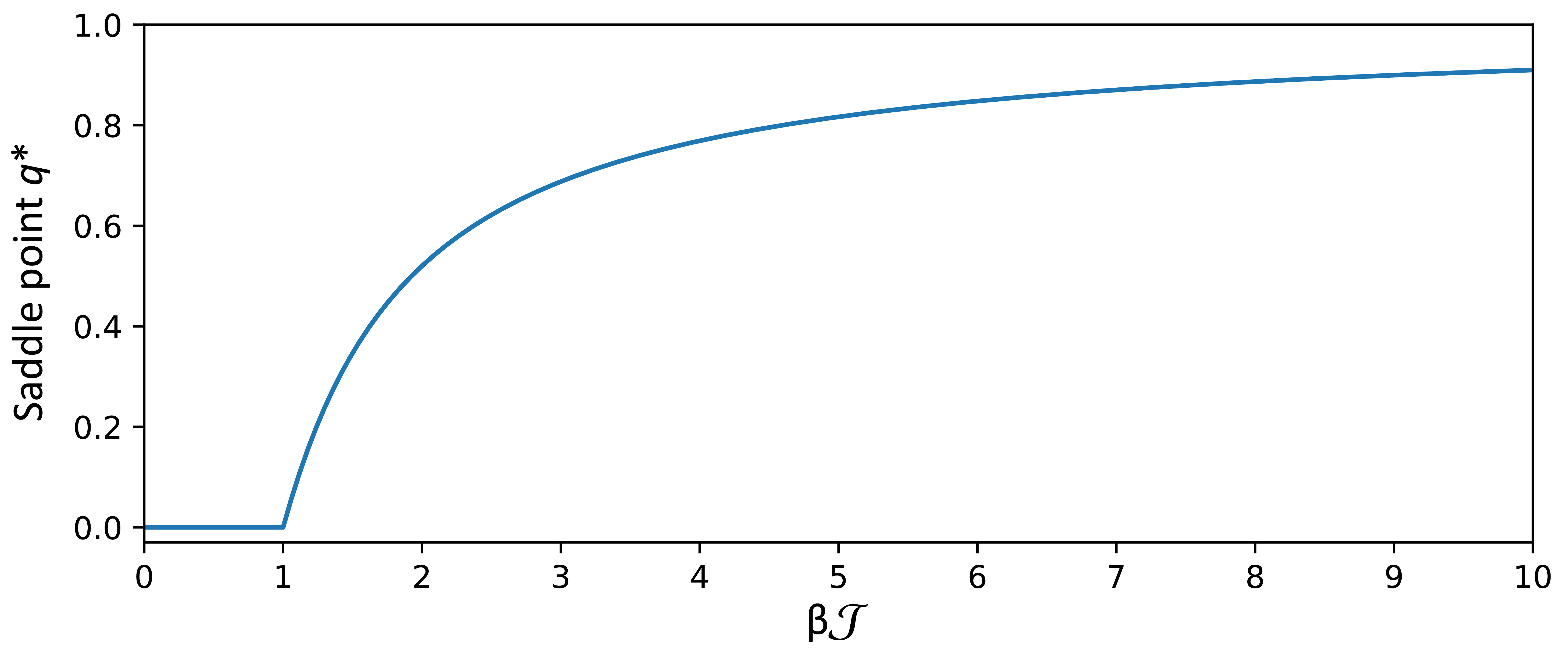}
    \caption{Replica-symmetric solution, showing the saddle point overlap value $q^*$ as function of temperature. Replica order (nonzero $q^*$) appears at $\beta \J=1$.}
    \label{fig:qRS}
\end{figure}

The quadratic term in $\F_\text{full}$ is simply
\begin{equation}
    -\beta\F_\text{quad}/n=-s\frac{\beta^2 \J^2}{2}-s(s-1)\frac{\beta^2 \J^2}{2}q^2.
\end{equation}
This adds up to a total replica-symmetric free energy of
\begin{equation}
    -\beta \F_{RS}/n=-s\frac{\beta^2 \J^2}{2}-s(s-1)\frac{\beta^2 \J^2}{2}q^2+   
    s{\beta^2\J^2}(1-q)+\log \int_0^\infty \frac{rdr}{ 2q\beta^2\J^2}\exp \left(-\frac 1{4q\beta^2\J^2}r^2+s\log I_0(r) \right).
\end{equation}
Taking the $s \to 0$ limit, this becomes
\begin{equation}
    -\beta \F_{RS}[q]/ns=-\frac{\beta^2 \J^2}{2}+\frac{\beta^2 \J^2}{2}q^2+   
    {\beta^2\J^2}(1-q)+\int_0^\infty \frac{rdr}{ 2q\beta^2\J^2}\exp \left(-\frac 1{4q\beta^2\J^2}r^2 \right)\log I_0(r).
\end{equation}
From here it is a matter of maximizing $\F_{RS}$ with respect to $q$. We call this saddle point value $\qs$, which depends on $\beta \J$. Figure~\ref{fig:qRS} shows a plot of this dependence. Note that $q^*$ increases continuously from zero like in the replica symmetric solution for the Sherrington-Kirkpatrick model, rather than discontinuously jumping as in the $p$-spherical model or structural glasses~\cite{Castellani2005stf}. This behavior is typical of systems with full replica symmetry breaking and no 1RSB-exact phase~\cite{Mezard2009ipa}.

\subsection{Replicon analysis}
\label{subsec:Replicon}
To analyze the stability of the replica-symmetric solution derived above, we will expand the free energy around the replica symmetric saddle point. The Hessian of the free energy has positive eigenvalues, corresponding to modes called \textit{replicons}~\cite{Fischer1991sg,DeDominicis2006rfa}, which we show to spontaneously break replica symmetry at temperatures below $\beta\J=1$.

Elements of the Hessian, i.e.~$\frac{\partial^2 \F}{\partial q_{\rho\sigma}\partial q_{\tau\upsilon}}$, have four replica indices. We will work in the space where $q$ is symmetric, $q_{\rho\sigma}=q_{\sigma\rho}$, and where $q^{xx}_{\rho\sigma}=q^{yy}_{\rho\sigma}\equiv q_{\rho\sigma}$. With this in mind, there are three cases of interest for this second derivative: $\rho\sigma$ and $\tau\upsilon$ are the same pair, $\rho\sigma$ and $\tau\upsilon$ have one element in common (we denote this case $\rho\sigma,\rho\tau$) and $\rho\sigma$ and $\tau\upsilon$ are completely distinct.

In all cases, the second derivative of $\F_\text{quad}$ can be written as
\begin{equation}
    -\beta \frac{\partial^2 \F_{quad}}{\partial q_{\rho\sigma}\partial q_{\tau\upsilon}}=-8n(\beta \J)^2 \delta_{\rho\tau}\delta_{\sigma\upsilon}.
\end{equation}
Switching to angle notation, the second derivatives of $\F_{\theta}$ can be written as
\begin{equation}
    -\beta\frac{\partial^2 \F_{\theta}}{\partial q_{\rho\sigma}\partial q_{\tau\upsilon}}=16N\beta^4\J^4 \left(\Eq{\cos(\theta_\rho-\theta_\sigma)\cos(\theta_\tau-\theta_\upsilon)}-\Eq{\cos(\theta_\rho-\theta_\sigma)}\Eq{\cos(\theta_\tau-\theta_\upsilon)}\right),
\end{equation}
where
\begin{equation}
    \Eq{\bullet}=\frac{\int d^s\theta \bullet \exp\Big({2(\beta \J)^2}\sum_{\alpha,\beta=1}^sq_{\alpha\beta}\cos(\theta_\alpha-\theta_\beta)\Big)}{\int d^s\theta \exp\Big({2(\beta \J)^2}\sum_{\alpha,\beta=1}^sq_{\alpha\beta}\cos(\theta_\alpha-\theta_\beta)\Big)}.
    \label{eq:Eqdef}
\end{equation}
We are interested in evaluating this second derivative at the replica-symmetric saddle point $q_{\rho\sigma}=\frac 12 \qs+\frac 12 (1-\qs)\delta_{\rho\sigma}$.

The saddle point equations for $\qs$ guarantee that at that saddle point, for all $\rho\neq \sigma$, $\Eq{\cos(\theta_\rho-\theta_\sigma)}$ is equal to $\qs$. However, the correlation term $\Eq{\cos(\theta_\rho-\theta_\sigma)\cos(\theta_\tau-\theta_\upsilon)}$ requires a more careful derivation, which we will provide in the next subsection.

\subsubsection{Correlation functions of $\cos(\theta)$}
Although many of the methods in this section can be generalized to arbitrary values of $q$, we will focus mainly on the special saddle point $\qs$ whose Hessian we are interested in. There, Eq.~\eqref{eq:Eqdef} becomes
\begin{equation}
    \Eqstar{\bullet}=\frac{\int d^s\theta \bullet \exp\Big({(\beta \J)^2}\sum_{\alpha\beta}\qs\cos(\theta_\alpha-\theta_\beta)\Big)}{\int d^s\theta \exp\Big({(\beta \J)^2}\sum_{\alpha\beta}\qs\cos(\theta_\alpha-\theta_\beta)\Big)},
    \label{eq:EqStardef}
\end{equation}
and by the same Hubbard-Stratonovich manipulations as earlier, the denominator can be written as
\begin{equation}
\begin{split}
    Z_{\theta}
    &=\int \frac{d^s\theta}{(2\pi)^s}\exp\Big({(\beta \J)^2}\qs\sum_{\alpha,\beta=1}^s\cos(\theta_\alpha)\cos(\theta_\beta)+\sin(\theta_\alpha) \sin (\theta_\beta)\Big)\\
    &=\int \frac{dXdY}{4\pi \qs \beta^2\J^2}\frac{d^s \theta}{(2\pi)^s} \exp \left(-\frac 1{4\qs\beta^2\J^2}(X^2+Y^2)+\sum_{\alpha=1}^s\left(X\cos \theta_\alpha+Y\sin \theta_\alpha\right) \right).
\end{split}
\end{equation}
We now compute the correlation functions of the $\theta$'s. To start, we calculate the variance of $\cos(\theta_\rho-\theta_\sigma).$ Breaking the numerator of Eq.~\eqref{eq:EqStardef} into an integral over $\theta_\rho$, an integral over $\theta_\sigma$, and an integral over the other $s-2$ angle variables, we have
\begin{equation}\begin{split}
    \Eqstar{\cos^2(\theta_\rho-\theta_\sigma)} = \frac{1}{Z_\theta} \int &\frac{dXdY}{4\pi \qs\beta^2\J^2}\frac{d\theta_\rho d\theta_\sigma d^{s-2} \theta}{(2\pi)^s} \\
    &\times\exp \left(-\frac 1{4\qs\beta^2\J^2}(X^2+Y^2)+\sum_{\alpha=1}^s\left(X\cos \theta_\alpha+Y\sin \theta_\alpha\right) \right)\cos^2(\theta_\rho-\theta_\sigma).
\end{split}\end{equation}
We can evaluate the angle integral exactly, by using trigonometric identities to reexpress the integrand as a sum of terms that each factor over the $\theta$s. Making a change of basis $\theta_\alpha\rightarrow\theta_\alpha + \arctan(Y/X)$, and defining $r=\sqrt{X^2+Y^2}$, we have
\begin{equation}
    \int \frac{d\theta_\rho d\theta_\sigma d^{s-2} \theta}{(2\pi)^s} \exp \left(\sum_{\alpha=1}^s\left(r\cos \theta_\alpha\right) \right)\left(\frac{1}{2} + \frac{1}{2} \cos 2\theta_\rho \cos 2\theta_\sigma + \frac{1}{2} \sin 2\theta_\rho\sin 2\theta_\sigma\right)= \frac{1}{2} I_0(r)^s + \frac{1}{2} I_0(r)^{s-2}I_2(r)^2.
\end{equation}
Here we again use the definitions of the modified Bessel functions to evaluate the cosine integrals, while the integrals involving sines are zero by symmetry. This gives us
\begin{equation}
    \Eqstar{\cos^2(\theta_\rho-\theta_\sigma)} = \frac{1}{Z_\theta} \int \frac{rdr}{2\qs\beta^2\J^2}\exp\left(-\frac{r^2}{4\qs\beta^2\J^2}\right)\left(\frac{1}{2}I_0(r)^s + \frac{1}{2}I_0(r)^{s-2}I_2(r)^2\right).
\end{equation}
Taking $s\to 0$, where $Z_\theta\to1$, this gives us
\begin{equation}
    C_1 \equiv \Eqstar{ \cos^2(\theta_\rho-\theta_\sigma)}-\Eqstar {\cos(\theta_\rho-\theta_\sigma)}^2=\int_0^\infty \frac{rdr}{2\qs\beta^2\J^2}\exp\left(-\frac{r^2}{4\qs\beta^2\J^2}\right)\left(\frac{1}{2} + \frac{1}{2}I_0(r)^{-2}I_2(r)^2\right)-{\qs}^2.
\end{equation}

Next, we evaluate the product $\Eqstar{\cos(\theta_\rho-\theta_\sigma)\cos(\theta_\rho-\theta_\tau)}$. The object in the integral becomes 
\begin{equation}
    \int \frac{d\theta_\rho d\theta_\sigma d\theta_\tau d^{s-3} \theta}{(2\pi)^s} \exp \left(\sum_{\alpha=1}^s\left(r\cos \theta_\alpha\right) \right) \cos^2\theta_\rho \cos\theta_\sigma \cos\theta_\tau = \frac{1}{2} I_0(r)^{s-3}I_2(r)I_1(r)^2+\frac{1}{2}I_0(r)^{s-2}I_1(r)^2,
\end{equation}
where again we drop any terms that have sines in them because they integrate out to 0. After taking the $s\to 0$ limit, we have
\begin{equation}\begin{split}
    C_2 &\equiv \Eqstar{\cos(\theta_\rho-\theta_\sigma)\cos(\theta_\rho-\theta_\tau)} - \Eqstar{\cos(\theta_\rho-\theta_\sigma)}\Eqstar{\cos(\theta_\rho-\theta_\tau)}\\
    &=\int_0^\infty \frac{rdr}{2\qs\beta^2\J^2}\exp\left(-\frac{r^2}{4\qs\beta^2\J^2}\right)\left(\frac{1}{2}I_0(r)^{-3}I_2(r)I_1(r)^2 + \frac{1}{2}I_0(r)^{-2}I_1(r)^2\right) - {\qs}^2.
\end{split}\end{equation}

In the case of all indices different, we use the same approach to find
\begin{equation}\begin{split}
    C_3 &\equiv \Eqstar{\cos(\theta_\rho-\theta_\sigma)\cos(\theta_\tau-\theta_\upsilon)} - \Eqstar{\cos(\theta_\rho-\theta_\sigma)} \Eqstar{\cos(\theta_\tau-\theta_\upsilon)}\\
    &= \int_0^\infty \frac{rdr}{2\qs\beta^2\J^2}\exp\left(-\frac{r^2}{4\qs\beta^2\J^2}\right) I_0(r)^{-4}I_1(r)^4 - {\qs}^2.
\end{split}\end{equation}

\subsubsection{The spectrum of the Hessian}
\begin{figure}
    \centering
    \includegraphics[width=0.35\textwidth]{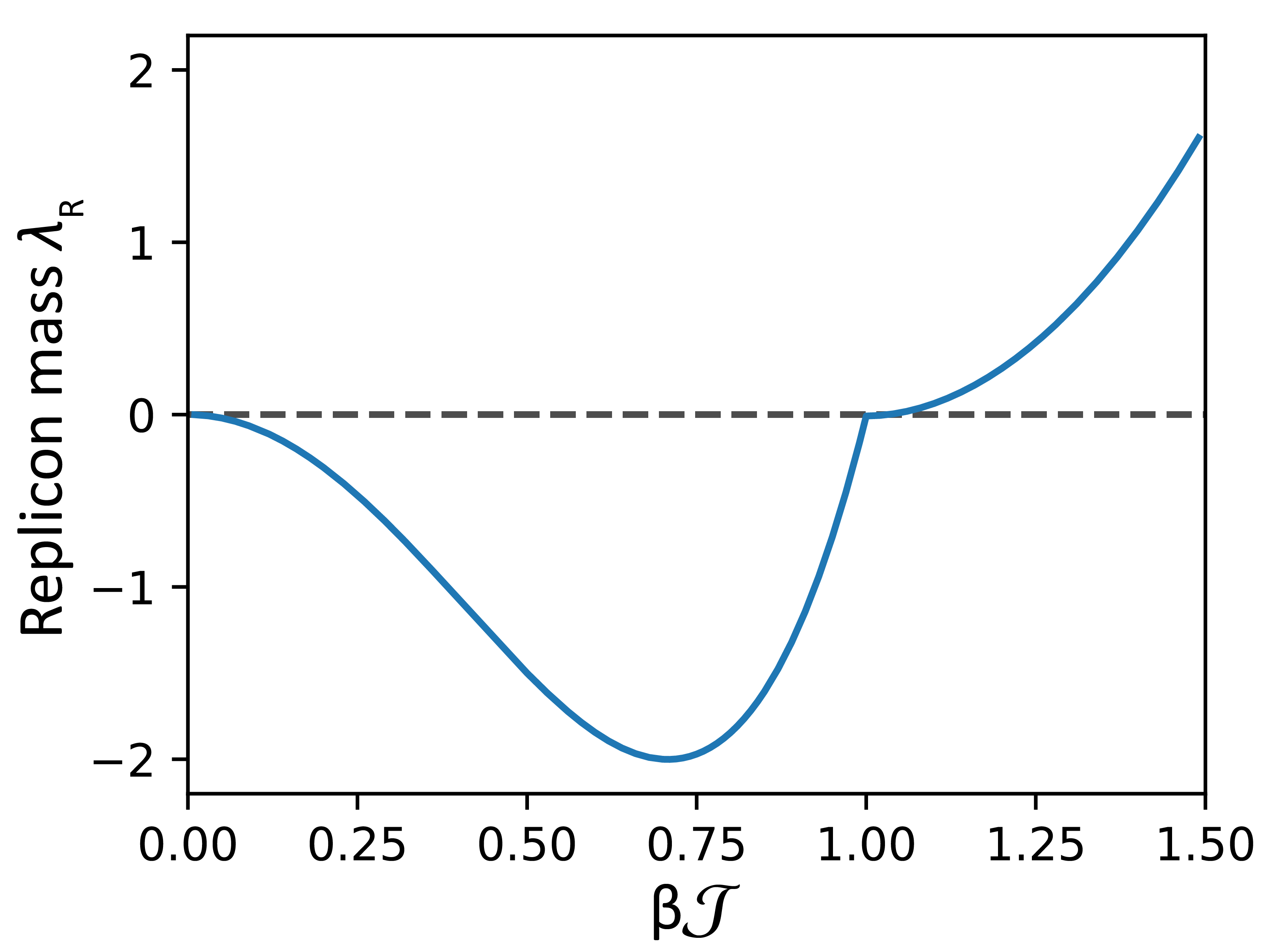}
    \caption{Stability of the replica-symmetric solution. Replicon mass (eigenvalue of the Hessian of the free energy) reveals instability of the replica-symmetric solution at $\beta\J=1$. This is exactly the same temperature at which replica order appears.}
    \label{fig:repliconMass}
\end{figure}

Synthesizing the above, we find the elements of the Hessian in all three cases:
\begin{equation}\begin{split}
    &-\frac{\beta}{n} \frac{\partial^2\F_{\textrm{full}}}{\partial q_{\rho\sigma}\partial q_{\rho\sigma}} \equiv M_1 = -8(\beta \J)^2+16(\beta \J)^4C_1, \\
    &-\frac{\beta}{n} \frac{\partial^2 \F_{\textrm{full}}}{\partial q_{\rho\sigma}\partial q_{\rho\tau}} \equiv M_2 = 16(\beta \J)^4C_2, \\
    &-\frac{\beta}{n} \frac{\partial^2 \F_{\textrm{full}}}{\partial q_{\rho\sigma}\partial q_{\tau\upsilon}} \equiv M_3 = 16(\beta \J)^4C_3.
\end{split}\end{equation}
Given the elements of the Hessian in this form, finding the eigenvalues is a standard problem in replica theory (see, for instance, section 6.3 of \cite{DeDominicis2006rfa} or section 3.3 of \cite{Fischer1991sg}).
The resulting eigenvalues of the Hessian, corresponding to longitudinal and replicon modes resp., are thus
\begin{equation}\begin{split}
    &\lambda_L = M_1-4M_2+3M_3,\\
    &\lambda_R = M_1-2M_2+M_3,
\end{split}\end{equation}
with degeneracies $s$ and $(s-3)s/2$ respectively. These eigenvalues can be calculated in terms of $\beta\J$ and shall be evaluated at the saddle point $q=\qs$, which itself is a function of $\beta \J$ as discussed in Subsec.~\ref{subsec:repAction}. Evaluating the eigenvalues for various values of $\beta\J$ at the saddle point value $\qs$, we find that the longitudinal modes are always stable ($\lambda_L<0$ everywhere). The dependence of the replicon mass $\lambda_R$ on $\beta\J$ is shown in Fig.~\ref{fig:repliconMass} and reveals that the replicon mode becomes unstable ($\lambda_R>0$) precisely when replica order appears at $\beta J=1$. This instability means that replica symmetry is spontaneously broken as soon as replica order appears.

\subsection{The RSB free energy}
\label{subsec:RSBAction}
In this subsection, we evaluate $\F[q_{\alpha\beta}]$ for $q_{\alpha\beta}$ satisfying the Parisi RSB ansatz. In other words, we are generalizing the work in Subsec.~\ref{subsec:repAction} to hold for overlap matrices $q_{\alpha\beta}$ that have an ultrametric structure.
To build intuition for the Parisi ansatz, we will start by considering the case where $s>1$. In this case, following the conventions in Ref.~\cite{Parisi2020tos}, we have blocks of size $m_{k}=1$ inside blocks of size $m_{k-1}>1$ inside blocks of size $m_{k-2}>m_{k-1}$ all the way until a block of size $m_{-1}=s$.
The value of $q_{\alpha\beta}$ when $\alpha$ and $\beta$ are in the same block of size $m_{i-1}$ but not the same block of size $m_{i}$ is $\frac{1}{2} q_i$. 
As an example, we will write out the matrix when $k=1$, $m_{0}=3$, $m_{-1}=s=9$:
\begin{equation}
    q_{\alpha\beta}=\frac 12\begin{pmatrix}
    1&q_1&q_1&q_0&q_0&q_0&q_0&q_0&q_0\\
    q_1&1&q_1&q_0&q_0&q_0&q_0&q_0&q_0\\
    q_1&q_1&1&q_0&q_0&q_0&q_0&q_0&q_0\\
    q_0&q_0&q_0&1&q_1&q_1&q_0&q_0&q_0\\
    q_0&q_0&q_0&q_1&1&q_1&q_0&q_0&q_0\\
    q_0&q_0&q_0&q_1&q_1&1&q_0&q_0&q_0\\
    q_0&q_0&q_0&q_0&q_0&q_0&1&q_1&q_1\\
    q_0&q_0&q_0&q_0&q_0&q_0&q_1&1&q_1\\
    q_0&q_0&q_0&q_0&q_0&q_0&q_1&q_1&1
    \end{pmatrix}.
\end{equation}
The probability of two different replicas being the in the same block is $\frac{m_0-1}{s-1}=\frac{3-1}{9-1}=\frac 14$.

Since the $s\to0$ limit must be taken to find the free energy, things become stranger when $s<1$. In this case, the ordering of $m$s is flipped: we have blocks of size $m_{k}=1$ inside blocks of size $m_{k-1}<1$ inside blocks of size $m_{k-2}<m_{k-1}$ all the way until a block of size $m_{-1}=s$. While it may seem strange to consider putting a larger block inside of a smaller block, the limiting result is known to be well defined~\cite{Charbonneau2023sgt}.
Note that overlap probabilities like $\frac{m_0-1}{s-1}$ are still between 0 and 1.

In this subsection, we evaluate the free energy in the RSB scenario. $\F_\text{quad}$ follows straightforwardly as
\begin{equation}
    -\beta \F_\text{quad}/n=-s\frac{(\beta \J)^2}{2}\left(1+\sum_{i=0}^k (m_{i-1}-m_{i})q_i^2\right).
\end{equation}
$\F_\theta$ requires more careful consideration. The key manipulation is analogous to the one performed in the replica symmetric case: adding in fictitious fields $X^\alpha,Y^\alpha$. In subsection \ref{subsec:repAction} we introduced a single pair $X,Y$ which is identical on all replicas. For the RSB scenario considered here, they are instead chosen to have covariance $2\beta^2\J^2 q_{\alpha\beta}$. One way to achieve this set of covariances is to slightly expand the set of variables: instead of just having a pair $X^\alpha, Y^\beta$ for each replica, we will have a series $\tilde X^\alpha_i,\tilde Y^\alpha_i$. The variables $\tilde X^\alpha_i,\tilde Y^\alpha_i$ are uniform on a given block of size $m_i$, and is a pair of independent variables with variance $2\beta^2 \J^2 (q_{i+1}-q_i)$. 

With this setup, the key ingredient is $g_i(X_i,Y_i)$. This is the partition function of a block of size $m_i$ with effective fields from lower levels totaling $X_i=\sum_{j=0}^{i}\tilde X_j$ and $Y_i=\sum_{j=0}^{i}\tilde Y_j$. This partition function is a standard quantity in $k$-step RSB analysis \cite{Parisi2020tos}, but is most often encountered in cases where the spin is a scalar and thus $g_i$ depends on only a single fictitious field $X_i$. Here we straightforwardly generalize this to functions of two fictitious fields $X_i$ and $Y_i$. As in previous subsections, it is convenient to define $r_i=\sqrt{X_i^2+Y_i^2}$, noting that $g_i$ will depend only on $r_i$.
If we define $\qb$ to be the restriction of $q_{\alpha \beta}$ to a block of size $m_i$ then
\begin{equation}
    g_i(X_i,Y_i)=\int \frac{d^{m_i} \theta}{(2\pi)^{m_i}}\exp\left(2\beta^2\J^2\sum_{\alpha,\beta=1}^{m_i}\qb\cos(\theta_\alpha-\theta_\beta)+\sum_{\alpha=1}^{m_i}\left(X_i\cos(\theta_\alpha)+Y_i\sin(\theta_\alpha)\right)\right).
\end{equation}
The total free energy can then be expressed as
\begin{equation}
    -\beta \F/n=-\beta \F_{\textrm{quad}}/n +\log g_{-1}(0,0).
\end{equation}

Starting with the $i=k$ case of a single replica per block, we have
\begin{equation}
    g_k(X_k,Y_k)=\exp\left(\beta^2 \mathcal J^2(1-q_k)\right)I_0(r_k).
\end{equation}
Now we derive a recursion for general $g_i$. Noting that each block of size $m_{i-1}$ is made out of $\frac{m_{i-1}}{m_i}$ blocks of size $m_i$ and that $X_i,Y_i$ are Gaussian random variables with variance $2\beta^2\J^2 q_{i}$, we have the recursion:
\begin{equation}
    g_{i-1}(X_{i-1},Y_{i-1})=\exp\left[\beta^2\J^2(q_{i}-q_{i-1})\left(\frac{\partial^2}{\partial X_i^2}+\frac{\partial^2}{\partial Y_i^2}\right)\right] \left(g_{i}(X_i',Y_i')\right)^{\frac{m_{i-1}}{m_{i}}}\vline_{X_i',Y_i'=X_{i-1},Y_{i-1}}.
\end{equation}
In words, to go from blocks of size $m_i$ to $m_{i-1}$ we raise $g_i$ to the power $(m_i-1)/m_i$ and then ``smear'' out the field through a diffusion operator by an amount proportional to $q_i-q_{i-1}$. 
This gives the recursion relation
\begin{equation}\begin{split}\label{eq:kRSB_recursion}
    g_{i-1}(X_{i-1},Y_{i-1})=\int_{-\infty}^\infty\int_{-\infty}^\infty \frac{dX'_idY'_i}{4\pi\beta^2\J^2(q_{i}-q_{i-1})}\exp\left(-\frac{(X'_i-X_{i-1})^2+(Y'_i-Y_{i-1})^2}{4\beta^2\J^2(q_{i}-q_{i-1})}\right)\left(g_{i}(r'_i)\right)^{\frac{m_{i-1}}{m_{i}}}\\
    =e^{-\frac{r_{i-1}^2}{4\beta^2\J^2(q_{i}-q_{i-1})}}\int_0^\infty \frac {r'_i dr'_i}{2\beta^2\J^2(q_{i}-q_{i-1})}\exp\left(-\frac{r_i'^2}{4\beta^2\J^2(q_{i}-q_{i-1})}\right) I_0\left(\frac{r_{i-1}r'_i}{2\beta^2\J^2(q_{i}-q_{i-1})}\right)\left(g_{i}(r'_i)\right)^{\frac{m_{i-1}}{m_{i}}}.
\end{split}\end{equation}
The $\theta$ part of $\F$ is simply the logarithm of the partition function of one $s\times s$ block with zero applied field: $n \log g_{-1}(0)$ (which will always be proportional to $s$). The total free energy is thus
\begin{equation}
    -\beta \F_{\textrm{full}}/n=-s\frac{(\beta \J)^2}{2}\left(1+\sum_{i=0}^k (m_{i-1}-m_{i})q_i^2\right)+\log g_{-1}(0).
\end{equation}

One can prove the presence of full replica symmetry breaking by analyzing the $k\to\infty$ limit of the above free energy. As an alternative approach, we can use the observation derived in Sec.~\ref{subsec:repAction} that the overlap parameter increases continuously from zero at the phase transition. For such systems, one can perform a power series expansion of the action in the matrix $q_{\alpha\beta}$~\cite{DeDominicis2006rfa,Crisanti2010tsm}, yielding 
\begin{equation}
    -\beta\F_{\textrm{full}}/n=\frac 12 \tau \Tr q^2+\frac 16w\Tr q^3+\frac u{12}\sum_{\alpha\beta}\left(q_{\alpha\beta}\right)^4+\dots,
\end{equation}
where, in our case, the expansion coefficients are
\begin{equation}
\begin{split}
\tau&=4\left((\beta \J)^4-(\beta \J)^2\right),\\
w&=16 (\beta \J)^6,\\
u&=19 (\beta \J)^8.
\end{split}
\end{equation}
This expansion is known~\cite{DeDominicis2006rfa,Crisanti2010tsm} to correspond to full RSB and an overlap distribution with density $\frac{2u}{w}=\frac{19}{8}\beta^2\J^2$ from zero until $Q_1$, with $Q_1$ satisfying $uQ_1^2-wQ_1+\tau=0$, and the rest of the mass at $Q_1$.

\subsection{The 1RSB solution}
\label{subsec:1RSB}
In Subsec.~\ref{subsec:Replicon}, we showed that the replica symmetric solution is unstable for $\beta \J>1$. In Subsec.~\ref{subsec:RSBAction} we derived the free energy assuming a general replica symmetry breaking ansatz for our system. Here, we will maximize that free energy subject to the 1RSB ($k=1$) ansatz: an $s\times s$ matrix composed of $m\times m$ blocks, with overlap $q_1$ for different replicas in the same block and $q_0$ for replicas in different blocks. 

\begin{figure}
    \centering
    \includegraphics[width=0.65\textwidth]{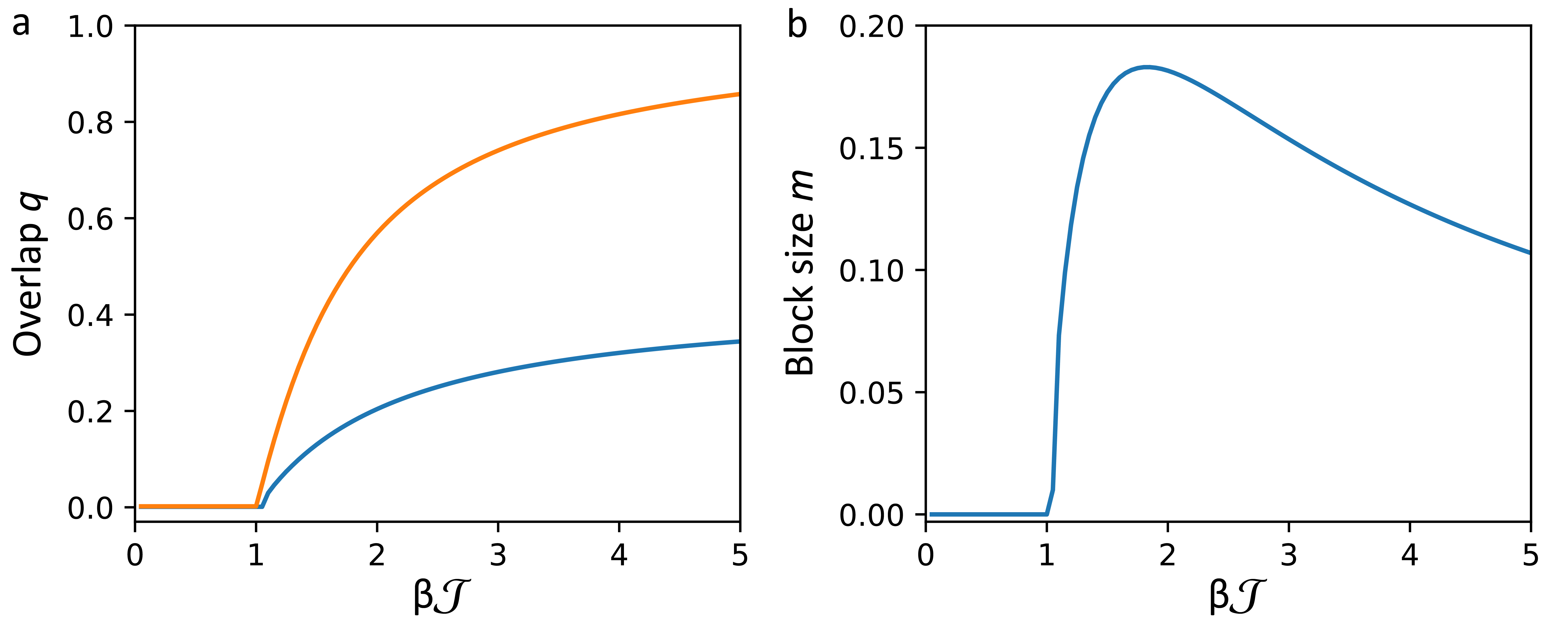}
    \caption{Solutions for 1-step RSB. (a) Saddle-point overlap values $q_0$ (blue) and $q_1$ (orange) as function of temperature. RSB order appears at $\beta\J=1$. (b) Saddle-point value for the 1RSB block size parameter $m$ as function of temperature, displaying the same RSB ordering feature.}
    \label{fig:1RSB}
\end{figure}

For the 1RSB ansatz, the general recursion Eq.~\eqref{eq:kRSB_recursion} simplifies to
\begin{equation}\begin{split}
    &g_{1}(r)=\exp\left(\beta^2 \mathcal J^2(1-q_1)\right)I_0(r),\\
    &g_0(r)=e^{-\frac{r^2}{2(q_1-q_0)}}\int_0^\infty \frac {r'dr'}{2\beta^2\J^2(q_{1}-q_{0})}\exp(-\frac{r'^2}{4\beta^2\J^2(q_{1}-q_{0})}) I_0\left(\frac{rr'}{2\beta^2\J^2(q_{1}-q_{0})}\right)\left(g_{1}(r')\right)^{m},\\
    &g_{-1}(0)=\frac 1m \int_0^\infty  \frac {dr'}{2\beta^2\J^2 q_0}\exp(-\frac{r'^2}{4\beta^2\J^2 q_0})\log \left(g_{0}(r')\right),
\end{split}
\end{equation}
and the total free energy is
\begin{equation}
    -\beta \F_{\textrm{full}}/sn=- \frac{1}{2} \beta^2\J^2\left(1+(m-1)q_1^2-mq_0^2\right)+\log g_{-1}(0).
\end{equation}
We can numerically maximize the free energy in terms of $m,q_0,q_1$ at any temperature $\beta\J$. The results are shown in Fig.~\ref{fig:1RSB}. We see that $q_0\neq q_1$ at $\beta\J\geq1$, which means replica symmetry is indeed broken.
Interestingly, $m$, which is a measure of how broken the symmetry is, never goes above $0.2$. In thermal equilibrium, a fraction $1-m$ (80-to-90 percent) of the probability mass of the Parisi order parameter distribution is in or near the ``goalposts," with the remainder in the nontrivial interior.  This fraction is consistent with the experimentally observed Parisi order parameter distribution in Fig.~4k and the numerical results in Sec.~\ref{sec:pt}.

\subsection{Isotropy of overlaps in the thermodynamic limit}
\label{subsec:nematicity}
Throughout this replica analysis we have assumed that $q_{\alpha\beta}^{xx}=q_{\alpha\beta}^{yy}$. Apart from the numerical motivation, one additional check we can perform is whether $q_{\alpha\beta}^{xx}\neq q_{\alpha\beta}^{yy}$ is ever preferred in the thermodynamic limit. This would be a case of overlap-anisotropy, since it corresponds to the system spontaneously choosing an axis ($x$ or $y$) in which the spins will all align, but not a particular direction ($+x$ or $-x$).

We study this question in a replica-symmetric ansatz with
\begin{equation}
\begin{split}
        \qx_{\alpha\beta}=(c_x \delta_{\alpha\beta}+q_x)/2,\\
        \qy_{\alpha\beta}=(c_y \delta_{\alpha\beta}+q_y)/2.
\end{split}
\end{equation}
In other words, the diagonal elements of the overlap matrix are $c_x+q_x,c_y+q_y$, while the off-diagonal elements are $q_x,q_y$. Normalization gives us the condition that $c_x+c_y+q_x+q_y=1$.

In the notation of Sec.~\ref{subsec:RSBAction}, we first need to calculate $g_0(X,Y)$. With the above anisotropic ansatz, this is
\begin{equation}
    g_0(X,Y)=\int_0^{2\pi}\frac{d\theta}{2\pi}\exp \left(X\cos\theta+Y\sin \theta+c_x\beta^2 \J^2 \cos^2\theta+c_y\beta^2 \J^2 \sin^2\theta\right),
\end{equation}
and $g_{-1}(0)$ is then given by
\begin{equation}
    \log g_{-1}(0)=\int_{-\infty}^\infty\int_{-\infty}^\infty \frac{dX'_idY'_i}{4\pi\beta^2\J^2\sqrt{q_xq_y}}\exp\left(-\frac{X'^2}{4\beta^2\J^2q_x}-\frac{Y'^2}{4\beta^2\J^2q_y}\right)\log \left(g_{0}(X'_i,Y'_i)\right).
\end{equation}
This results in a total action of
\begin{equation}\begin{split}
    -\beta \F_{\textrm{full}}/sn=-&\beta^2 \J^2\left(c_x^2+2q_xc_x+c_y^2+2q_yc_y\right)\\
    &+
    \int_{-\infty}^\infty\int_{-\infty}^\infty \frac{dX'_idY'_i}{4\pi\beta^2\J^2\sqrt{q_xq_y}}\exp\left(-\frac{X'^2}{4\beta^2\J^2q_x}-\frac{Y'^2}{4\beta^2\J^2q_y}\right)\log \left(g_{0}(X'_i,Y'_i)\right).
\end{split}\end{equation}
A numerical search for saddle points using the minimax principle \cite{Baldwin2023rtr,Chen2024fei} finds no overlap-anisotropic saddles (where $c_x\neq c_y$ or $q_x\neq q_y$) up to $\beta \J=16$. This provides evidence that the spin overlaps are always isotropic in the thermodynamic limit.

\section{Nonequilibrium mean-field simulations}
\label{sec:mf}
While the simulations described in Sec.~\ref{sec:pt} provide insight into equilibrium properties, our experiment is manifestly nonequilibrium. To account for the dissipation and the superradiant phase transition, we simulate the mean-field dynamics by unraveling the master equation into stochastic trajectories, each of which serves as a replica~\cite{Marsh2024ear}.

\subsection{Atom-only Lindblad master equation}
An atom-only Lindblad master equation is derived by adiabatically eliminating the cavity mode while taking into account the effect of cavity dissipation. We consider uniform loss of each mode at rate $2\kappa$, described by collapse operators $\sqrt{\kappa}\hat{a}_\mu$ for all modes $\mu$. Adiabatic elimination of the mode operators $\hat{a}_\mu$ using Eq.~\eqref{eq:adiabatic_mode} yields the master equation for the atomic density matrix $\rho$ written in the Gell-Mann basis (as introduced in Sec.~\ref{theory}),
\begin{equation}
    \frac{d}{dt}\rho = -i[\hat{H}_\text{atom-only},\rho]+\frac{\kappa}{|\Delta_C|}\sum_{ij=1}^n \Big( \big(J^{\mathrm{local}}_{ij}+J^{\mathrm{non}}_{ij}\big)D[\hat{\Lambda}_i^{(1)},\hat{\Lambda}_j^{(1)}] + \big(J^{\mathrm{local}}_{ij}-J^{\mathrm{non}}_{ij}\big)D[\hat{\Lambda}_i^{(4)},\hat{\Lambda}_j^{(4)}]  \Big),
\end{equation}
where $D[\hat{X},\hat{Y}]=2\hat{X}\rho \hat{Y}^\dag -\{Y^\dag X,\rho\}$ and $\hat{H}_\text{atom-only}$ is given by Eq.~\eqref{eq:HquantumKgellmann}. The Lindblad-Kossakowski matrices~\cite{Breuer2007tto} $J^c=J^{\mathrm{loc}}+J^{\mathrm{non}}$ and $J^s=J^{\mathrm{loc}}-J^{\mathrm{non}}$ are positive semidefinite, following from the outer product structure of the Green's function in Eq.~\eqref{eq:Greens}. We thus consider their eigenstate decompositions $J^{c/s} = \sum_{k=1}^n \lambda_k^{c/s} \mbf{v}_{c/s}^k(\mbf{v}_{c/s}^k)^\intercal$ with eigenvalues $\lambda_k^{c/s}\geq 0$ and eigenvectors $\mbf{v}_{c/s}^k$. This leads to the diagonalized master equation
\begin{equation}\label{eq:atomOnlyMaster}
    \frac{d}{dt}\rho = -i[\hat{H}_\text{atom-only},\rho]+ \sum_{k=1}^n \Big( D[\hat{C}_{ck}] + D[\hat{C}_{sk}] \Big),
\end{equation}
where the collapse operators $\hat{C}_{\alpha k}$ are given by
\begin{equation}
    \hat{C}_{c k}= \sqrt{\frac{\lambda_k^c \kappa}{|\Delta_C|} }\sum_{i=1}^n (\mathbf{v}^k_c)_i \hat{\Lambda}_i^{(1)},\quad \hat{C}_{s k}= \sqrt{\frac{\lambda_k^s \kappa}{|\Delta_C|} }\sum_{i=1}^n (\mathbf{v}^k_s)_i \hat{\Lambda}_i^{(4)}.
\end{equation}

We study the dynamics within individual experimental trials by unraveling the density matrix into quantum trajectories. Each trajectory describes a stochastic evolution of the system under continuous heterodyne measurement of the emitted cavity light. For a fixed disorder realization of the $J$ and $K$ matrices, each trajectory corresponds to an independent replica of the spin glass~\cite{Marsh2024ear}. The stochastic evolution generated by Eq.~\eqref{eq:atomOnlyMaster} for the unnormalized wavefunction is written in It\^o form as~\cite{Breuer2007tto}
\begin{equation}
    d\ket{\psi} = -i\hat{H}_\text{atom-only}\ket{\psi}dt-\sum_{k=1}^n\sum_{\alpha=c,s}\left(\tfrac{1}{2}\hat{C}_{\alpha k}^\dag \hat{C}_{\alpha k} dt -\ex{\hat{C}_{\alpha k}^\dag}\hat{C}_{\alpha k} dt -\hat{C}_{\alpha k}dW_{\alpha k}^* \right) \ket{\psi},
\end{equation}
where each $dW_{\alpha k}$ is the differential of an independent, complex Wiener process with unit variance. The corresponding differential equation for the normalized expectation value of an observable $\hat{A}$ is given by
\begin{equation}\label{eq:MF_EOM_general}
\begin{split}
    d\ex{\hat{A}} &= i\ex{[\hat{H}_\text{atom-only},\hat{A}]}dt + \sum_{k=1}^n\sum_{\alpha=c,s}\big( 2\ex{\hat{C}_{\alpha k}^\dag \hat{A} \hat{C}_{\alpha k}} - \ex{\hat{C}_{\alpha k}^\dag \hat{C}_{\alpha k} \hat{A}} - \ex{\hat{A}\hat{C}_{\alpha k}^\dag \hat{C}_{\alpha k}} \big)dt \\
    &\quad+\sum_{k=1}^n\sum_{\alpha=c,s}\left[\sqrt{2}\big( \ex{\hat{C}_{\alpha k}^\dag \hat{A}} - \ex{\hat{C}_{\alpha k}^\dag}\ex{ \hat{A}}\big)dW_{\alpha k} + \sqrt{2}\big( \ex{\hat{A}\hat{C}_{\alpha k}} - \ex{\hat{A}}\ex{\hat{C}_{\alpha k}}\big)dW_{\alpha k}^*\right]. 
\end{split}
\end{equation}

We now compute the first-order equations of motion for each spin using Eq.~\eqref{eq:MF_EOM_general}. The mean-field approximation is used to decouple expectation values of products as $\ex{\GM{\mu}_i\GM{\nu}_j}=\ex{\GM{\mu}_i}\ex{\GM{\nu}_j}$ in order to achieve a closed system of equations. This approximation becomes valid in the limit of large collective spins and away from the superradiant phase transition where quantum fluctuations could play a significant role. Our experiment uses collective spins of size approximately $2.3\times 10^5$ atoms, satisfying the first requirement. However, evolution through the superradiant phase transition is a key driving force in the spin organization and may not be fully captured at the mean-field level. Regardless, the mean-field dynamics provide a first-order approximation to the true spin dynamics of the experiment. The equations are given by
{\allowdisplaybreaks
\begin{align}\label{eqn:MF_EOM}
    d\ex{\GM{1}_i} &= 2\left(E_r+\frac{\Omega^2}{8\Delta_A}\right) \ex{\GM{2}_i}dt - \Big\{ \sum_{k=1}^n \ex{\hat{T}^s_{ik}} \ex{\GM{7}_i} +\mathrm{c.c.}\Big\}dt +\ex{\GM{7}_i}dS_{si} -J^s_{ii}\ex{\GM{1}_i}\frac{\kappa dt}{|\Delta_C|}\\
    d\ex{\GM{2}_i} &= -2\left(E_r+\frac{\Omega^2}{8\Delta_A}\right)\ex{\GM{1}_i}dt + \Big\{ \sum_{k=1}^n\Big[2\ex{\hat{T}^c_{ik}}\ex{\GM{3}_i} - \ex{\hat{T}^s_{ik}} \ex{\GM{6}_i} \Big] +\mathrm{c.c.}\Big\}dt -2\ex{\GM{3}_i}dS_{ci}+\ex{\GM{6}_i}dS_{si} \\&\quad-(4J_{ii}^c+J_{ii}^s)\ex{\GM{2}_i}\frac{\kappa dt}{|\Delta_C|} \\
    d\ex{\GM{3}_i} &= -\Big\{ \sum_{k=1}^n\Big[2\ex{\hat{T}^c_{ik}}\ex{\GM{2}_i} + \ex{\hat{T}^s_{ik}} \ex{\GM{5}_i} \Big] +\mathrm{c.c.}\Big\}dt +2\ex{\GM{2}_i}dS_{ci}+\ex{\GM{5}_i}dS_{si}\\&\quad-(4J_{ii}^c+J^s_{ii})\ex{\GM{3}_i}\frac{\kappa dt}{|\Delta_C|}-\sqrt{3}J_{ii}^s\ex{\GM{8}_i}\frac{\kappa dt}{|\Delta_C|}\\
    d\ex{\GM{4}_i} &= 2\left(E_r+\frac{\Omega^2}{8\Delta_A}\right)\ex{\GM{5}_i}dt +\Big\{ \sum_{k=1}^n \ex{\hat{T}^c_{ik}}\ex{\GM{7}_i} +\mathrm{c.c.}\Big\} dt -\ex{\GM{7}_i} dS_{ci}-J_{ii}^c\ex{\GM{4}_i}\frac{\kappa dt}{|\Delta_C|}\\
    d\ex{\GM{5}_i} &=-2\left(E_r+\frac{\Omega^2}{8\Delta_A}\right)\ex{\GM{4}_i} dt -\Big\{\sum_{k=1}^n \Big[\ex{\hat{T}^c_{ik}}\ex{\GM{6}_i} - \ex{\hat{T}^s_{ik}} \big(\ex{\GM{3}_i}+\sqrt{3}\ex{\GM{8}_i}\big) \Big] +\mathrm{c.c.}\Big\} dt \\&\quad +\ex{\GM{6}_i}dS_{ci}-\big(\ex{\GM{3}_i}+\sqrt{3}\ex{\GM{8}_i}\big)dS_{si} -(J^c_{ii}+4J^s_{ii})\ex{\GM{5}_i}\frac{\kappa dt}{|\Delta_C|} \\
    d\ex{\GM{6}_i} &= \Big\{\sum_{k=1}^n \Big[\ex{\hat{T}^c_{ik}}\ex{\GM{5}_i} + \ex{\hat{T}^s_{ik}} \ex{\GM{2}_i} \Big] +\mathrm{c.c.}\Big\} dt -\ex{\GM{5}_i}dS_{ci}-\ex{\GM{2}_i}dS_{si} -(J^c_{ii}+J^s_{ii})\ex{\GM{6}_i}\frac{\kappa dt}{|\Delta_C|} \\
    d\ex{\GM{7}_i} &= -\Big\{\sum_{k=1}^n \Big[\ex{\hat{T}^c_{ik}}\ex{\GM{4}_i} - \ex{\hat{T}^s_{ik}} \ex{\GM{1}_i} \Big] +\mathrm{c.c.}\Big\} dt +\ex{\GM{4}_i}dS_{ci}-\ex{\GM{1}_i}dS_{si} -(J^c_{ii}+J^s_{ii})\ex{\GM{7}_i}\frac{\kappa dt}{|\Delta_C|}\\
    d\ex{\GM{8}_i} &= -\sqrt{3}\Big\{\sum_{k=1}^n \ex{\hat{T}^s_{ik}}\ex{\GM{5}_i} +\mathrm{c.c.}\Big\} dt +\sqrt{3}\ex{\GM{5}_i}dS_{si}-J^s_{ii}(\sqrt{3}\ex{\GM{3}_i}+3\ex{\GM{8}_i})\frac{\kappa dt}{|\Delta_C|}.
\end{align}
}
Above, we use $\hat{T}^c_{ik} = J_{ik}^c\GM{1}_k + K_{ik}\GM{4}_k$ and $\hat{T}^s_{ik}=J_{ik}^s\GM{4}_k + K_{ik}\GM{1}_k$ for brevity. Finally, the stochastic steps $dS_{\alpha i}$, where $\alpha=c,s$, are given by
\begin{equation}
    dS_{\alpha i} = \sqrt{\frac{2\kappa}{|\Delta_C|}}\sum_{k=1}^n \sqrt{\lambda_k^\alpha }\Im[dW_{\alpha k}](\mathbf{v}^k_\alpha )_i.
\end{equation}

We numerically evolve the equations of motion given by Eq.~\eqref{eqn:MF_EOM} to simulate mean-field trajectories. The system is initialized near the normal state, for which $\ex{\GM{3}_i +\frac{1}{\sqrt{3}}\GM{8}_i}=4N_i/3$ for all $i$, where $N_i$ is the number of atoms in node $i$, while $\ex{\GM{k}_i}=0$ for all $k\neq3,8$. We consider uniform $N_i=N=2.3\times10^5$ atoms per node as measured experimentally. A small perturbation is added to each initial spin state, as is typically required in mean-field approaches to break symmetry. To perturb an initial spin state, note that any product state for an individual node may be described by a quantum state $\ket{\Psi_i}=\ket{\psi_i}^{\otimes N_i}$ where $\ket{\psi_i}$ is a three-dimensional quantum state vector describing a single spin within the node. The normal state is described by $\ket{\psi_i}=(1,0,0)^\intercal$. We perturb the normal state of each node by considering initial states $\ket{\psi_i}=(1-x_i,y_i,z_i)^\intercal$, where $x_i$, $y_i$, and $z_i$ are complex, normally distributed random variables of standard deviation 0.01. 

The spins are driven through the superradiant phase transition following the same pump schedule of the experiments, given by
\begin{equation}
    \Omega^2(t)=\begin{cases}
        \Omega_c^2 \left(\frac{t}{1~\text{ms}}\right) &\mbox{$0\leq t < 1.25 \text{ms}$}\\
        2.5\Omega_c^2 &\mbox{$1.25~\text{ms}\leq t \leq 1.25~\text{ms}+500$~$\mu$s.}
    \end{cases}
\end{equation}
Above, recall that $\Omega^2(t)$ is proportional to the transverse pump power and appears as a prefactor multiplying the $J^{c/s}$ and $K$ matrices. The quantity $\Omega^2_c$ is the semiclassical threshold pump power for the superradiant phase transition, given by the condition $\frac{1}{N}\left[2E_r+\frac{\Omega_c^2}{4\Delta_A}\right] = \max_i \lambda_i$, where $\lambda_i$ are the eigenvalues of the matrix $M$ given by Eq.~\eqref{eq:thresholdmatrix}. As in the experiment, we infer the final spin configuration by integrating the phase of the emitted cavity light during the last $500$~$\mu$s of the trajectory. Within the framework of this mean-field atom-only theory, the phase of the emitted light at each node is given by the (complex) phase of $\ex{\GM{1}_i}+i\ex{\GM{4}_i}$. See also Sec.~\ref{sec:Efield} for the form of the emitted field. Each trajectory yields a single spin configuration. We simulate 1000 trajectories per disorder instance to calculate an overlap distribution.

\begin{figure}[htbp]
    \centering
    \includegraphics[width=\textwidth]{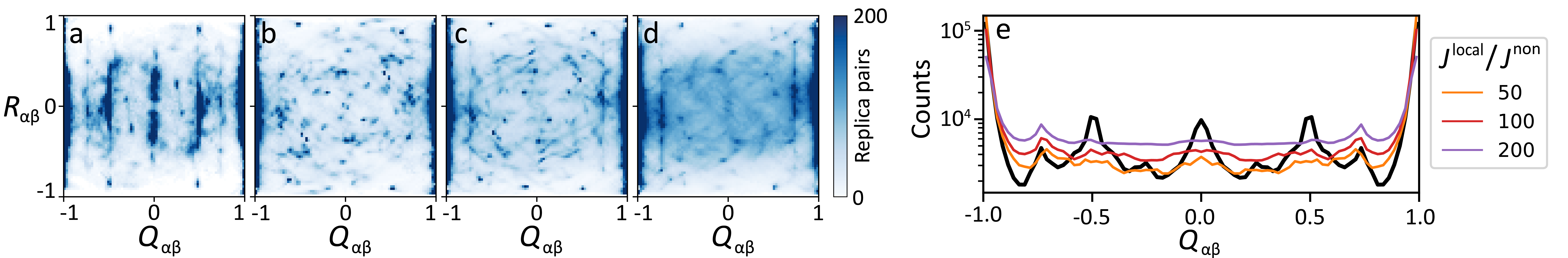}
    \caption{Strength of local interaction. (a) Experimental Parisi order parameter distribution, reproduced from Fig.~\ref{fig4}e. (b-d) Numerical simulations of Parisi order parameter distributions versus relative strength of the local interaction, $J^\text{local}/J^\text{non}$: (b) 50; (c) 100; (d) 200. Panel (c) is identical to Fig.~\ref{fig4}j. (e) Comparison of 1D marginal histograms for the distributions in (a-d). The black line is the experimental marginal from Fig.~\ref{fig4}k.}
    \label{fig:MFscale}
\end{figure}

The mean-field approximation allows for a tractable description of the nonequilibrium dynamics in the experiment. However, it does not capture well the level of adiabaticity when traversing the superradiant phase transition, as has been noted in simulations of the spin-$1/2$ limit of the system~\cite{Marsh2024ear}. We counteract this effect in our mean-field trajectories by allowing for a rescaling of a single global parameter: the relative strength of the local versus nonlocal cavity-mediated interaction, $J^\text{local}$ and $J^\text{non}$, respectively. Experimentally, the ratio of these interactions strengths is $J^\text{local}/J^\text{non}\approx 10$. To qualitatively match the experimental overlap distributions, we instead use $J^\text{local}/J^\text{non}=100$ for the simulation results shown in Fig.~\ref{fig4}. This value is chosen using a coarse-grained optimization to match the measured and simulated Parisi order parameter distributions, which we summarize in Fig.~\ref{fig:MFscale}. The Parisi order parameter distribution shows some dependence on the chosen ratio, though its general features remain the same. This is also demonstrated in the marginal distribution in Fig.~\ref{fig:MFscale}e: The general features, such as the goal posts and continuous support for $-1<Q<1$, remain present throughout. The Hellinger distance to the experimental marginal distribution (shown in black) is minimized for $J^\text{local}/J^\text{non}=100$ and motivates our choice for this ratio for all simulations.

\subsection{Evolution of the Parisi order parameter distribution}\label{sec:MF_Parisi}
\begin{figure}[t!]
    \centering
    \includegraphics[width=\textwidth]{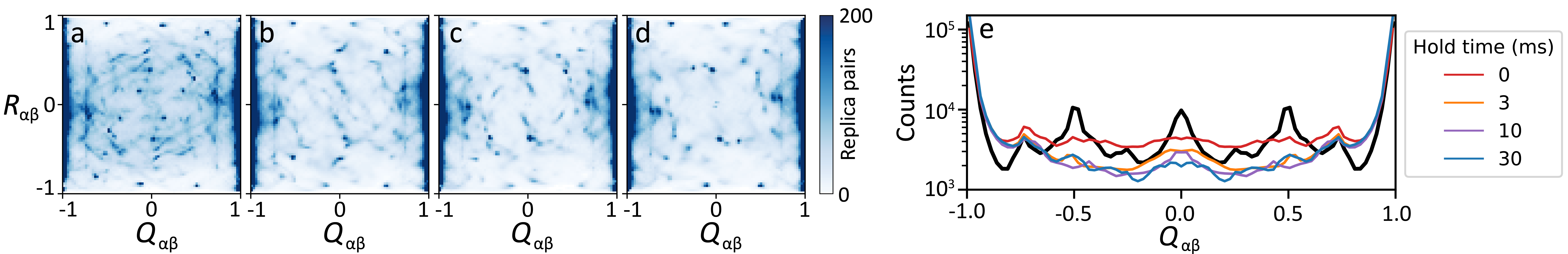}
    \caption{Evolution of the Parisi order parameter distribution in mean-field simulations. (a) Parisi order parameter distribution for the ramp schedule used in the experiment. This panel is identical to Fig.~\ref{fig4}j and Fig.~\ref{fig:MFscale}c. (b-d) Parisi order parameter distributions for a ramp schedule with an additional hold before spin readout. The additional holding time is: (b) 3~ms; (c) 10~ms; (d) 30~ms. (e) Comparison of 1D marginal histograms for the distributions in (a-d). The black line is the experimental marginal from Fig.~\ref{fig4}k.}
    \label{fig:MFstability}
\end{figure}

We additionally use mean-field trajectories to probe the stability of the Parisi order parameter distribution over longer time periods than explored experimentally. We simulate a modified ramp schedule where the system is held in the superradiant regime for a variable time before performing the quenched readout of the spin configuration. The modified ramp schedule is given by
\begin{equation}
    \Omega^2(t)=\begin{cases}
        \Omega_c^2 \left(\frac{t}{1~\text{ms}}\right) &\mbox{$0\leq t < 1.25~\text{ms}$}\\
        1.25 \Omega_c^2 &\mbox{$1.25~\text{ms}\leq t < 1.25~\text{ms} + t_\text{hold}$}\\
        2.5\Omega_c^2 &\mbox{$1.25~\text{ms} + t_\text{hold}\leq t \leq 1.25~\text{ms}+t_\text{hold}+500$~$\mu$s.}
    \end{cases}
\end{equation}
The simulations in Fig.~\ref{fig:MFscale} and in the main text correspond to $t_\text{hold}=0$~ms. The time evolution of the simulated Parisi order parameter distribution is shown in Fig.~\ref{fig:MFstability}a-d for $t_\text{hold}=0$, 3, 10, 30~ms, respectively. We observe no drastic change in the features of the overlap distribution: continuous support with interior peaks is retained for all hold times, providing evidence of a stable RSB phase. The marginal $Q$ distribution, shown in Fig.~\ref{fig:MFstability}e, allows for a more quantitative assessment of the evolution. Support in the $-1<Q<1$ interior is reduced by a factor $\sim$3 over the range of simulated hold times. Simultaneously, we observe that the average interaction energy goes down (not shown). These could be evidence of aging effects or dissipative cooling, the latter of which has been studied before in this confocal cavity system~\cite{Marsh2021eam}, and will be explored further in future work.

\subsection{Finite-size scaling}\label{MFfinitesize}
We now demonstrate that glassy overlap distributions persist to larger system size for the nonequilibrium, driven-dissipative spin glass.  (We had previously  shown this for the equilibrium model in Sec.~\ref{sec:pt}.) Additional mean-field simulations are performed for system sizes $n=8$, 16, 32, 64, and 128 to probe the nonequilibrium steady states of larger systems. For each $n$, 200 disorder realizations of SK $J$ and $K$ matrices are sampled as described in Sec.~\ref{sec:pt}, with the relative weight between local and nonlocal cavity-mediated interactions still fixed at $J^\text{local}/J^\text{non}=100$ as in Sec.~\ref{sec:MF_Parisi}. We perform 200 mean-field trajectories per disorder realization to sample final spin configurations and generate overlap distributions. The pump schedule is modified to ramp linearly to $2\Omega_c^2$ over a variable time $t_{\mathrm{ramp}}$ before again holding for a $500~\mu$s period during which the spin configuration is measured:
\begin{equation}
    \Omega^2(t)=\begin{cases}
        2\Omega_c^2 t/t_{\mathrm{ramp}} &\mbox{$0\leq t < t_{\mathrm{ramp}} $}\\
        2\Omega_c^2 &\mbox{$t_{\mathrm{ramp}} \leq t \leq t_{\mathrm{ramp}} +500$~$\mu$s.}
    \end{cases}
\end{equation}
The time taken to ramp through the transition affects the distribution of steady states found in the mean-field dynamics. Longer ramps produce steady states of lower energy on average, while shorter ramps lead to higher energy. We thus consider $t_{\mathrm{ramp}}=4$~ms and $t_{\mathrm{ramp}}=40$~ms to explore both regimes.

\begin{figure}[p]
    \centering
    \includegraphics[width=0.9\textwidth]{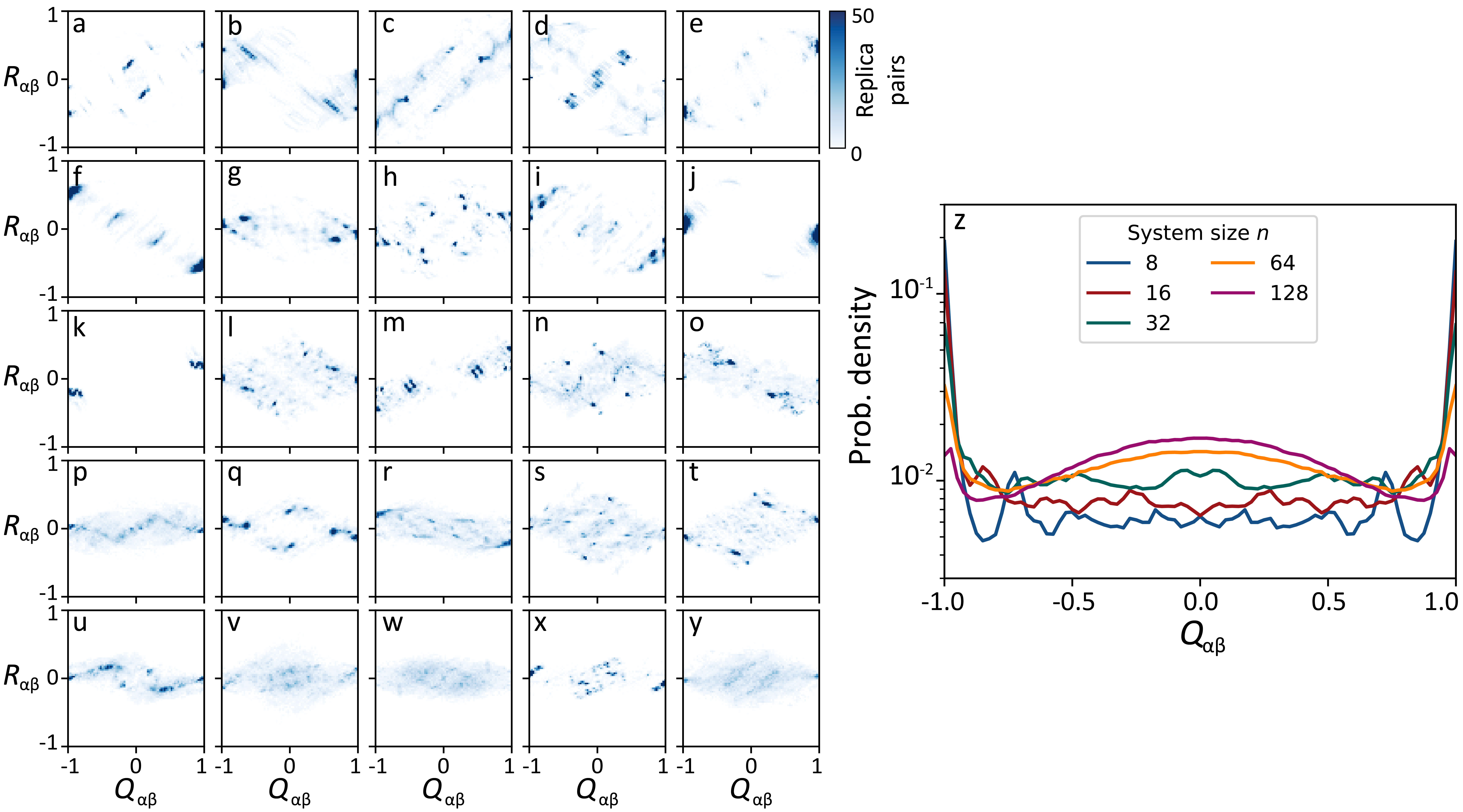}
    \caption{Finite-size scaling of the driven-dissipative overlap distribution, using mean-field trajectory simulations with a fixed ramp time $t_\text{ramp}=4$~ms. (a-y) Example overlap distributions for varying system sizes. Five random disorder instances are shown per system size: (a-e), $n=8$; (f-j), $n=16$; (k-o), $n=32$; (p-t), $n=64$; (u-y), $n=128$. (z) Comparison of the 1D marginal of the Parisi order parameter distribution for different system sizes.}
    \label{fig:MFscaling4}
\end{figure}

\begin{figure}[p]
    \centering
    \includegraphics[width=0.9\textwidth]{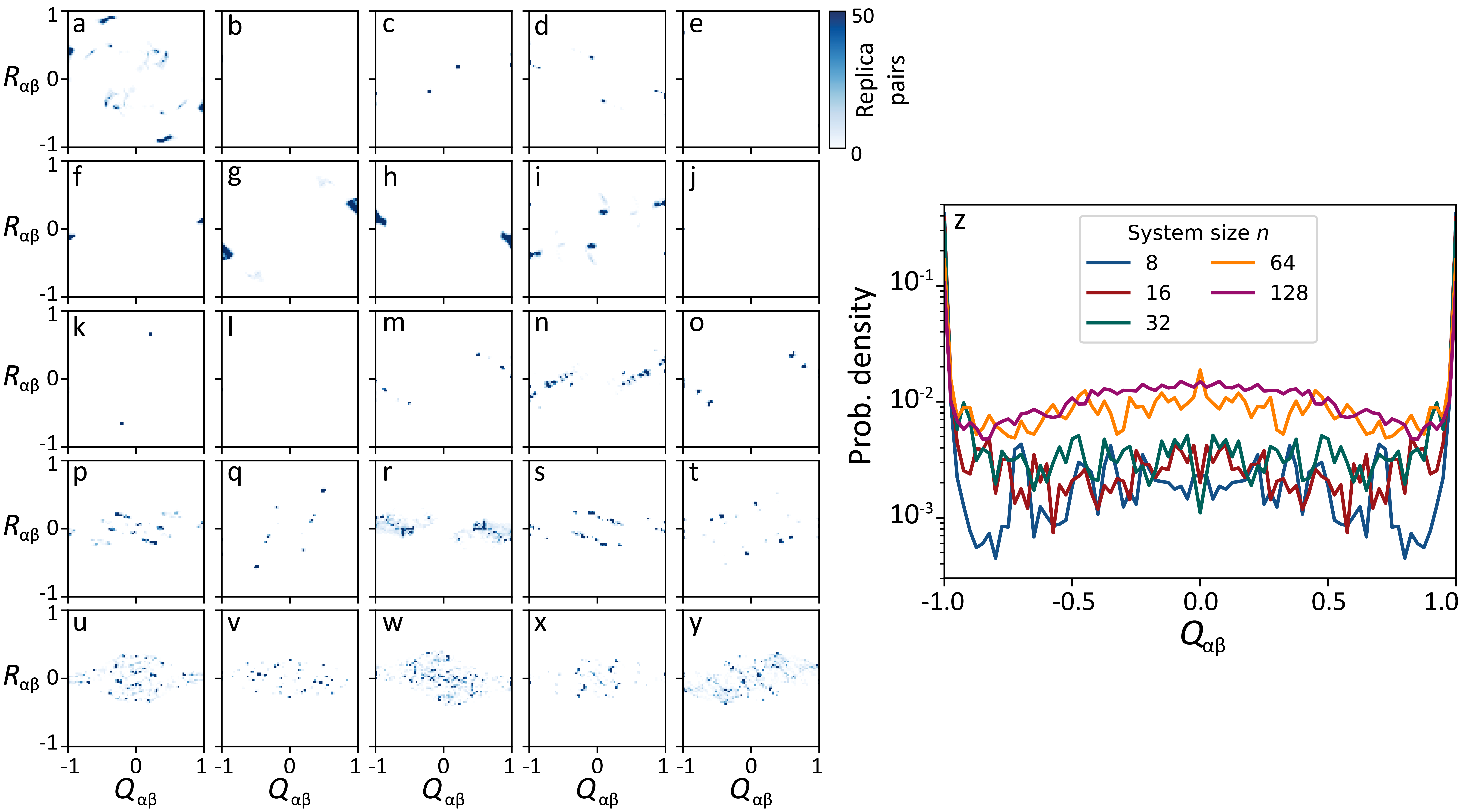}
    \caption{Finite-size scaling of the driven-dissipative overlap distribution, using mean-field trajectory simulations with a fixed ramp time $t_\text{ramp}=40$~ms. (a-y) Example overlap distributions for varying system sizes. Five random disorder instances are shown per system size: (a-e), $n=8$; (f-j), $n=16$; (k-o), $n=32$; (p-t), $n=64$; (u-y), $n=128$. (z) Comparison of the 1D marginal of the Parisi order parameter distribution for different system sizes.}
    \label{fig:MFscaling40}
\end{figure}

We first consider the overlap distributions for mean-field trajectories with $t_{\mathrm{ramp}}=4$~ms. Figure~\ref{fig:MFscaling4}a--y illustrates overlap distributions for five out of the two hundred disorder realizations for each system size studied. The primary indicator of RSB remains for all sizes: Interior peaks of the overlap distribution are found in random locations that vary between disorder realizations. For the largest system sizes, these peaks appear more blurred. This behavior can be understood from the expected proliferation of nearly degenerate low-energy spin configurations in larger systems. Due to shrinking gaps between low-energy configurations,  the degree of adiabaticity through the transition decreases for a fixed $t_{\mathrm{ramp}}$, and this leads to a higher energy density of the spin configurations. For each system size, the overlap distributions for all 200 disorder instances are combined into a Parisi distribution. Figure~\ref{fig:MFscaling4}z shows the 1D marginal for $Q$. These Parisi distributions show continuous support across the interior for all system sizes studied, again indicative of RSB. The marginal for $n=8$ is similar to that shown in Fig~4k of the main text, and displays the same finite-size effects in the form of secondary peaks at $Q=\pm0.75$. The boundary peaks (i.e., `goalposts') decrease in height with increasing system size. As with the individual overlap distributions, this can be understood from the decrease in energy gaps as system size grows. RSB persists to the largest system size studied.

As shown in Fig.~3 of the main text, a lower energy-density ensemble can be obtained by reducing the rate at which the pump power is increased through the transition. Figure~\ref{fig:MFscaling40} summarizes the results for a ten-fold slower ramp ($t_{\mathrm{ramp}}=40$~ms). All qualitative features of the overlap distribution remain: Interior peaks can be found in random locations for each disorder instance shown in Fig.~\ref{fig:MFscaling40}a--y. Comparing to Fig.~\ref{fig:MFscaling4}, for a fixed system size the interior peaks are sharper. For $n=8$, this can result in the interior peaks disappearing altogether, as also demonstrated in Fig.~3h of the main text. The marginals of the Parisi distributions are shown in Fig.~\ref{fig:MFscaling40}z for the system sizes studied. The key features of RSB remain: prominent goalposts with full support across the interior. For the smallest system size this support is more ragged, since overlap distributions typically have fewer interior peaks. Averaging over a larger number of disorder instances would yield a smoother distribution. Due to the plethora of low-energy states for the largest system size, the support is more continuous and higher (by approximately one order of magnitude). The continuous distribution is not perfectly flat but features a hump near $Q=0$. The exact shape of the Parisi distribution can reveal information about the distribution of glassy spin configurations, which here is clearly not of the form expected from the Gibbs measure.  That is not surprising, given the driven-dissipative nature of the simulation.  

Summarizing, signatures of RSB are thus found to be robust features of the mean-field nonequilibrium dynamics across system sizes and ramp times, each spanning an order of magnitude in scale. These dynamics are a close approximation to those realized experimentally, derived directly from the driven-dissipative quantum master equation. The mean-field results presented here thus provide direct evidence that the driven-dissipative spin glass phase persists to large system sizes and that RSB can be observed at the $n=8$ system size  studied experimentally. 

\section{Bootstrap error analysis}
\label{sec:boot}
The statistical error in magnetization and overlap distributions incurred from having a finite number of experimental replicas is estimated using a bootstrap analysis. We focus here on the overlap distributions; analysis of the magnetization distributions follows identically. Bootstrap samples for a given disorder realization of the $J$ and $K$ matrices are generated by sampling with replacement from the set of measured spin states. The size of the bootstrap sample is equal to the number of measured replica spin states; 900 replicas for the ferromagnetic data in Fig.~2 or 100 replicas per $J$ matrix for the data in Figs.~3 and 4. We then compute the 2D overlap distribution $p_J^{(k)}(Q,R)$ of that bootstrap sample to arrive at a bootstrapped distribution, where $k$ indexes the bootstrap samples. This distribution is compared to the overlap distribution $p_J(Q,R)$ generated directly from the measured set of replicas. Fluctuations in the distributions $p_J^{(k)}(Q,R)$ from one bootstrap sample $k$ to the next are used to determine a bootstrap error estimate on the measured distribution $p_J(Q,R)$. We use 100 bootstrap samples in the results that follow.  

A quantitative error estimate is generated using the Hellinger distance between the measured and bootstrapped overlap distributions. The Hellinger distance is
\begin{equation}
    d^{(k)}=1-\sum_{Q,R}\sqrt{p_J(Q,R)p_J^{(k)}(Q,R)},
\end{equation}
and is zero for identical, normalized distributions, and one for completely nonoverlapping distributions. This distance thus represents a fractional difference between the measured and bootstrapped overlap distributions. Hellinger distances depend on the number of bins in the distribution. Choosing a larger number of bins requires a greater number of measured replicas to achieve the same level of convergence to the large-system-size limit. We choose to use 80 bins per dimension for all overlap and magnetization distributions. This results in an estimated average convergence to within 5\% for the measured overlap distributions in Figs.~3 and 4, with a standard deviation of two percentage points over all $N_J=123$ disorder realizations of the $J$ matrix. The two ferromagnetic systems in Fig.~2, each with 900 replicas, are both estimated to have converged within 4\% of their overlap distributions and 2\% of their magnetization distributions.

Bootstrap analysis is also used to estimate the convergence of the aggregate overlap and magnetization distributions. While the distributions converge in the limit of many disorder realizations, our experimental ensemble contains a finite number of such realizations. 
The measured aggregate distribution in Fig.~4e of the main text is constructed to be the average over disorder realizations, $p(Q,R)=\sum_{i=1}^{N_J} p_{J_i}(Q,R)/N_J$. A bootstrap sample can be constructed by sampling with replacement $N_J$ times from the collection of $p_{J_i}(Q,R)$ distributions. That is, a bootstrap sample is generated from a random sample $B_k$ of size $N_J$ that contains possibly repeated disorder realization indices. The bootstrapped aggregate distribution is then constructed as $p^{(k)}(Q,R)=\sum_{i\in B_k}p_{J_i}(Q,R)/N_J$. Fluctuations in the $p^{(k)}(Q,R)$ distributions between bootstrap samples $k$ now indicate the level of convergence in the measured aggregate distribution. The Hellinger distance between measured $p(Q,R)$ and bootstrapped $p^{(k)}(Q,R)$ distributions is again used to extract a metric of convergence. We find that the aggregate overlap distribution of Fig.~4e is estimated to have converged to within $2\%$ of the distribution corresponding to an average over all possible disorder realizations. Similarly, the aggregate magnetization is estimated to be within $3\%$ of the full disorder-averaged distribution.     

\section{Effective temperature for ferromagnetic data}

The effective temperatures for the ferromagnetic systems shown in Fig.~2 of the main text are found via a Maxwell-Boltzmann fit using the simplified interaction energy in Eq.~\eqref{eq:angleOnly}. These are normalized using an estimated critical temperature $T_c$. We first describe the form of the Maxwell-Boltzmann distribution used for the fit before describing how $T_c$ is estimated from equilibrium numerical methods.  

\subsection{Derivation of Maxwell-Boltzmann distribution}

The Maxwell-Boltzmann distributions for the ferromagnetic systems are amenable to a low-temperature analysis by local expansion around its two (spin-flip-symmetry-related) basins of attraction in the free-energy landscape. We now derive its generic form by considering the density of states near these global minima. Taylor expansion of Eq.~\eqref{eq:angleOnly} near the global minimum configuration $\bm{\theta}_0$ leads to
\begin{equation}\label{eq:TaylorExp}
E(\bm{\theta}_0 + \bm{\epsilon}) = E_\text{gs} + \bm{\epsilon}^\intercal H \bm{\epsilon} + O(\bm{\epsilon}^3),
\end{equation}
where $E_\text{gs}=E(\bm{\theta}_0)$ is the ground state energy, and $H$ is the Hessian matrix evaluated at $\bm{\theta}_0$. This expansion is valid in proximity to the ground state, where $|\bm{\epsilon}|\ll 1$. This condition is satisfied when the temperature $T$ of the Maxwell-Boltzmann distribution is sufficiently small so that states far from the ground state are not populated.

\begin{figure}[b!]
    \centering
    \includegraphics[width=0.7\linewidth]{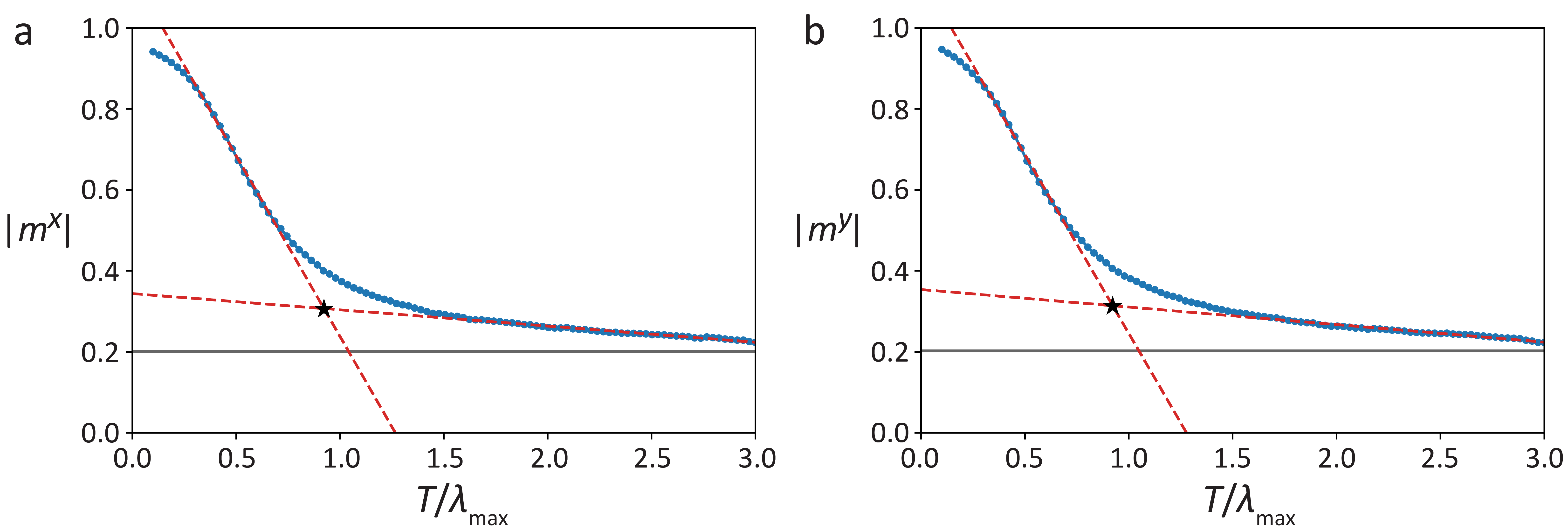}
    \caption{Estimation of $T_c$ for finite-sized \xFM and \yFM systems. (a) The absolute magnetization for the \xFM state as a function of temperature is found via parallel-tempering simulations. The temperature is normalized to the maximum magnitude eigenvalue of the measured $J$ matrix, $\lambda_\text{max}$. Two separate linear, least-squares fits are performed in the regions $T/\lambda_{\text{max}}=[0.4,0.7]$ and $[1.5,3]$, shown as red dashed lines. Their intersection, marked by a star, yields an estimated $T_c=0.92 \cdot \lambda_{\text{max}}$. The gray horizontal line shows the remnant magnetization ${\propto} 1/\sqrt{n}$ found in the paramagnetic phase for our finite-size system. (b) The same analysis applied to the \yFM system also yields the estimate $T_c=0.92 \cdot\lambda_{\text{max}}$ in terms of the maximum eigenvalue of the measured \yFM $J$ matrix.}
    \label{fig:Tc}
\end{figure}

We now calculate the density of states at a fixed energy level for a general $n$-vertex spin network system described by Eq.~\eqref{eq:TaylorExp}. Integration over the configuration space at a given energy $E$ yields
\begin{equation}
    g(E) = \int  \delta\left(E - E_\text{gs} - \bm{\epsilon}^\intercal H \bm{\epsilon}\right) d^n \bm{\epsilon}
    = \frac{(E - E_\text{gs})^{n/2 - 1}}{\sqrt{\det(H)}} \int \delta\left(1 - \bm{u}^\intercal \bm{u}\right) d^n \bm{u}, 
\end{equation}
where we used the fact that the Hessian is a positive definite matrix to perform a change of the integration variable. The resulting integral calculates the surface area of an $n$-dimensional unit sphere, which is
\begin{equation}
    \int \delta\left(1 - \bm{u}^\intercal \bm{u}\right) d^n \bm{u} = \frac{2\pi^{n/2}}{\Gamma(n/2)}.
\end{equation}
The density of states is thus given by
\begin{equation}
    g(E) = \frac{2}{\Gamma(n/2)}\sqrt{\frac{\pi^n}{\det(H)}} (E - E_\text{gs})^{n/2 - 1}.
\end{equation}
The Maxwell-Boltzmann distribution at temperature $T$ is now found by weighting the density of states by the associated Boltzmann-factor and normalizing the distribution:
\begin{equation}
    f(E, T)
    = \frac{(E - E_\text{gs})^{n/2 - 1}}{\Gamma(n/2)T^{n/2}} e^{- (E - E_\text{gs})/T}.
\end{equation}
This is the functional form used in the fits of Fig.~2 of the main text to extract the temperature of the ferromagnetic systems.

\subsection{Estimation of the ferromagnetic $T_c$}

The critical temperature of the paramagnet-to-ferromagnet transition is estimated via parallel-tempering Monte Carlo using the energy in Eq.~\eqref{eq:angleOnly}. We compute the average absolute $m^x$ and $m^y$ magnetizations after the local gauge transformations described in Sec.~\ref{sec:gauge}. The magnetizations serve as an order parameter as the temperature is varied. A nonanalytic kink in the magnetization occurs at $T_c$ in the thermodynamic limit. For our finite-size system, there is only a crossover region between two approximately linear regimes. We estimate the crossover temperature $T_c$ by performing linear fits on either side of the crossover region and finding the intersection point of the fit results. This would occur at $T_c$ in the large-$n$ limit. The fit results in Fig.~\ref{fig:Tc} show that both the \xFM and \yFM have an estimated crossover temperature $T_c=0.92\lambda_{\text{max}}$, where $\lambda_{\text{max}}$ is the magnitude of the eigenvalue of the measured $J^{\text{non}}$ matrix with largest absolute value. The dependence on the maximum eigenvalue is standard for mean-field models~\cite{Cherrier2003rot}, as this eigenvalue sets the overall energy scale. The appearance of the absolute value reflects the interaction form $S_i^xS_j^x - S_i^yS_j^y$: For \yFM there is an additional minus sign which must be taken into account. We conservatively estimate 10\% uncertainty in this estimate of $T_c$,  accounting for the fact that this is a crossover. The error reported in $T/T_c$ in the main text is a combination of this uncertainty in $T_c$ and the fit uncertainty of $T$. 

\section{Experimental ensemble of $J$-matrices}

We now discuss the ensemble of $J$-matrices used for the disorder average in Fig.~4e. First, we ensure no two $J$-matrices are identical. Since the $J$ (and $K$) matrices are functions of vertex position, we directly compare position configurations between disorder instances. As long as at least one vertex location is different between configurations, the resulting $J$ matrices will be distinct. For each pair of configurations we thus find the most displaced vertex and note its displacement. Across all configuration pairs, these displacements are larger than $4.5$~$\mu$m$>\sigma_A$. This shows that none of the disorder realizations are identical.

\begin{figure}[t!]
    \centering
    \includegraphics[width=0.7\textwidth]{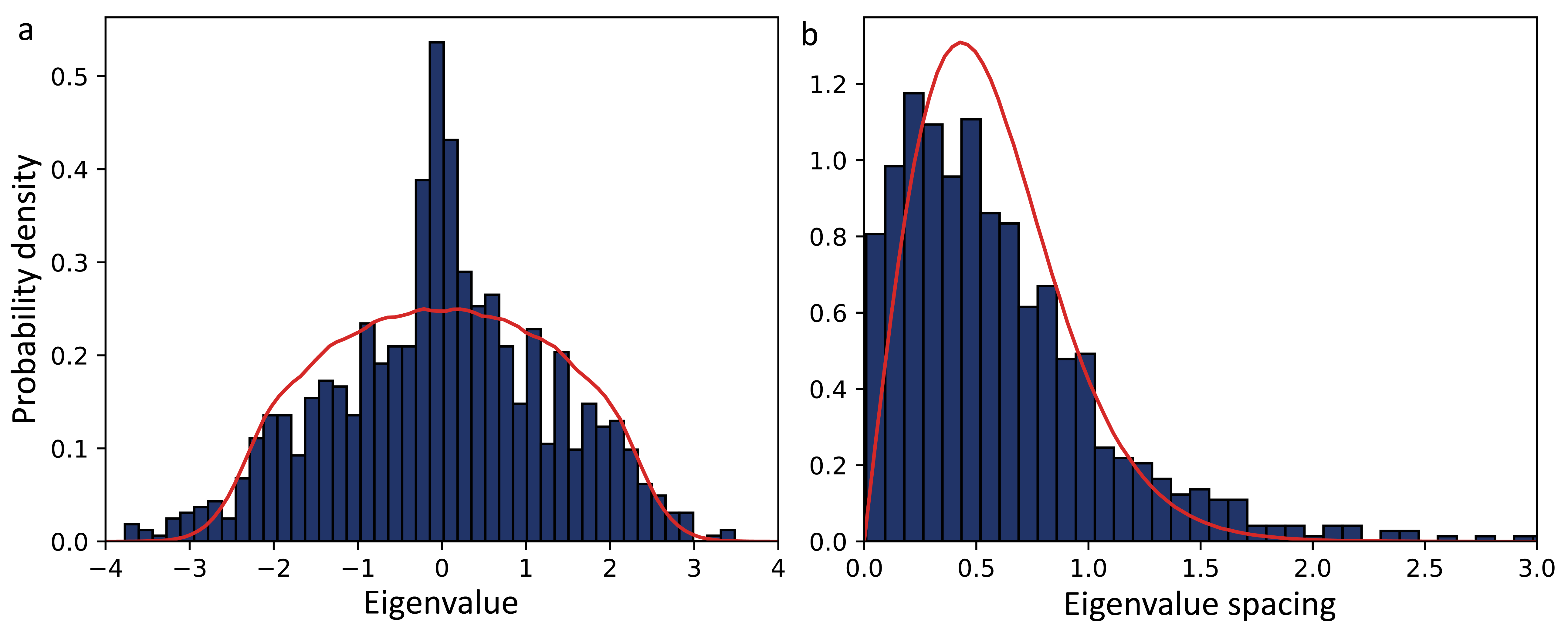}
    \caption{Experimental ensemble of interaction matrices. (a) Eigenvalue spectrum and (b) level-spacing statistics for the $J$-matrices calculated from the observed position configurations. Red lines show the expected results for a fully random ensemble, corresponding to $n=8$ finite-size versions of the Wigner semicircle and Wigner surmise, respectively.}
    \label{fig:jensemble}
\end{figure}

To assess the randomness of the ensemble, we calculate the eigenvalue spectrum and level-spacing distribution for the ensemble of interaction matrices. These are shown in Fig.~\ref{fig:jensemble}, together with an 8-by-8 symmetric Gaussian orthogonal ensemble for comparison. The experimental ensemble is close to fully random, with the most notable feature being an over-representation of eigenvalues near zero. This is a feature of the confocal cavity interaction, see  Ref.~\cite{Marsh2021eam} for an in-depth study.

\section{Quantitative analysis of ultrametricity}

To provide evidence for emergent ultrametricity, we analyze the configuration-space structure of the experimental ensemble of states. For ultrametric spaces, the \textit{strong} triangle inequality holds: $D_{\alpha\beta}\leq \max\{D_{\alpha\gamma},D_{\beta\gamma}\}$ for all $\alpha,\beta,\gamma$. This can be quantitatively assessed by calculating the $\mathcal{K}$-correlator~\cite{Katzgraber2009uac}, which is defined for any triplet of replicas as $\mathcal{K} = \left(D_\text{max} - D_\text{med}\right)/\sigma(D)$, where $D_{\text{max}(\text{med})}$ is the maximum (median) of the three distances between replicas, and $\sigma(D)$ is the standard deviation of the distribution of all distances. For an ultrametric space, the probability distribution of this correlator is $p(\mathcal{K})=\delta(\mathcal{K})$. Finite-system sizes introduce some probability for finding small but nonzero values of $\mathcal{K}$, and ultrametricity in such systems has been studied by performing system-size scaling~\cite{Katzgraber2009uac,Katzgraber2012upo}.

\begin{figure}
    \centering
    \includegraphics[width=0.7\textwidth]{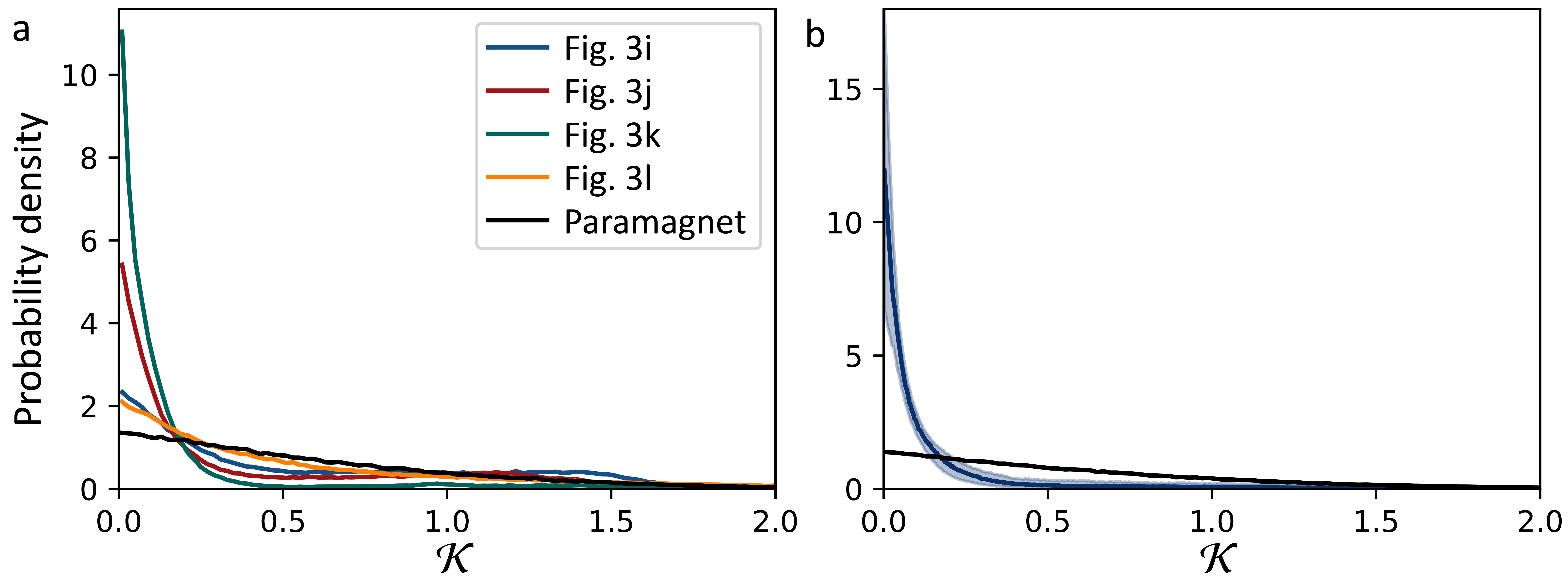}
    \caption{Quantitative analysis of ultrametric structure. (a) Probability distribution for the $\mathcal{K}$-correlator for the ensembles in Fig.~3 of the main text. A paramagnetic (non-ultrametric) ensemble is shown for comparison. (b) Average probability distribution for the $\mathcal{K}$-correlator across all disorder instances. The shaded area indicates the interquartile range around the median (solid line).}
    \label{fig:ultrametricity}
\end{figure}

Figure~\ref{fig:ultrametricity}a presents the probability distributions for the experimental ensembles in Fig.~3 of the main text. All triplets of replicas are included in the distribution. As a reference for finite-system-size effects, we show $p(\mathcal{K})$ for a paramagnet, i.e., a phase where each state is completely random, sampling 100 states. For the fastest ramp (blue) the system is at high temperature and the distribution is qualitatively comparable to that of the paramagnet. For intermediate ramp rates (red and green) we observe a drastic departure from paramagnetic behavior: The distribution is much stronger peaked at zero, and values of $\mathcal{K}$ greater than $0.5$ are significantly suppressed. The slowest, and hence coldest, ensemble (orange) is a single thermodynamic state. Because it does not sample much of the configuration space, its $p(\mathcal{K})$ looks like a paramagnet's. As a single figure-of-merit, we quote the mean $\langle \mathcal{K}\rangle$: This is $0.57$ for the paramagnet, and (i) $0.56$, (j) $0.39$, (k) $0.16$, and (l) $0.63$ for the experimental ensembles in those panels of Fig.~3.  

We perform a similar analysis of all 123 disorder instances included in this work. To show the aggregate behavior, Fig.~\ref{fig:ultrametricity}b plots the median (solid blue line) and the interquartile range (shaded area) across all probability distributions. Again the paramagnet is shown for comparison (black line), highlighting the departure of our experimental ensembles from that state with no ultrametric structure. The mean of $\mathcal{K}$ is $\langle \mathcal{K}\rangle=0.22$, with a standard deviation of $0.14$ over the different disorder instances (cf.~the paramagnetic value of $\langle \mathcal{K}\rangle=0.57$).

\begin{figure}
    \centering
    \includegraphics[width=\textwidth]{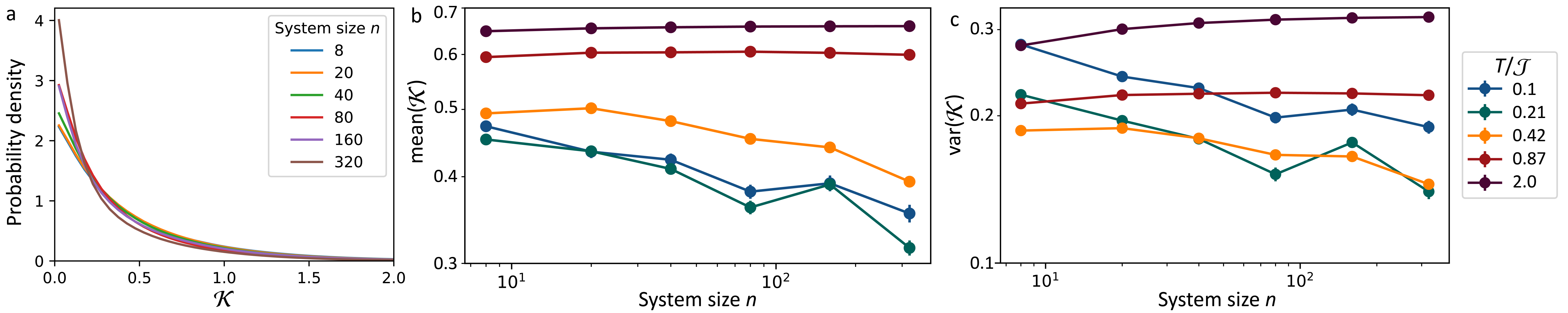}
    \caption{Numerical equilibrium finite-size scaling analysis of ultrametricity via the $\mathcal{K}$-correlator. (a) Probability distributions for the $\mathcal{K}$-correlator for varying system size at $T/\J=0.21$ from parallel tempering Monte Carlo. (b,c) Scaling of the (b) mean and (c) variance of $\mathcal{K}$ with system size. At low temperatures, the downward sloping data suggest a divergence for $\mathcal{K}\rightarrow0$. Error bars follow from bootstrap analysis. }
    \label{fig:ultrametricity_scaling}
\end{figure}

Though the experimental evidence of ultrametricity is limited to $n=8$, we can perform a numerical system-size scaling using the parallel-tempering Monte-Carlo framework described in Sec.~\ref{sec:pt}. We follow the prescriptions from Ref.~\cite{Katzgraber2009uac}: After equilibrating the system for $10^4 n$ Monte-Carlo sweeps, we record its spin configuration every $100 n$ further sweeps for a total of 1000 configurations. A probability distribution for the $\mathcal{K}$-correlator is produced for each disorder instance by randomly sampling $10^5$ triplets of states, and we use 500 disorder instances per system size. Figure~\ref{fig:ultrametricity_scaling}a shows example distributions for $\mathcal{K}$ for six system sizes, all at $T/\J=0.21$ and indicates a trend to an increasingly sharp distribution at $\mathcal{K}=0$.

To perform more quantitative finite-size-scaling analysis we reduce each $p(\mathcal{K})$ distribution to a mean and variance. These quantities are shown in Fig.~\ref{fig:ultrametricity_scaling}b,c as function of system size, for various temperatures. For high temperatures, where the system realizes a paramagnet, no ultrametricity is expected and neither the mean nor variance go to zero as system size is increased. At low temperatures, where there is RSB, both the mean and variance gradually decrease, suggesting that $p(\mathcal{K})\rightarrow\delta(\mathcal{K})$ in the thermodynamic limit, exactly as expected for an ultrametric space. The decrease is not strictly monotonic at the coldest temperatures, which could potentially be caused by  a lack of thermal equilibrium in the simulations at the largest system sizes. We note that the demonstrated slow convergence to the large-system-size limit is similar to results found in other numerical studies of ultrametricity; see, e.g., Refs.~\cite{Katzgraber2009uac,Katzgraber2012upo}, which also found a 20-30\% decrease in the variance per decade in system size.

\begin{figure}
    \centering
    \includegraphics[width=0.7\textwidth]{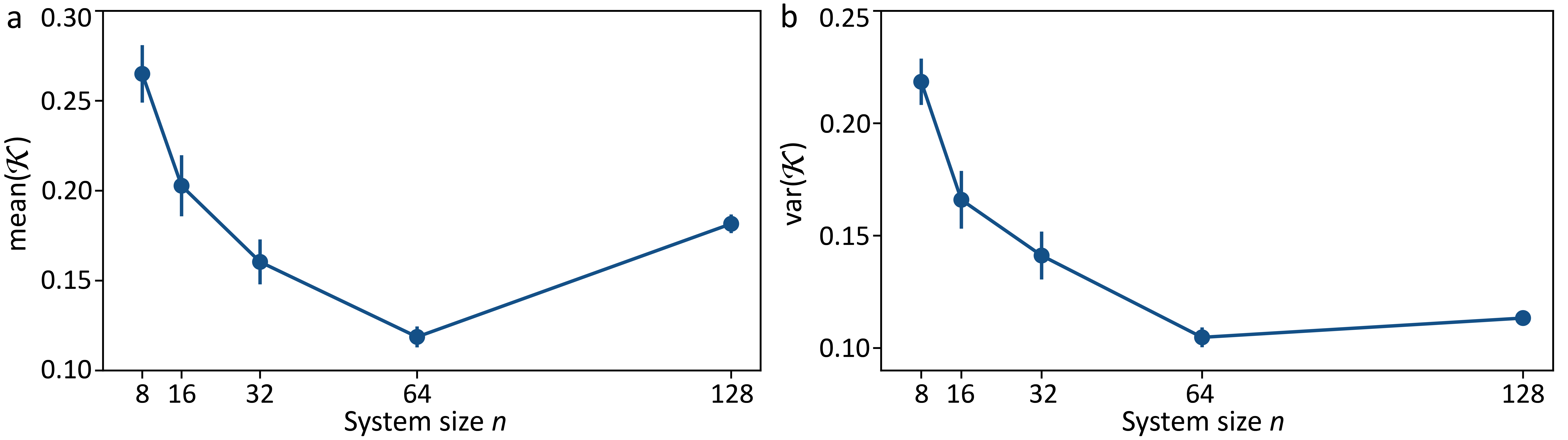}
    \caption{Numerical nonequilibrium finite-size scaling analysis of ultrametricity using the mean-field trajectory simulations.  (a,b) Scaling of the (a) mean and (b) variance of $\mathcal{K}$ with system size. The downward trending data suggest a divergence for $\mathcal{K}\rightarrow0$, as expected for an ultrametric system. Error bars follow from bootstrap analysis.}
    \label{fig:mf_ultrametricity_scaling}
\end{figure}

Beyond this equilibrium finite-size scaling of ultrametricity, we can also analyze ultrametricity using the inherently nonequilibrium mean-field trajectory simulations described in Sec.~\ref{sec:mf}. Each individual trajectory yields one spin configuration, and we calculate the $\mathcal{K}$-correlator for all triplets of configurations. We focus on the configurations produced by the mean-field simulations of Sec.~\ref{MFfinitesize} for the slowest ramp rate ($t_\text{ramp}=40$~ms), and use all 200 disorder instances to assemble $p(\mathcal{K})$ for each system size. As with the equilibrium simulations above, we reduce the probability distributions to their mean and variance. The results are shown in Fig.~\ref{fig:mf_ultrametricity_scaling}a,b and display a downward trend as system size increases, as expected of a system hosting RSB. Non-monotonicity for the largest system size can be explained by the effect noted in Sec.~\ref{MFfinitesize}: While the energy density is fixed by the ramp rate, the density of low-lying states grows as the system size is increased.  A much slower ramp is needed for the largest simulated size to further reduce the mean and variance of the K distribution. Nonetheless these results  further indicate that ultrametricity persists in the large system-size limit of this driven-dissipative system.

\begin{figure}[p]
    \centering
    \includegraphics[width=0.9\textwidth]{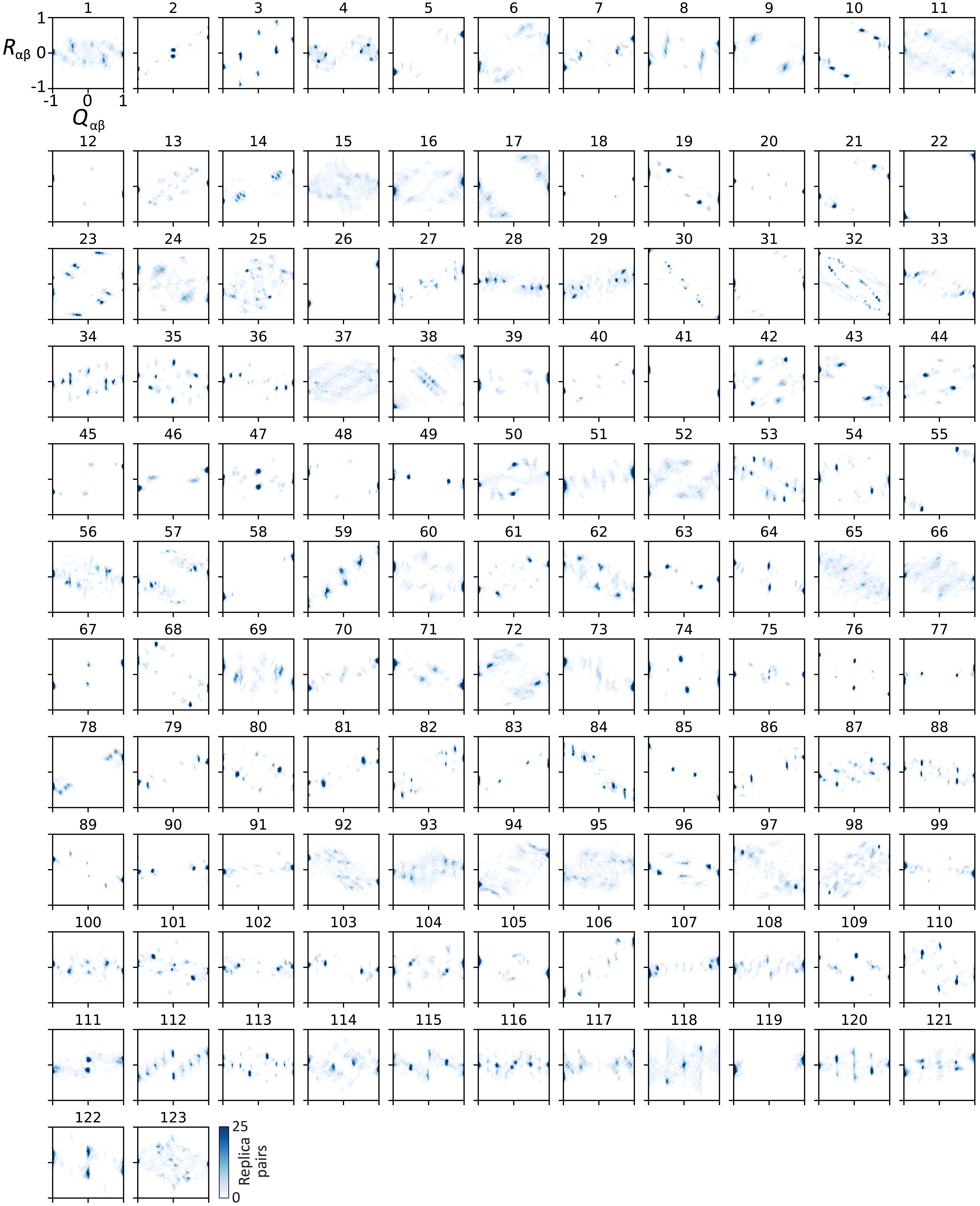}
    \caption{All experimentally measured spin-overlap distributions.}
    \label{fig:alloverlaps}
\end{figure}

\section{Comprehensive set of spin overlaps}
\label{sec:alloverlaps}
Figure~4e of the main text presents the experimental Parisi order parameter distribution from 123 disorder instances.  The spin-overlap distribution for each of these instances is shown in  Fig.~\ref{fig:alloverlaps}. Of these, panels 1 through 4 are presented in the main text as Figs.~4a--d, respectively.  Note that we ignore overlap components $q^{xy}_{\alpha\beta}$ and $q^{yx}_{\alpha\beta}$ throughout because we believe they play no role as order parameters.  Recall that each disorder instance is realized by choosing a new set of positions for the atomic vertices, leveraging the control discussed in Sec.~\ref{sec:bec}. All sets of positions and the resulting interaction matrices can be found in the data repository.

\end{document}